\documentclass{aa}  

\usepackage{graphicx}
\usepackage{txfonts}
\usepackage{amsmath}
\usepackage[normalem]{ulem}
\usepackage{subcaption}
\usepackage{hyperref}
\usepackage{xcolor}
\usepackage{ulem}
\usepackage{dblfloatfix}
\usepackage[export]{adjustbox}

\def\rsun{{~R}_{\odot}}
\def\msun{{~M}_{\odot}}
\def\zsun{{~Z}_{\odot}}

\newcommand{\kms}{\ensuremath{\,\rm{km}\,\rm{s}^{-1}}}

\newcommand{\Msun}{\ensuremath{\rm \,M_\odot}}

\newcommand{\Zsun}{\ensuremath{\,Z_\odot}}

\bibpunct[; ]{(}{)}{;}{a}{}{;}

\newcommand{\logteff}{\ensuremath{\,{\rm log} (T_{\rm eff} / {\rm K})}}
\newcommand{\logL}{\ensuremath{\,{\rm log} (L / L_{\odot})}}
\newcommand{\yr}{\ensuremath{\,\rm yr}}

\usepackage{array}
\newcolumntype{L}[1]{>{\raggedright\let\newline\\\arraybackslash\hspace{0pt}}m{#1}}
\newcolumntype{C}[1]{>{\centering\let\newline\\\arraybackslash\hspace{0pt}}m{#1}}
\newcolumntype{R}[1]{>{\raggedleft\let\newline\\\arraybackslash\hspace{0pt}}m{#1}}

\usepackage{hyperref}
\hypersetup{
    colorlinks = true,
    citecolor ={blue}
}

\begin{document} 

   \title{Massive donors in interacting binaries: effect of metallicity}

   \author{Jakub Klencki\inst{1}
          \and
          Gijs Nelemans\inst{1,2,3}      
          \and
          Alina G. Istrate\inst{1}
          \and 
          Onno Pols\inst{1}
   }

   \institute{
   Department of Astrophysics/IMAPP, Radboud University, P O Box 9010, NL-6500 GL Nijmegen, The Netherlands\\
          \email{j.klencki@astro.ru.nl}
         \and
          Institute of Astronomy, KU Leuven, Celestijnenlaan 200D, B-3001 Leuven, Belgium
         \and
         SRON, Netherlands Institute for Space Research, Sorbonnelaan 2, NL-3584 CA Utrecht, The Netherlands
   }
   \date{Received Sep 12, 2018; accepted ???}

  \abstract
   { 
   
   Metallicity is known to significantly affect the radial expansion of a massive star: the lower the metallicity, the 
more compact the star, especially during its post-main sequence evolution. Our goal is to study this effect in the 
context of binary evolution.
 
   Using the stellar-evolution code MESA, we computed evolutionary tracks of massive stars at six different metallicities 
between $1.0\zsun$ and $0.01\zsun$. We explored variations of factors known to affect the radial expansion of massive 
stars (e.g., semiconvection, overshooting, or rotation). Using observational constraints, we find support for
an   evolution in which already at a metallicity $Z \approx 0.2 \zsun$ massive stars remain relatively compact ($\sim 100 
\rsun$) during the Hertzprung-gap (HG) phase and most of their expansion occurs during core-helium burning (CHeB).
   
   Consequently, we show that metallicity has a strong influence on the type of mass transfer evolution in binary 
systems.  At solar metallicity, a case-B mass 
transfer is initiated shortly after the end of the main sequence, and a giant donor is almost always a rapidly expanding HG star. 
However, at lower metallicity, the parameter space for mass transfer from a more evolved, slowly expanding CHeB star 
increases dramatically. This means that envelope stripping and formation of helium stars in low-metallicity environments 
occurs later in the evolution of the donor, implying a shorter duration of the Wolf-Rayet phase (even by an order 
of magnitude) and higher final core masses. This metallicity effect is independent of the effect of 
metallicity-dependent stellar winds. 
   
   At metallicities $Z \leq 0.04 \zsun$ , a significant fraction of massive stars in binaries with periods longer than $100$ 
days engages in the first episode of mass transfer very late into their evolution, when they already have a well-developed CO core.
The remaining lifetime ($\lesssim 10^4$ yr) is unlikely to be long enough to strip the entire H-rich 
envelope. Cases of unstable mass transfer leading to a merger would produce CO cores that spin fast at the 
moment of collapse.
   
   We find that the parameter space for mass transfer from massive donors ($> 40 \msun$) with outer convective envelopes 
is extremely small or even nonexistent. We briefly discuss this finding in the context of the formation of binary black 
hole mergers. 

   }
   
   \maketitle

\section{Introduction}
\label{sec:intro}
 
Massive stars ($> 8 \msun$) play a vital role in the evolution of the Universe and various branches of astronomy. 
They provide feedback and chemical enrichment in star-forming galaxies and stellar clusters, they are responsible for 
supernovae and nucleosynthesis of heavy elements, and they lead to the formation of neutron stars and black 
holes \citep[e.g.,][]{Langer2012}. One key factor that significantly affects the evolution and fate of massive stars is 
wind mass-loss \citep{Smith2014}. In particular, the strength of line-driven winds of hot massive stars is 
metallicity dependent, as is indicated by observations and predicted theoretically \citep{Puls1996,Vink2001,Puls2008}. This 
introduces 
a relation between metallicity and the evolution of massive stars. As a result, it is believed that certain 
stellar-origin phenomena are predominantly or even exclusively associated with low-metallicity environments, for 
instance, massive stellar black holes \citep[BHs,][]{Zampieri2009,Mapelli2009,Belczynski2010,Spera2015}, single stellar 
origin long gamma ray bursts \citep[GRBs,][]{Yoon2006}, chemically homogeneous stars 
and binaries \citep{deMink2009,deMink2016,Marchant2017}, or pair-instability supernovae 
\citep{Heger2002,Heger2003,Farmer2019}. Metallicity is likely to play a role in the formation efficiency of 
compact binary mergers and gravitational-wave sources \citep[e.g.,][]{Klencki2018b,Giacobbo2018,Chruslinska2019}

Metallicity also influences stellar radii.
On the main sequence (MS), stars at low metallicity 
remain more compact than their higher metallicity counterparts \citep[the difference can be especially 
large for stars with masses $ \gtrsim 50 \msun$;][]{Sanyal2017}. The sizes of more evolved post-MS 
massive stars can be influenced by their metallicity even more significantly \citep{Brunish1982,Baraffe1991}. When 
 stellar evolution tracks are compared
across different metallicities, increasingly more massive-star models remain 
relatively compact during the phase of rapid Hertzprung-Gap (HG) expansion as the metallicity ($Z$)
decreases, see for example, \citet[][models at $Z = 
0.0088$, $0.0047$, and $0.0021$]{Brott2011}, \citet[][$Z = 0.002$]{Georgy2013}, \citet[][$Z = 0.0004$]{Groh2019}, 
\citet[][$Z = 0.0034$]{Schootemeijer2019}, or \citet[][$Z = 0$, Pop-III ]{Marigo2001}. Such stars 
regain thermal equilibrium and start the core-helium  burning phase (CHeB) as B- or A-type blue supergiants (BSGs, $R 
\sim 100 \rsun$) and typically reach the sizes of red supergiants (RSGs, $R \sim 1000 \rsun$) only at the final stages 
of CHeB or during even more advanced evolutionary phases. 

The way in which a star changes its radius during its evolution is especially important when it is a member of a binary 
system: any Roche-lobe overflow (RLOF) and mass-transfer phase is usually associated with a phase of expansion of 
the donor star. The vast majority of massive stars are formed in binary and higher-order systems 
\citep{Sana2012,Duchene2013}, in which, more often than not, they will at some point engage in RLOF \citep{Sana2013}. For 
this reason, it is essential to place the evolution of single massive stars into the binary perspective and to analyze the 
models from the point of view of such stars becoming donors in mass transfer episodes. 
Because of the effect of metallicity on the radial evolution of stars, two binary systems at different metallicities 
with otherwise identical initial parameters can enter mass transfer at very different ages, with the donor 
stars being at different evolutionary stages and leading to different fates. For instance, \citet{deMink2008} pointed out 
that the small sizes of HG stars at very low metallicity ($Z = 0.00001$) can allow for many more cases of case C mass 
transfer when the donor stars have well-developed carbon-oxygen (CO) cores. This evolutionary route may be required 
in some binary formation scenarios of long GRBs \citep[see the discussion in][and Sec.~\ref{sec.disc_SN_GRB_BBH}]{Wolf2007}. 
\citet[][and in much more detail in Klencki et al. in prep.]{Klencki2019} suggested that mass transfer from a 
slowly expanding CHeB donor (typical for subsolar metallicity) can relax to a nuclear-timescale evolution instead of 
stripping the entire envelope of the giant donor on a much shorter thermal timescale, as is the case of HG donors 
\citep[typical for solar metallicity,][]{Kippenhahn1967}.

In this paper we argue that because of the effect it has on the stellar radii, metallicity can have significant 
implications for the evolution of massive binaries at various orbital periods and deserves a detailed study. To this 
end, we employ the MESA stellar evolution code \citep{Paxton2011,Paxton2013,Paxton2015} to compute stellar tracks of 
massive stars between $10$ and 
$80\msun$ for six metallicities between $Z = 0.017 = \zsun$ and $Z = 0.00017 = 0.01 \zsun$.
We investigate several variations in the input physical parameters that are known to affect the radial expansion of 
massive stars (e.g., semiconvection, overshooting, and rotation) and constrain the models by observations of supergiants in 
the Small Magellanic Cloud (SMC) and in the Milky Way.

The paper is organized as follows. In Section 2 we describe our computational method and physical 
assumptions. In Section 3 we present our results: stellar evolutionary tracks, and the inferred binary parameter ranges 
for different cases of mass transfer. In Section 4 we compare our reference model and its variations to the 
observational constraints. In Section 5 we discuss the robustness of our results as well as the implications of our 
findings. We conclude in Section 6. A detailed study of the actual mass transfer phases is presented in a separate paper 
(Klencki et al. in prep.).

\section{Stellar models: physical ingredients}

\label{sec:comp_method}
We employed the MESA stellar evolution code of \cite{Paxton2011,Paxton2015,Paxton2019}\footnote{ 
MESA version r11554,  \url{http://mesa.sourceforge.net/}.}.
We modeled convection using the mixing-length theory \citep{BohmVitense1958} with a 
mixing-length parameter $\alpha = 1.5$, and we adopted the Ledoux criterion for convection. We accounted for semiconvection 
following \citet{Langer1983} with a relatively high efficiency parameter $\alpha_{\rm SC} = 100.0$, as guided 
by the recent results of \citet[][]{Schootemeijer2019}. While the value of $\alpha_{\rm SC}$ 
has been shown to have limited effect on the MS evolution, its effect on the post-MS expansion in radius 
can be substantial (see Sec.~\ref{sec.disc_Z_trend} and App.~\ref{sec:app_mixing}). In Sec.~\ref{sec.res_model_var} we 
also explore variations with $\alpha_{\rm SC} = 1.0$ and $10.0$.

We accounted for convective overshooting above the hydrogen-burning core by applying the step overshooting formalism 
with an overshooting length $\sigma_{\rm ov}$ of 0.345 pressure scale heights based on the calibration of \cite{Brott2011} 
for a $16\Msun$ star. This also agrees with the best fits to the SMC supergiant population obtained by 
\citet{Schootemeijer2019} for $\sigma_{\rm ov}$ of 0.33. While there is increasing evidence for 
a relation between $\sigma_{\rm ov}$ and the stellar mass \citep[e.g.,][also Grin et al., private 
communication]{Claret2018}, a quantitative calibration for the massive stars is still lacking. We assumed the same 
amount of overshooting above the helium-burning core. For consistency with 
\citet{Schootemeijer2019}, we assumed no overshooting associated with burning shells and convective zones in stellar 
envelopes. In Sec.~\ref{sec.res_model_var} we also explore variations with the lower convective overshooting 
efficiency calibrated by \citet{Choi2016}.

We modeled stellar winds following \cite{Brott2011}. For hot ($T_{\rm eff} > 25 \, kK $) and hydrogen-rich ($X_{\rm S} > 
0.7$) stars, we adopted the mass-loss rates from \citet{Vink2000,Vink2001}. For hydrogen-poor stars ($X_{\rm S} < 
0.4$), we applied the wind mass-loss as calculated by \citet{Hamann1995}, divided by 10 \citep{Yoon2006}. For stars with 
intermediate hydrogen abundances ($0.4 < X_{\rm S} < 0.7$), we linearly interpolated between the two. For cooler stars 
($T_{\rm eff} < 25 \, kK $), we took the maximum of the two wind mass-loss rates as calculated from the methods of 
\citet{Vink2001} and \citet{Nieuwenhuijzen1990}. All these mass-loss rates 
for hydrogen-rich and hydrogen-poor stars include a 
metallicity dependence of $\dot{M} \propto (Z/\zsun)^{0.85}$ \citep{Vink2001,Vink2005}.

For the value of solar metallicity, we assumed $\Zsun = 0.017$ \citep{Grevesse1998}. Given a metallicity $Z$, we calculated 
the initial helium abundance as $Y$ as $Y =Y_{\rm proto} + (Y_{\odot} - Y_{\rm proto})Z/\Zsun,$ where $Y_{\odot} = 0.28$ 
\citet{Grevesse1996} and $Y_{\rm proto} = 0.249$ \citep{PlanckColl2016}. The hydrogen abundance follows as $X = 
1 - Y - Z$. Relative abundances of other elements were assumed to be same as in \citet{Grevesse1996}. 

We avoided using the MLT++ option in MESA \citep{Paxton2013} in our reference model. While MLT++ typically helps 
with many numerical problems in evolving a massive star until the Wolf-Rayet (WR) stage, it also artificially reduces 
the stellar radii during the giant phase and therefore affects the predictions for mass transfer evolution. We consider 
a variation with MLT++ in Sec.~\ref{sec.res_model_var} (and also App.~\ref{sec:App_mltpp}). We used nuclear reaction 
networks provided with MESA: \texttt{basic.net} for H and He burning, and \texttt{co\_burn.net} for C and O burning.

We computed models in the mass range between $10$ and $80\Msun$ in steps of $2 \msun$ (below $20 \msun)$ or $2.5\Msun$ 
(above $20 \msun$) at six different metallicities: $Z$ = 0.017, 0.0068, 0.0034, 0.0017, 0.00068, 
and 0.00017 (which correspond to $1.0 \zsun$, $0.4 \zsun$, $0.2 \zsun$, $0.1 \zsun$, $0.04 \zsun$, and $0.01 \zsun$). Most 
of our models are nonrotating, but we also explored rotating models with the initial rotation rate set to 40\% of the 
critical value ($\Omega / \Omega_{\rm crit} = 0.4$). In these models we included the effects of Eddington-Sweet circulation,
secular shear instabilities, and the Goldreich-Schubert-Fricke instability, with an efficiency factor $f_c = 1/30$ \citep[see 
the calibration to nitrogen enrichment in rotating stars and the references in][]{Heger2000,Brott2011}.

There is no one common termination condition for all our models.
Because the goal is to study single stellar tracks in the context of binary evolution and massive interacting binaries, 
we are interested in the evolution of single stars roughly until they 
have reached their largest radius. By that point, if they were members of binary systems, they would usually have gone 
into RLOF. In some cases, typically at low metallicity or for the low-mass end of our grid, stars continue to expand 
even during the late evolutionary stages of advanced burning and reach their maximum sizes near the very end of their 
lives. In these cases we evolved our models until the onset of oxygen burning. At this stage, a massive star 
is only about a year away from core collapse (depending on the mass), and there is no time left for any 
significant change in its size. 
In the remaining cases, the expansion of stars is quenched earlier through strong stellar winds. When a significant 
fraction of the envelope mass has been lost, the evolution in the Hertzprung-Russel (HR) diagram turns around toward 
higher effective temperatures. From that point onward, as increasingly more mass is lost from the envelope, the star 
continues to decrease in radius to eventually become a WR star. This part of the evolution is not relevant in the case 
of interacting binaries: either a RLOF would have occurred earlier, or it is not going to occur at all. In these cases 
we only evolved the models until the final turnaround in the HR diagram has been reached (taking into account the 
possibility of blue loops). The resulting mass transfer and post-mass transfer evolution is discussed in Klencki et al. 
(in prep.).

The MESA inlists (input files) used in this work are available at \url{http://cococubed.asu.edu/mesa_market/inlists.html}.
 All the output files, in particular stellar tracks, can be found at \url{https://zenodo.org/communities/mesa}.

\section{Results}
\label{sec:results}

\subsection{Stellar tracks and radius evolution of massive stars}

\label{sec.res_tracks}

\begin{figure*}
\centering
    \includegraphics[width=0.7\textwidth]{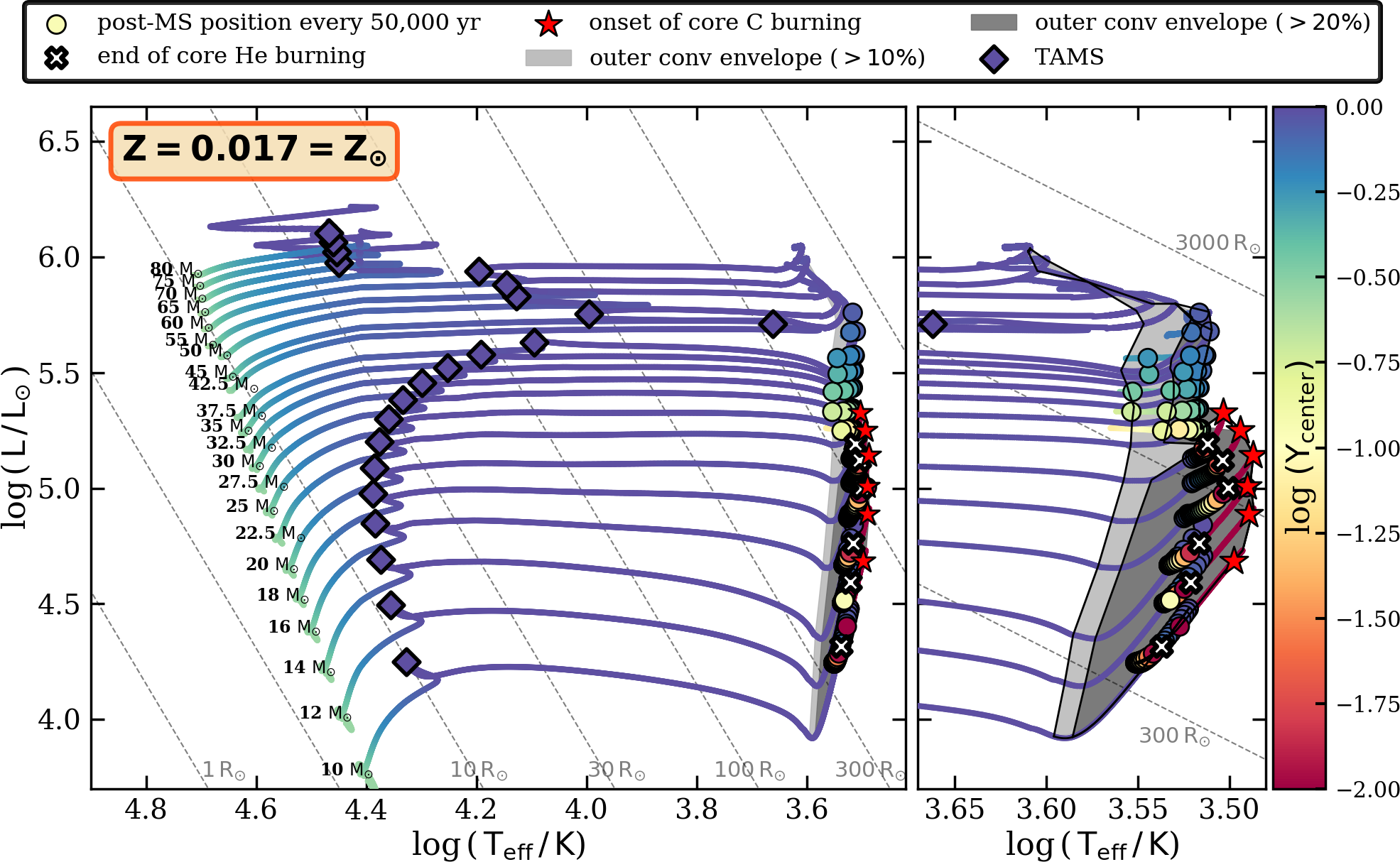}
    \includegraphics[width=0.7\textwidth]{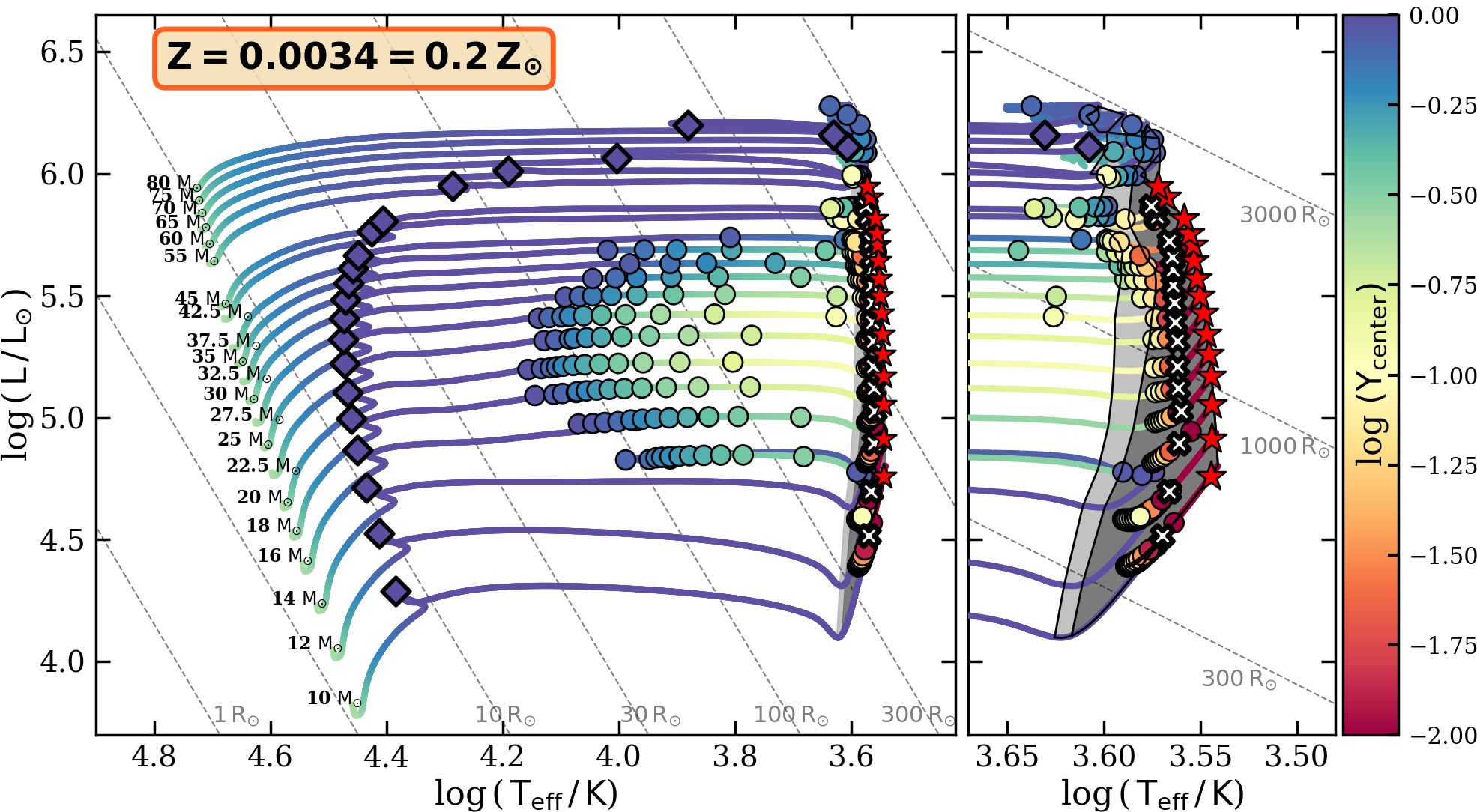}
    \includegraphics[width=0.7\textwidth]{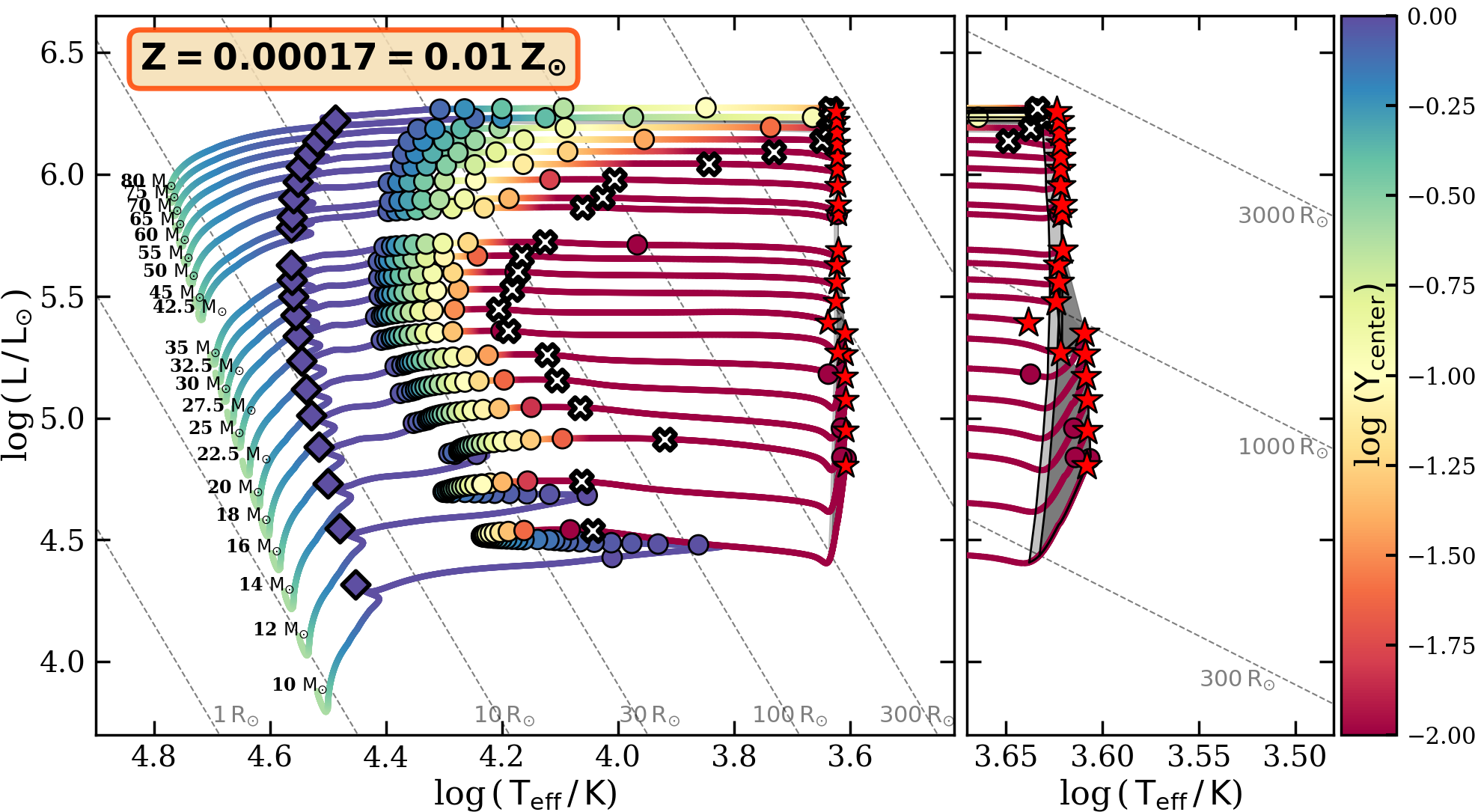}
    \caption{Evolutionary tracks of single stars of masses between $10$ and $80\Msun$ computed at three different 
metallicities: solar (top), SMC-like (middle), and very low metallicity $Z = 0.01 \zsun$ (bottom). See 
Fig.~\ref{fig.HRD_app} for tracks at three additional metallicities.  Filled circles mark the position of a 
star during its post-MS evolution taken every 50,000 years. Color indicates the central helium abundance ($Y_{\rm C}$, in 
logarithmic scale). With a cross we mark the end of core-helium burning ($Y_{\rm C} < 10^{-3}$), and with a red star we 
mark the onset of core-carbon burning ($X^{\rm carbon}_{\rm  C} < \, 0.95 \, X^{\rm carbon}_{\rm  C;max}$). Shaded 
regions mark stars with outer convective envelope layers of at least 10\% (or 20\%) of the mass of the entire star. 
For tracks that turn around towards the hotter part of the HR diagram (and the WR 
regime) because of extensive mass loss, we only show the evolution until shortly after the maximum radius has been reached.
}
    \label{fig.HRD}
\end{figure*}

In Figure~\ref{fig.HRD} we show HR diagrams with evolutionary tracks of single stars with masses 
between $10$ and $80\Msun$ computed for solar ($Z = \zsun = 0.017$), subsolar (SMC-like, $Z = 0.2 \zsun = 0.0034$), 
and very low metallicity ($Z = 0.01 \zsun = 0.00017$). In the appendix (Fig.~\ref{fig.HRD_app}) we provide 
additional plots for intermediate metallicity values ($0.4 \zsun$, $0.1 \zsun$, and $0.04\zsun$). 
Only the evolution until the maximum radial expansion and the subsequent decrease in radius by 
$\sim 20\%$ is shown (i.e., excluding the final leftward evolution toward WR stars in the case of high-mass and 
high-metallicity models). Circles in Fig.~\ref{fig.HRD} mark the position of a star during its post-MS evolution taken every 
50,000 years. Thus, a clustering of circles corresponds to the relatively long-lived phase of core-helium burning 
(CHeB). It also indicates which part of the HR diagram is most likely to be occupied by a population of observed single stars 
(excluding MS stars), as predicted by our models. A lack of circles corresponds to the MS evolution and to phases 
of rapid expansion: the HG (between the end of MS and the onset of CHeB) and the helium HG (HeHG, between the end of 
CHeB and the onset of carbon burning). The track colors correspond to the central helium abundance.

When the location and evolution of CHeB stars in the HR diagram across different metallicities (i.e. the 
clustering of the circles) are compared, a metallicity trend in the post-MS radius evolution of 
massive stars is evident. The lower the metallicity, the smaller (and hotter) the massive stars at the end of the rapid HG 
expansion and during the subsequent CHeB phase. At solar metallicity (top panel), models below $\sim 65 \msun$ expand 
all the way to the red giant branch (RGB, ${\rm log}~ T_{\rm eff} \lesssim 3.7$) during the short-lived HG phase,
and almost their maximum radii at that stage ($R_{\rm HG;max} \sim 300-2000 \rsun$, depending on the 
mass). At this point, the models with masses $\lesssim 40\msun$ begin the slower core-helium burning evolution, whereas the 
more massive models quickly turn around toward the left in the HR diagram because of extensive mass loss. In the case of 
even more massive models ($M > 65 \msun$), strong winds quench expansion at smaller sizes (up to $\sim 100 \rsun$). At 
SMC-like metallicity (middle panel, $Z = 0.2\zsun$), the post-MS radius evolution of models between $\sim 16$ and $\sim 
40\Msun$ is significantly different to their solar metallicity counterparts. Instead of expanding all the way 
to the RGB during the HG phase, these models regain thermal equilibrium at much smaller radii ($\sim 100 \rsun$) as 
slowly evolving CHeB stars. At very low metallicity (bottom panel, $Z = 0.01\zsun$), the onset of the CHeB phase is 
located at even higher effective temperature (and smaller radius) for the entire mass range $10$-$80\msun$ , and there is 
only a very small gap in the HR diagram that is due to the HG phase.

The position of the terminal-age MS (TAMS) changes with metallicity as well: the lower the metallicity, the hotter 
and smaller the MS stars (with the exception of the most massive models at solar metallicity). The largest differences 
between metallicities appear at the high-mass end of our grid ($\gtrsim 40 \msun$). Notably, around the $40-50 \msun$ 
range at solar metallicity (top panel) and above $\sim 60 \msun$ at SMC-like metallicity (middle panel), models 
expand up to $\sim 1000 \rsun$  at the end of the MS. This is a signature of inflated envelopes, and this phenomenon 
was previously described by \citet{Kato1985} for very massive MS stars and by \citet{Petrovic2006} for WR stars. In such stars, 
the Eddington limit is locally exceeded sufficiently close to the top of their envelopes (in this case, at the iron 
opacity peak). To prevent $L_{\rm rad}$ from exceeding $L_{\rm Edd}$, the layers above expand substantially and the 
envelope becomes extremely diluted and extended in radius (i.e., inflated), with a density inversion formed at the top 
\citep[see][for details]{Grafener2012,Sanyal2015}.
The differences in position of the TAMS at the high-mass end between different 
metallicities are primarily associated with the relative strength of stellar winds and the metallicity dependence of the
envelope inflation in massive stars \citep{Sanyal2017}. The small sizes of our $M > 65 \msun$ models at solar metallicity are in 
general agreement with the models by \citet{Sanyal2015}, whose models never reach 
an effective temperature of $T_{\rm eff} \approx 15 kK$ ($\sim 4.2$ in log) above a certain mass.

In the case of very low metallicity (bottom panel, $Z = 0.01\zsun$), most of the models reach the end of CHeB 
while still in the blue part of the HR diagram ($\logteff \sim 4.1$). These models expand significantly during the 
subsequent contraction of the CO core and the HeHG phase and reach the RGB by the very end of their 
evolution, sometimes shortly ($\sim$ 1000 yr) before the final core collapse. This is also the case for $Z = 0.04 
\zsun$ models in Fig.~\ref{fig.HRD_app}, but to a lesser extent.

\begin{figure}
\centering
      \includegraphics[width=\columnwidth]{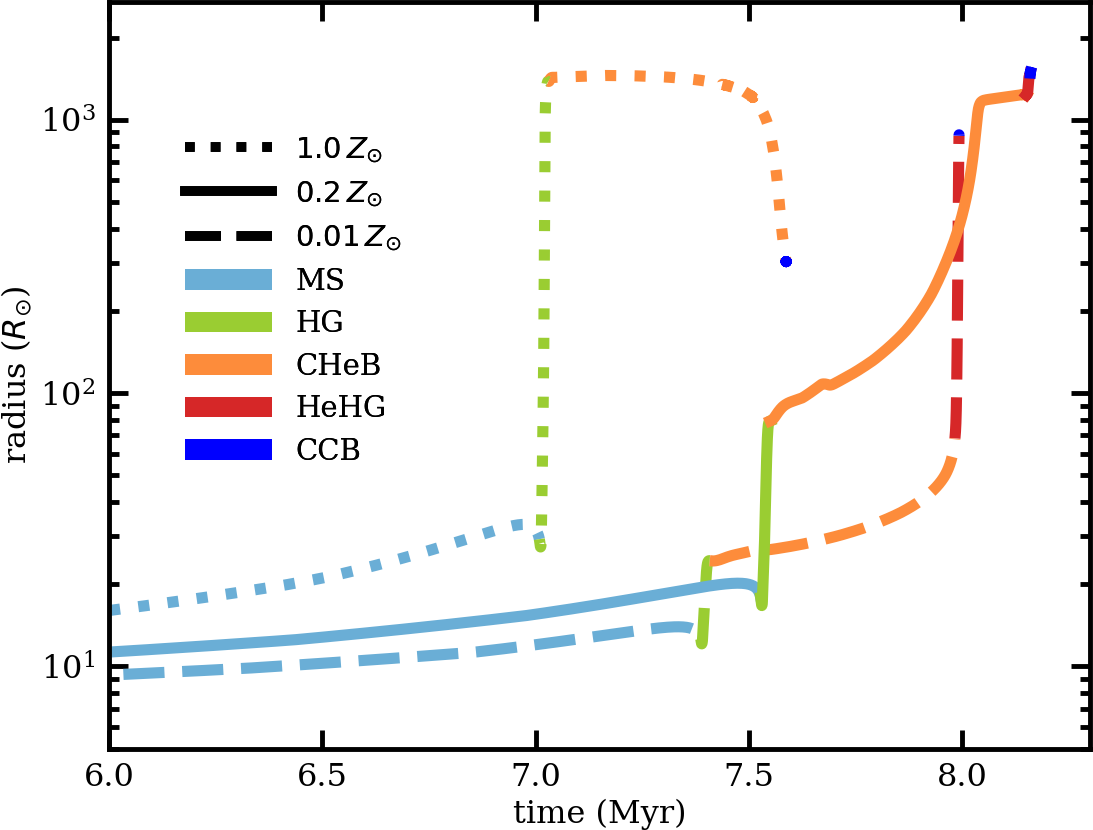}
    \caption{Radius evolution of a $25 \msun$ single star model at solar ($Z = 0.017 = \zsun$), SMC-like ($Z = 0.0034 
= 0.2\zsun$), and very low metallicity ($Z = 0.00017 = 0.01 \zsun$). The lines are color-coded according to different 
evolutionary stages.
} 
    \label{fig.define_cases}
\end{figure}

In summary, depending on the metallicity, a substantial expansion in radius can occur at a 
completely different evolutionary stage of a massive star. We illustrate this in Fig.~\ref{fig.define_cases}, 
where we show the time evolution of the radius of a $25 \msun$ model for three different metallicities.
At solar metallicity, most of the expansion occurs during the HG phase, when the star expands rapidly on a
thermal timescale. At subsolar SMC-like metallicity ($Z = 0.2\zsun$), most of the expansion takes place during CHeB, 
when the star is in thermal equilibrium and the radius changes slowly on a nuclear timescale. Very low metallicity 
stars expand significantly only after the end of CHeB during another phase of rapid out-of-equilibrium expansion: the HeHG 
phase. What follows is the core-carbon burning stage (CCB).

\subsection{Binary parameter ranges for different donor types}

\label{sec:res_binpar}

\begin{figure*}
\begin{tabular}{R{6.4cm}  C{5.5cm}  L{5.5cm}}
    \includegraphics[width=0.35\textwidth,height=150px]{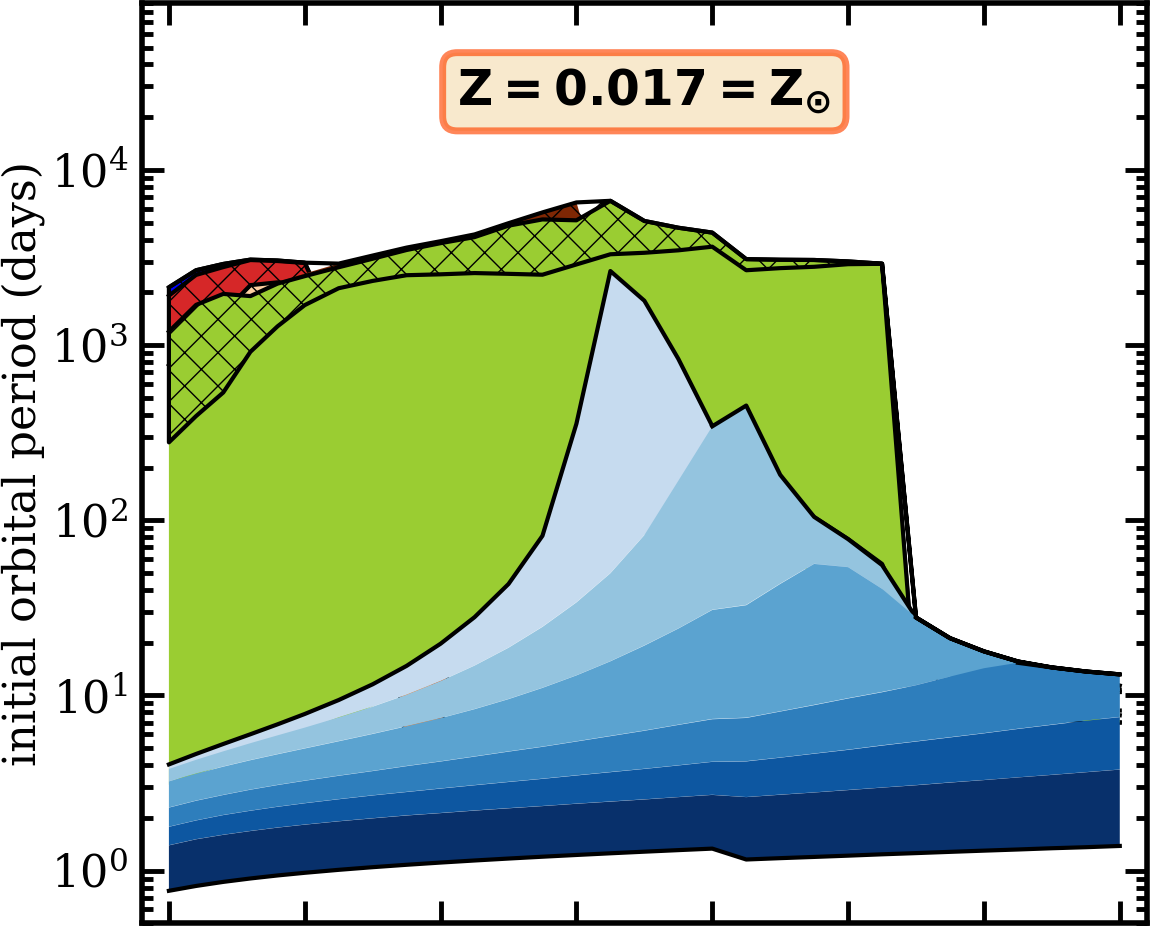} &
    \includegraphics[width=0.3\textwidth,height=150px]{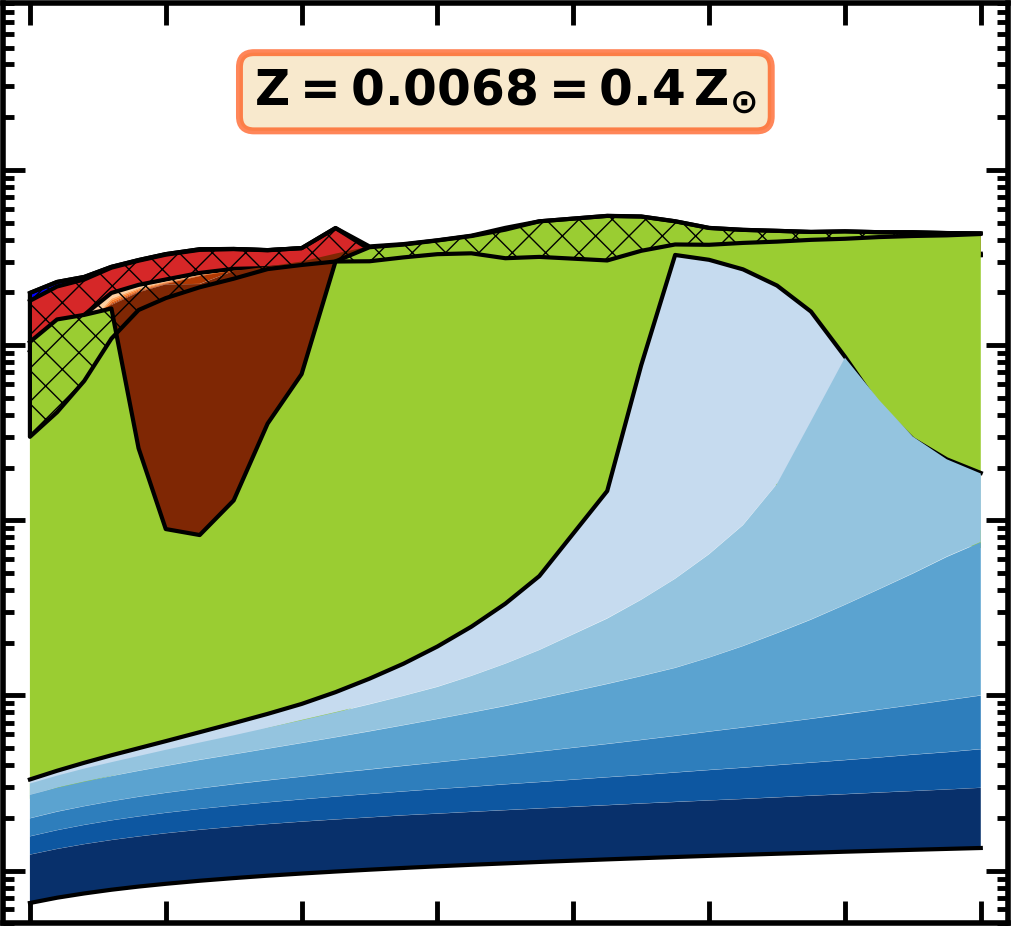} &
    \includegraphics[width=0.3\textwidth,height=150px]{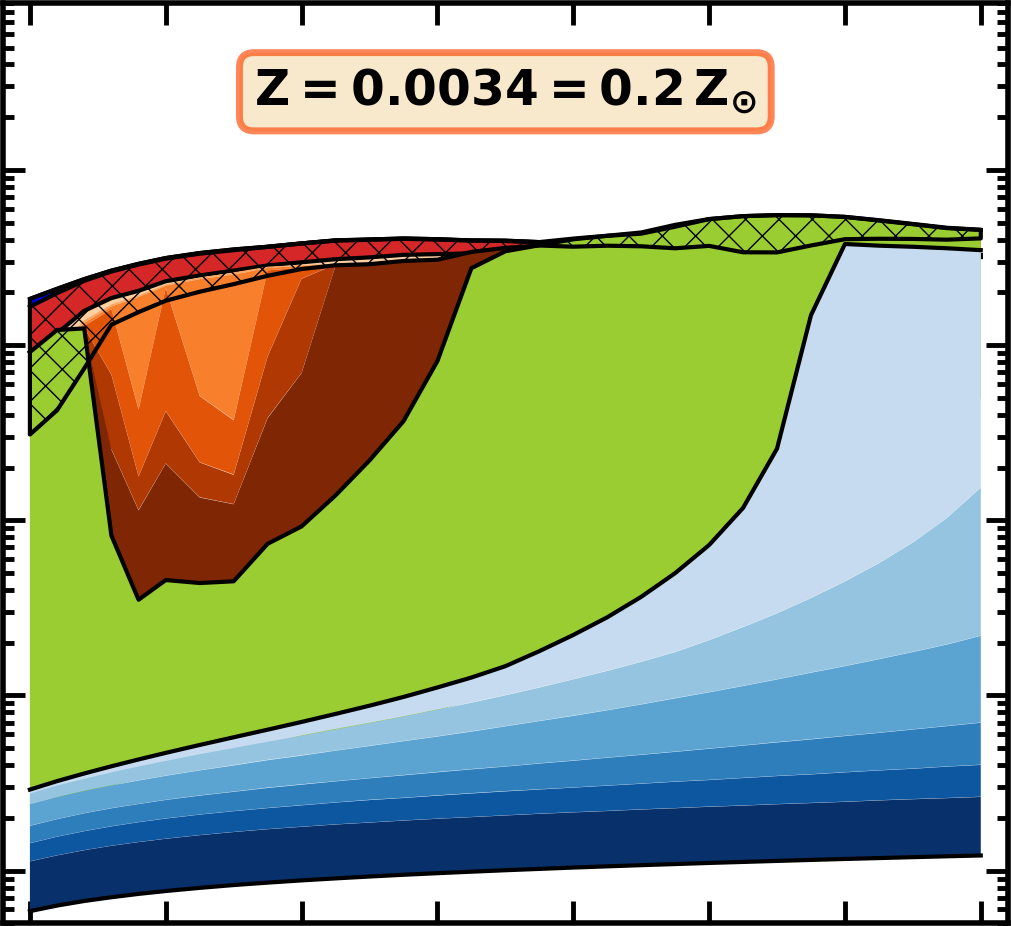} \\
    \includegraphics[width=0.35\textwidth,height=170px]{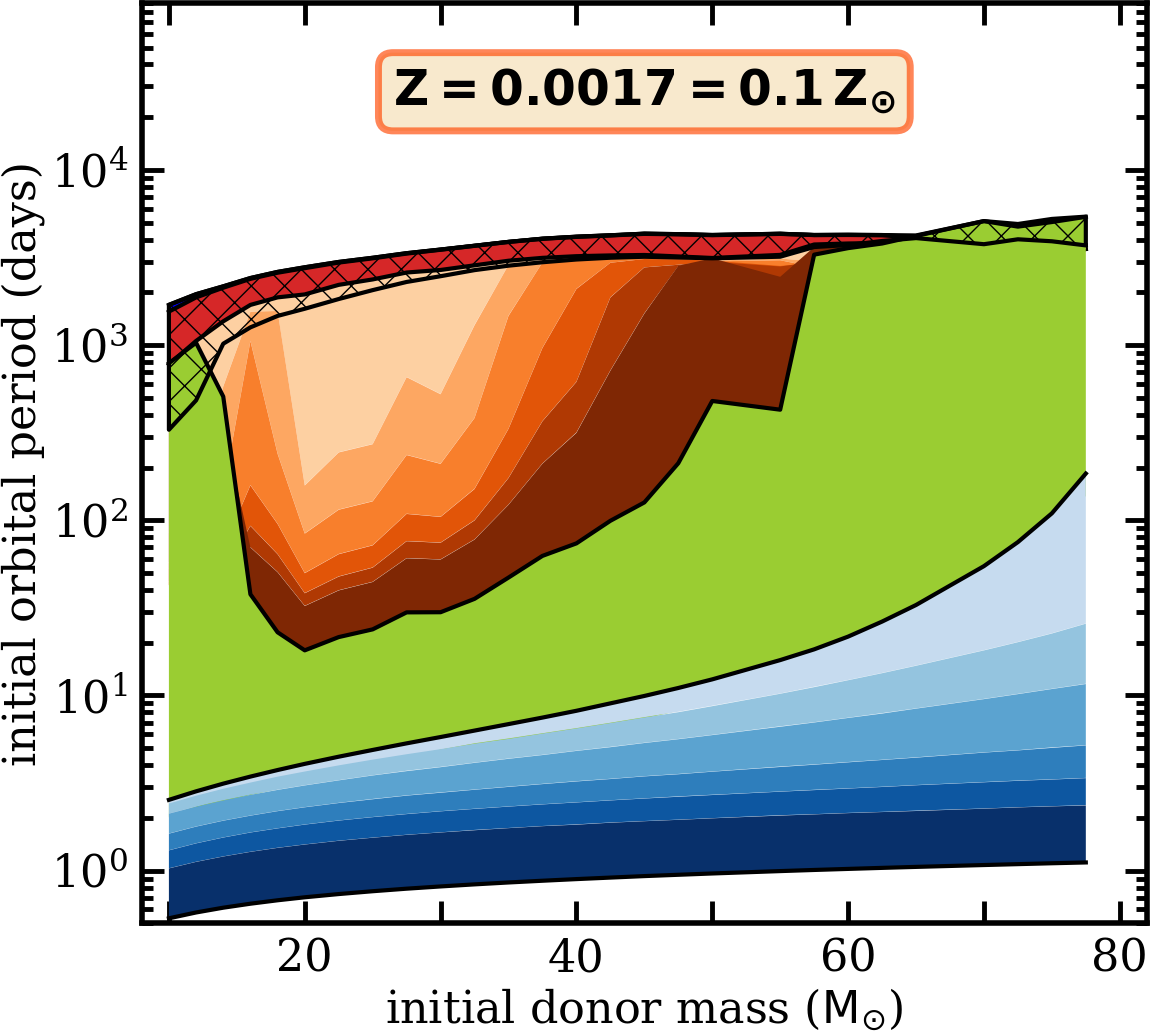} &
     \includegraphics[width=0.3\textwidth,height=170px]{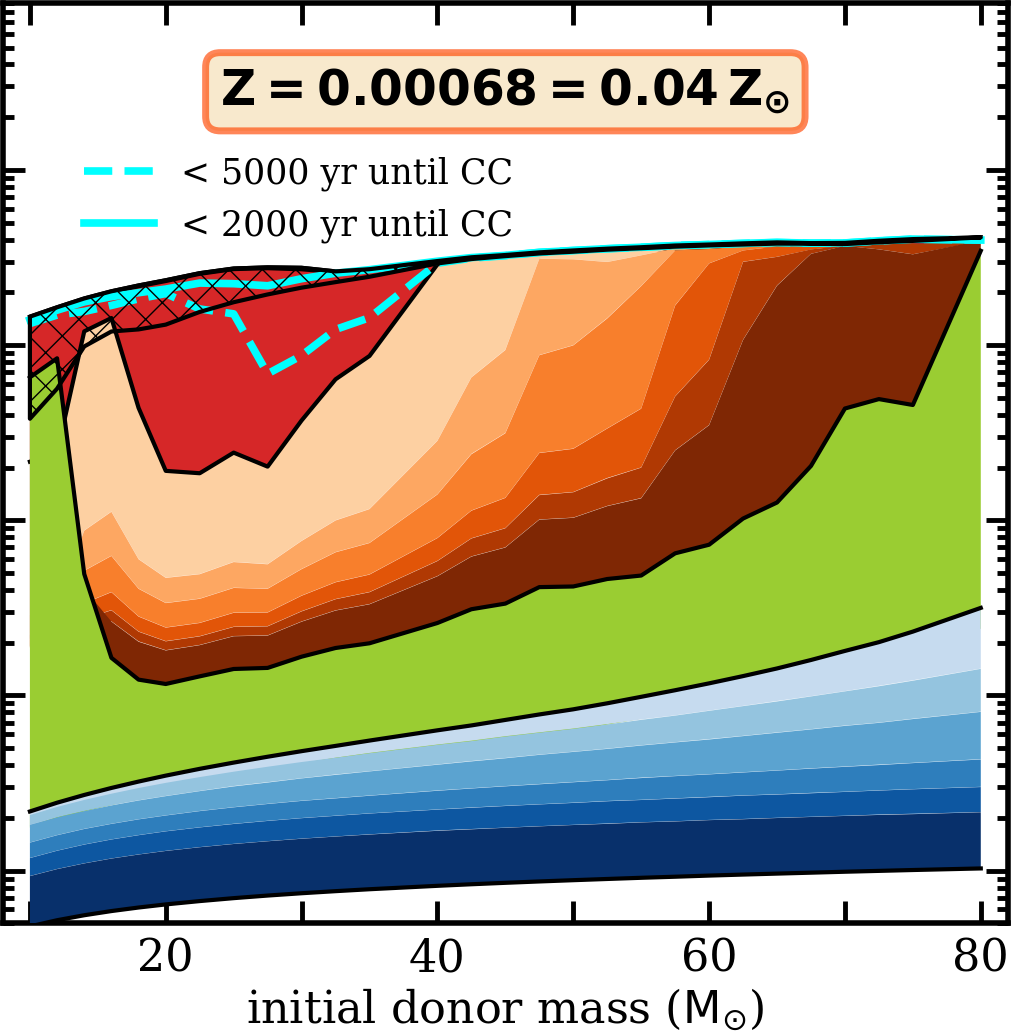} &
    \includegraphics[width=0.3\textwidth,height=170px]{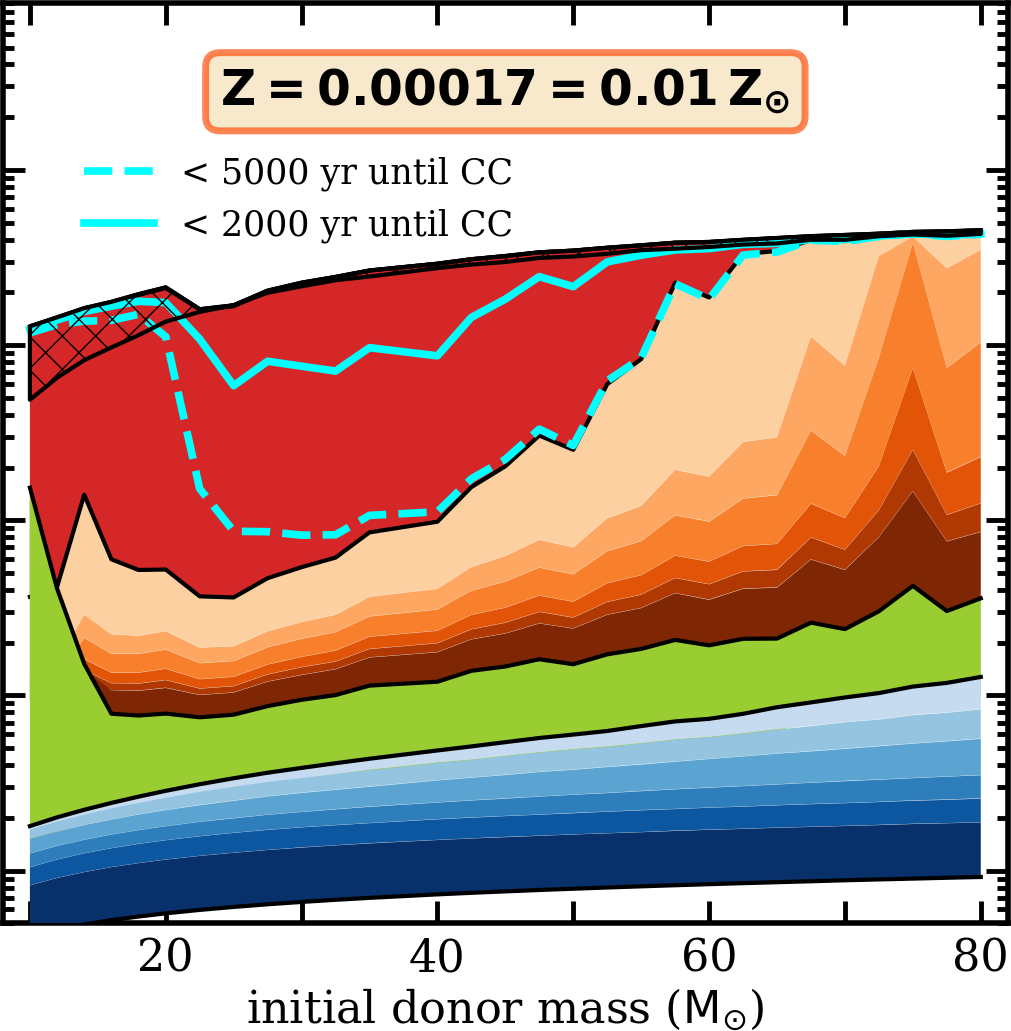} \\
\end{tabular}

\centering
\vspace{0.1cm}
\includegraphics[width=0.92\textwidth]{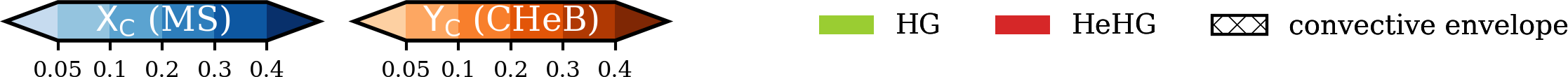}
\caption{Evolutionary state of a donor star at the point of RLOF in massive interacting binaries as a function of 
initial donor mass (at the ZAMS) and initial orbital period, estimated based on the single stellar tracks presented in 
Sec.~\ref{sec.res_tracks}. The companion is assumed to be twice less massive (initial mass ratio $q = 0.5$), and 
it can be viewed as either a less massive star or under additional assumptions about the previous 
evolution (see text) as a stellar BH. Different shades of blue (orange) correspond 
to different central abundances of hydrogen (helium), indicating how advanced the MS (CHeB) evolution is. Hatched 
regions indicate donors with outer convective envelopes (at least 10\% in mass coordinate). Cyan lines mark the
threshold above which the donors are only 2000 or 5000 years away from the core-collapse (only shown for 
the two lowest metallicities for clarity).}
\label{fig.bin_param_ranges}
\end{figure*}

In this section we take a binary perspective: we use the stellar tracks presented in the previous section to determine 
 the evolutionary state of a donor star at the onset of mass transfer depending on the binary 
orbital period and metallicity. In order to do so, we assume that our single stellar models
represent the evolution of the more massive component in a binary system until the eventual RLOF. We assume that the 
companion has half the mass of the primary star (mass ratio $q = 0.5$) and that its mass loss in winds is negligible 
compared to that of the primary. For simplicity, we assume that 
the amount of mass accreted by the companion from the wind is negligible and that its mass is constant. We then compute 
the effect of the wind mass loss of the primary on the binary 
orbit by assuming that all the mass that the primary loses carries specific angular momentum equal to the specific 
orbital angular momentum of the primary.

The less massive companion can be considered to be either a less massive MS star or a stellar black hole (BH). 
The latter case would mean that the now more massive stellar component is what used to be the secondary star in a 
zero-age binary, and that the star that used to be the actual primary has already collapsed into a BH. Before 
BH formation, the binary has most likely gone through a phase of mass transfer. In this case, the now more massive 
stellar component of the BH binary has previously been an accretor.

With these assumptions, we plot in Fig.~\ref{fig.bin_param_ranges} in different colors the evolutionary state of the 
massive donor star at the point of RLOF as a function of the initial donor mass (i.e., at the zero-age main sequence, ZAMS, not at 
RLOF) and initial orbital period for different metallicities. When the companion is assumed to be a stellar BH, the 
initial period roughly corresponds to the orbital period at the moment of the BH formation. We neglect 
the effect of spin-orbit coupling by tidal interactions, which would shrink the pre-RLOF orbits by a small factor 
\citep[see, e.g.,][]{Hurley2002} and shift the boundaries in Fig.~\ref{fig.bin_param_ranges} to somewhat higher initial 
orbital period values.

Several trends with metallicity are present in Fig.~\ref{fig.bin_param_ranges}. We describe them below.
\begin{enumerate}
\item The lower the metallicity, the more evolved the donor stars at the point of RLOF. In particular, at $Z = 
\zsun$ and $0.4\zsun$ , almost all cases of mass transfer from an evolved (post-MS) donor involve a rapidly expanding 
HG star (green). As the metallicity decreases, however, the parameter space for RLOF initiated by a slowly expanding 
CHeB supergiant increases (orange) and also includes increaslingly more massive donors. This is a natural consequence of the fact that
low-metallicity stars expand less strongly during the immediate post-MS expansion (HG phase) and can regain thermal 
equilibrium as much smaller CHeB stars ($\sim 100 \rsun$) than their higher metallicity counterparts ($\sim 
1000\rsun$), see Sect.~\ref{sec.res_tracks}.
We discuss the implications of this trend for the binary formation channel of stripped helium stars in 
Sec.~\ref{sec.disc_WR_formation}. The difference in post-MS radius expansion between low- and high -metallicity stars may also have significant consequences for the mass transfer evolution itself, see discussion in 
Sec.~\ref{sec.disc_mass_transfer}.

\item The lower the metallicity, the smaller the parameter space for RLOF from convective donors (marked with the hatched 
area in Fig.~\ref{fig.bin_param_ranges}). The parameter space for convective donors disappears almost completely in the case 
of the more massive ($M > 40\msun$) models at very low metallicity ($Z = 0.04\zsun$ and $Z = 0.01\zsun$). It is also 
quenched in the case of the most massive models ($M > 65\msun$) at solar metallicity because these models only attain relatively small sizes. 
Only a small fraction of massive BH binaries evolve through mass transfer initiated by a convective envelope donor. 
This fact may have important consequences for the common-envelope (CE) formation channel 
of double BH binary mergers (see also Sec.~\ref{sec.res_conv_env} and the discussion in Sec.~\ref{sec.disc_SN_GRB_BBH}).

\item The lower the metallicity, the larger the parameter space for a late RLOF, that is, RLOF from a star that is only 
several thousand years away from core collapse. For a metallicity between $0.017$ ($\zsun$) and $0.0017$ ($0.1\zsun$) ,
the parameter space for such a late RLOF is very small (even smaller than for convective donors) and limited to $M 
\lesssim 40 \msun$. However, for very low metallicities, that is, $Z = 0.04 \zsun$ and $0.01\zsun$, the probability of a 
late RLOF grows significantly, as highlighted by the cyan threshold lines. See Sec.~\ref{sec.disc_SN_GRB_BBH} for a 
further discussion.

 \item The lower the metallicity, the smaller the parameter space for mass transfer from MS donors. This is to a large 
extent a consequence of the formation of inflated envelopes in massive MS stars that evolve near the Eddington limit 
\citep{Sanyal2015}. The sets of tracks at higher metallicities ($1.0, \, 0.4,  \text{and} \, 0.2 \zsun$) all include a mass 
range in which the models become inflated and expand up to $\sim 1000\rsun$ at the end of the MS (which corresponds to 
initial orbital periods for RLOF of $\gtrsim 1000$ days in Fig.~\ref{fig.bin_param_ranges}). As the metallicity 
decreases, this mass range shifts toward higher masses \citep[see also][]{Sanyal2017}. As a result, for $Z \leq 
0.1 \zsun$ the envelopes of MS stars do not become inflated within the mass limits of our grid (up to $80 \msun$). 
It should be noted that normal not inflated MS stars at lower metallicity are also smaller than their higher 
metallicity counterparts.

\end{enumerate}

\subsection{Model variations}

\label{sec.res_model_var}

\begin{figure*}
\begin{tabular}{R{6.4cm}  C{5.5cm}  L{5.5cm}}
\begin{subfigure}{0.35\textwidth}
\includegraphics[width=\textwidth,height=150px]{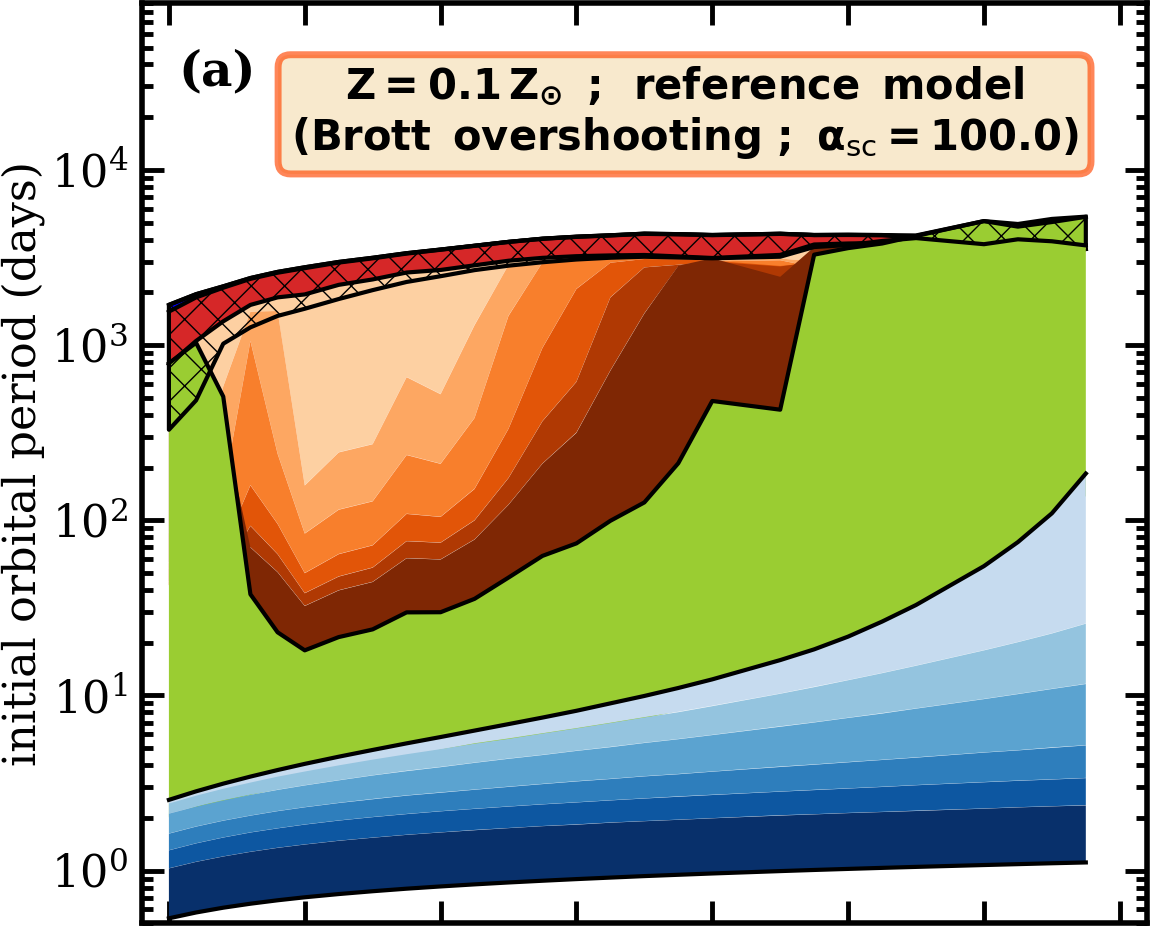}
\phantomsubcaption
\label{fig.bin_param_variations_a}
\end{subfigure} &
\begin{subfigure}{0.3\textwidth}
\includegraphics[width=\textwidth,height=150px]{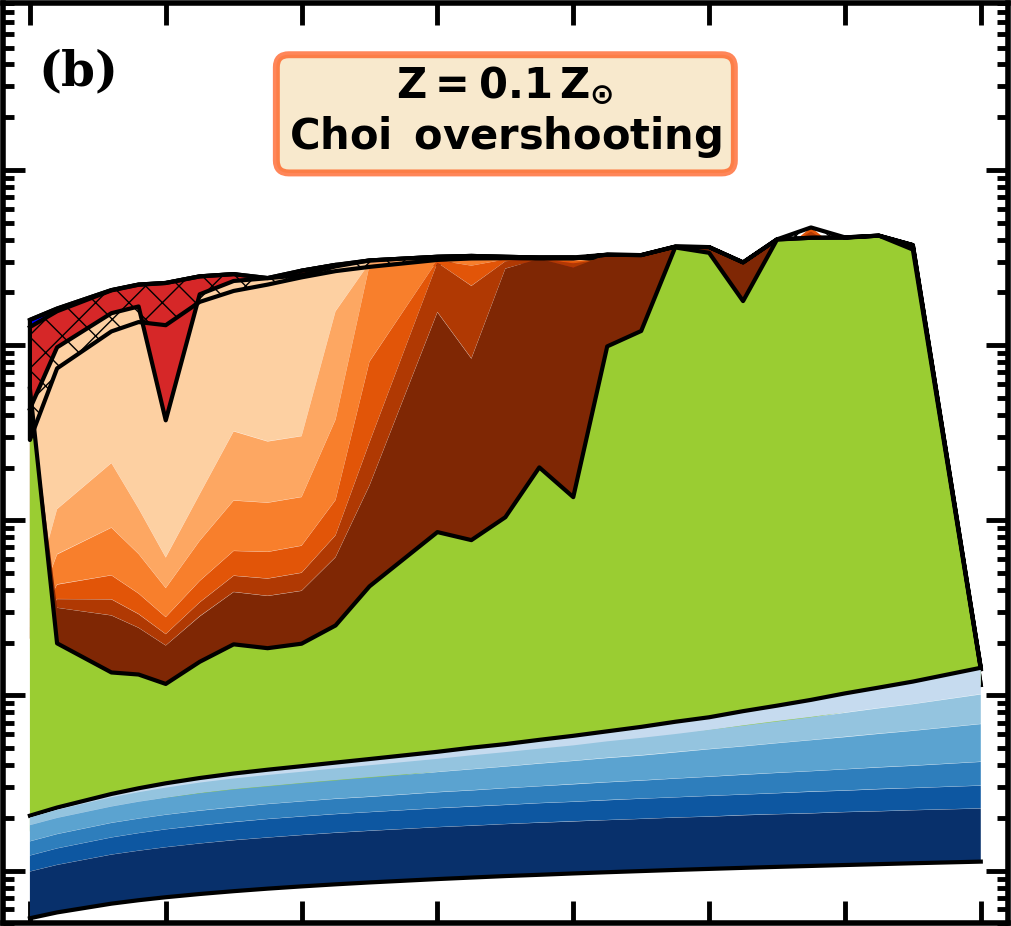} 
\phantomsubcaption
\label{fig.bin_param_variations_b}
\end{subfigure} &
\begin{subfigure}{0.3\textwidth}
\includegraphics[width=\textwidth,height=150px]{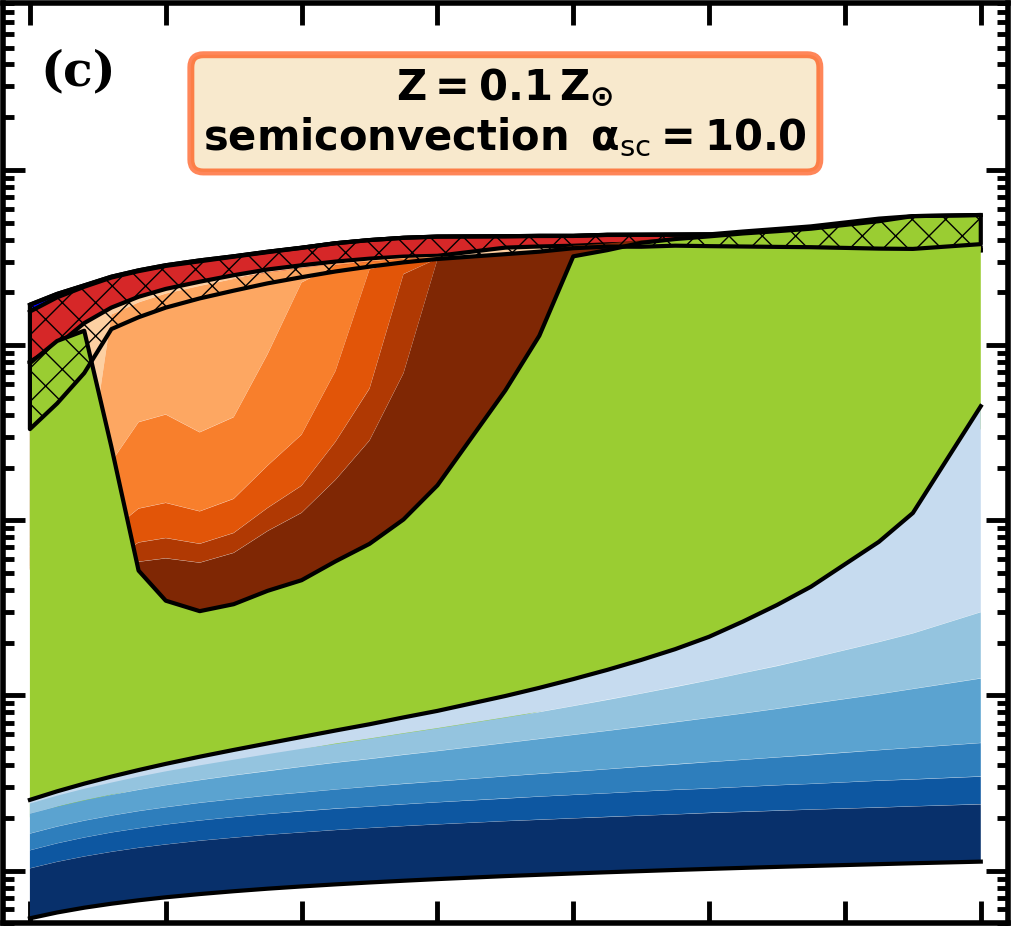} 
\phantomsubcaption
\label{fig.bin_param_variations_c}
\end{subfigure} \\
\begin{subfigure}{0.35\textwidth}
\includegraphics[width=\textwidth,height=170px]{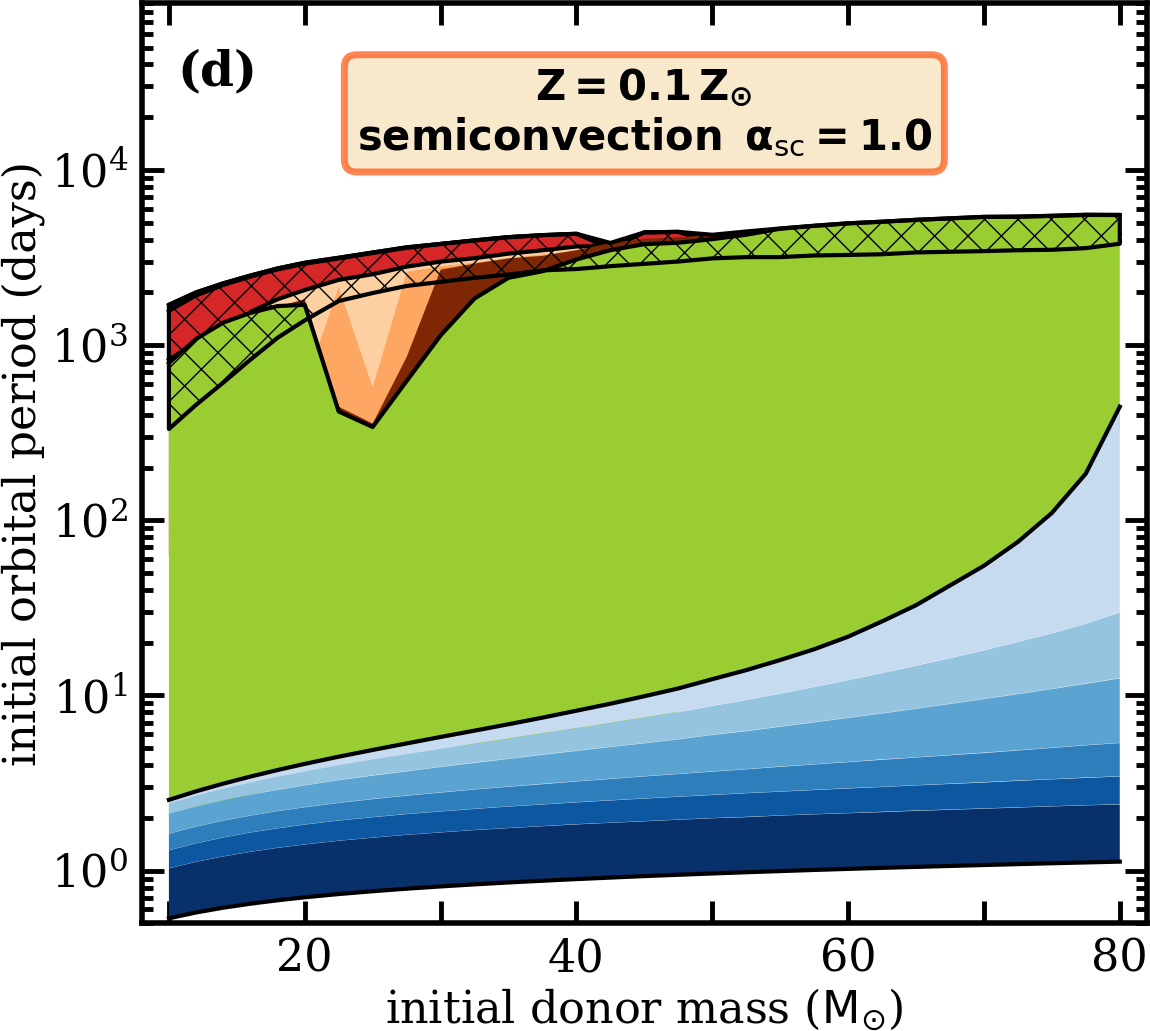}
\phantomsubcaption
\label{fig.bin_param_variations_d}
\end{subfigure} &
\begin{subfigure}{0.3\textwidth}
\includegraphics[width=\textwidth,height=170px]{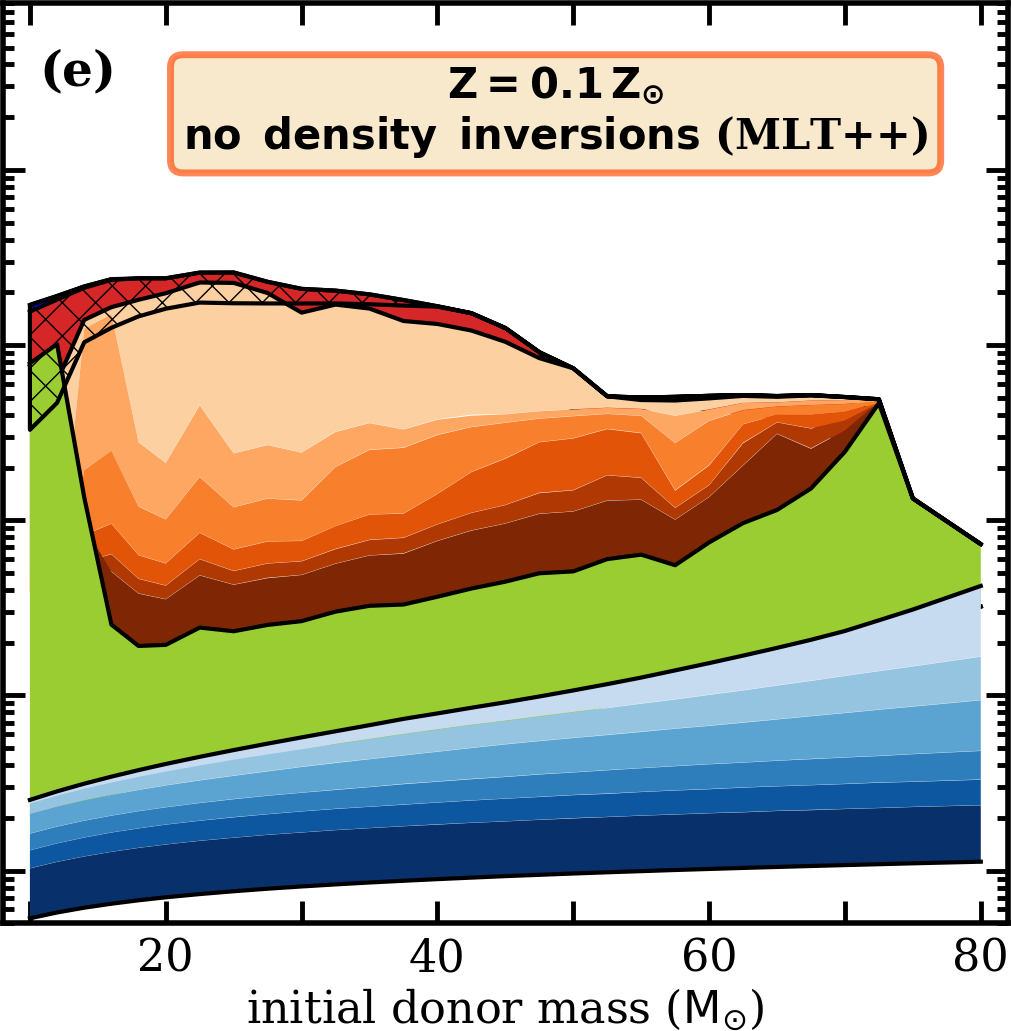} 
\phantomsubcaption
\label{fig.bin_param_variations_e}
\end{subfigure} &
\begin{subfigure}{0.3\textwidth}
\includegraphics[width=\textwidth,height=170px]{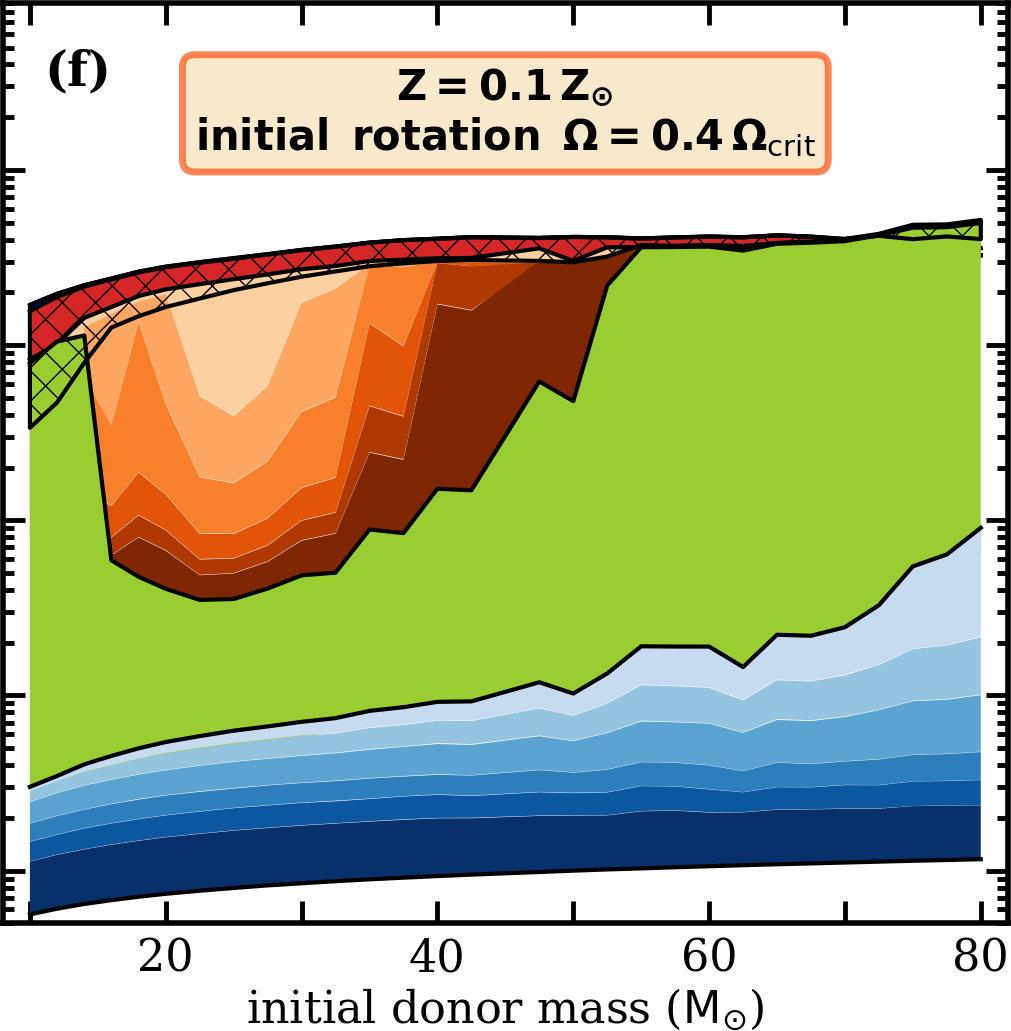}
\phantomsubcaption
\label{fig.bin_param_variations_f}
\end{subfigure} \\
\end{tabular}
\centering
\vspace{0.1cm}
\includegraphics[width=0.92\textwidth]{Figures_used/binary_par_ranges_legend.png}
\caption{Same as Fig.~\ref{fig.bin_param_ranges}, but computed for one particular metallicity ($Z = 0.0017 = 0.1\zsun$) 
and exploring several variations in the input parameters to the model with respect to the reference model presented in 
the previous sections (and in the top left panel): different overshooting parameterization \citep{Choi2016}, lower
efficiency of semiconvective mixing ($\alpha_{\rm sc} = 10$ or $\alpha_{\rm sc} = 1$), application of MLT++ treatment 
of 
convection in radiation-dominated superadiabatic envelope layers \citep{Paxton2013}, and initial
rotation velocity $\Omega = 0.4 \Omega_{\rm crit}$. In the case of $Z = 0.2\zsun$, for which observation of evolved 
supergiants SMC can be used to constrain the models, none of the model variations presented above can be excluded, except for the one 
with $\alpha_{\rm sc} = 1$;  see Sec.~\ref{sec:obs_b_vs_r}.} 
\label{fig.bin_param_variations}
\end{figure*}

In order to illustrate the uncertainty in binary parameter ranges for the different evolutionary states of donor 
stars presented in Fig.~\ref{fig.bin_param_ranges}, we explore in Fig.~\ref{fig.bin_param_variations} several 
variations in the input parameters for one particular metallicity: $Z = 0.0017 = 0.1\zsun$ (see 
Fig.~\ref{fig.app_bin_param_variations} in the Appendix for additional figures at different metallicities). Fig.~\ref{fig.bin_param_variations_a} corresponds to the reference set of stellar tracks presented in the previous sections,
computed using the assumptions described in Sec.~\ref{sec:comp_method}. In
particular, our reference model assumes step overshooting with $\sigma_{\rm ov;step} = 0.345$ \citep{Brott2011}, high 
efficiency of semiconvective mixing ($\alpha_{\rm sc} = 100.0$), no MLT++ approach, and nonrotating models. 

The model Fig.~\ref{fig.bin_param_variations_b} assumes a different parameterization of convective 
overshooting \citep[see Sec.~3.6.2 of][]{Choi2016}, one calibrated to reproduce the shape of the MS turnoff 
in the open cluster M67 \citep{Magic2010}. It is expressed according to the exponential formulation of overshooting 
\citep{Herwig2000} with the characteristic length $\sigma_{\rm ov;exp} = 0.016$ for the core overshooting. This value 
is roughly equivalent to $\sigma_{\rm ov;step} = 0.2$ in the step overshooting description. Additionally, we 
tested that the calibration of overshooting carried out by \citet{Brott2011} (based on the value of ${\rm log}(g)$ at the 
Henyey hook of $\sim16\msun$ stars in the LMC clusters NGC 2004 and N11) can be best fit in the exponential 
overshooting description with $\sigma_{\rm ov;exp} \approx 0.03$. Thus, the Choi exponential parameterization 
effectively results in smaller overshooting than the step overshooting parameterization employed in \citet{Brott2011} and in our 
reference model.

The model in Fig.~\ref{fig.bin_param_variations_c} assumes somewhat less efficient 
semiconvection ($\alpha_{\rm sc} = 10$) than in the reference model ($\alpha_{\rm sc} = 100$). Both values appear to be 
consistent with the tentative observational constraints \citep[][see also Sec.~\ref{sec:obs_b_vs_r}]{Schootemeijer2019}.
In Fig.~\ref{fig.bin_param_variations_d}  we show a model with even less efficient semiconvection, $\alpha_{\rm sc} = 
1$. This model is disfavored based on the above-mentioned observations, see also Sec.\ref{sec:obs_02Z}. 

\begin{figure*}
\centering
    \includegraphics[width=0.9\textwidth]{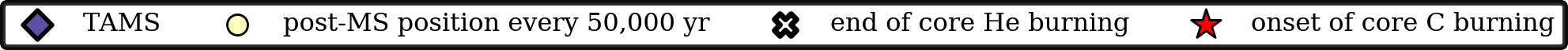}    
\begin{tabular}{R{0.45\textwidth} L{0.49\textwidth}}
\includegraphics[width=0.41\textwidth,height=210px]{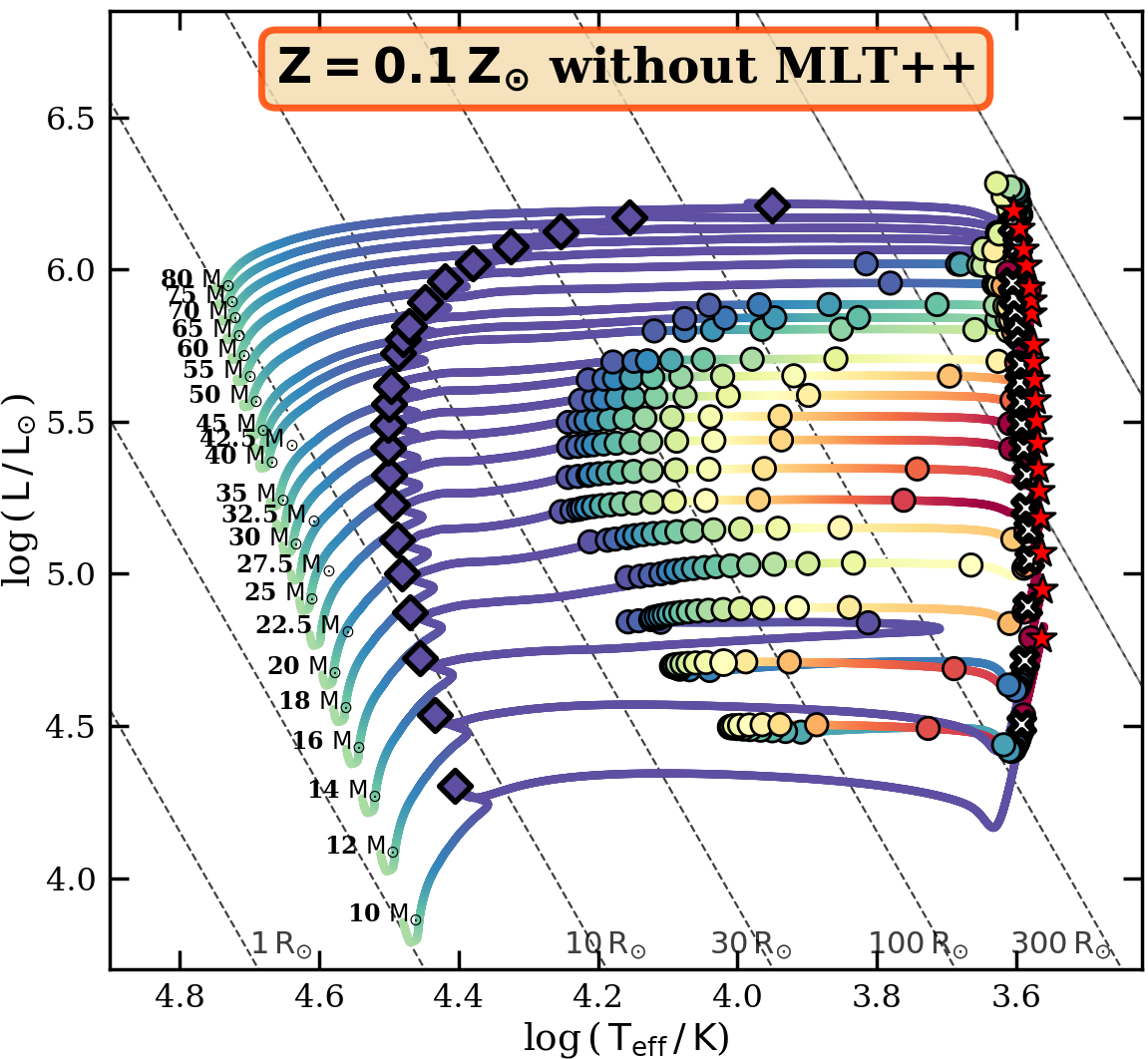} &
\includegraphics[width=0.45\textwidth,height=210px]{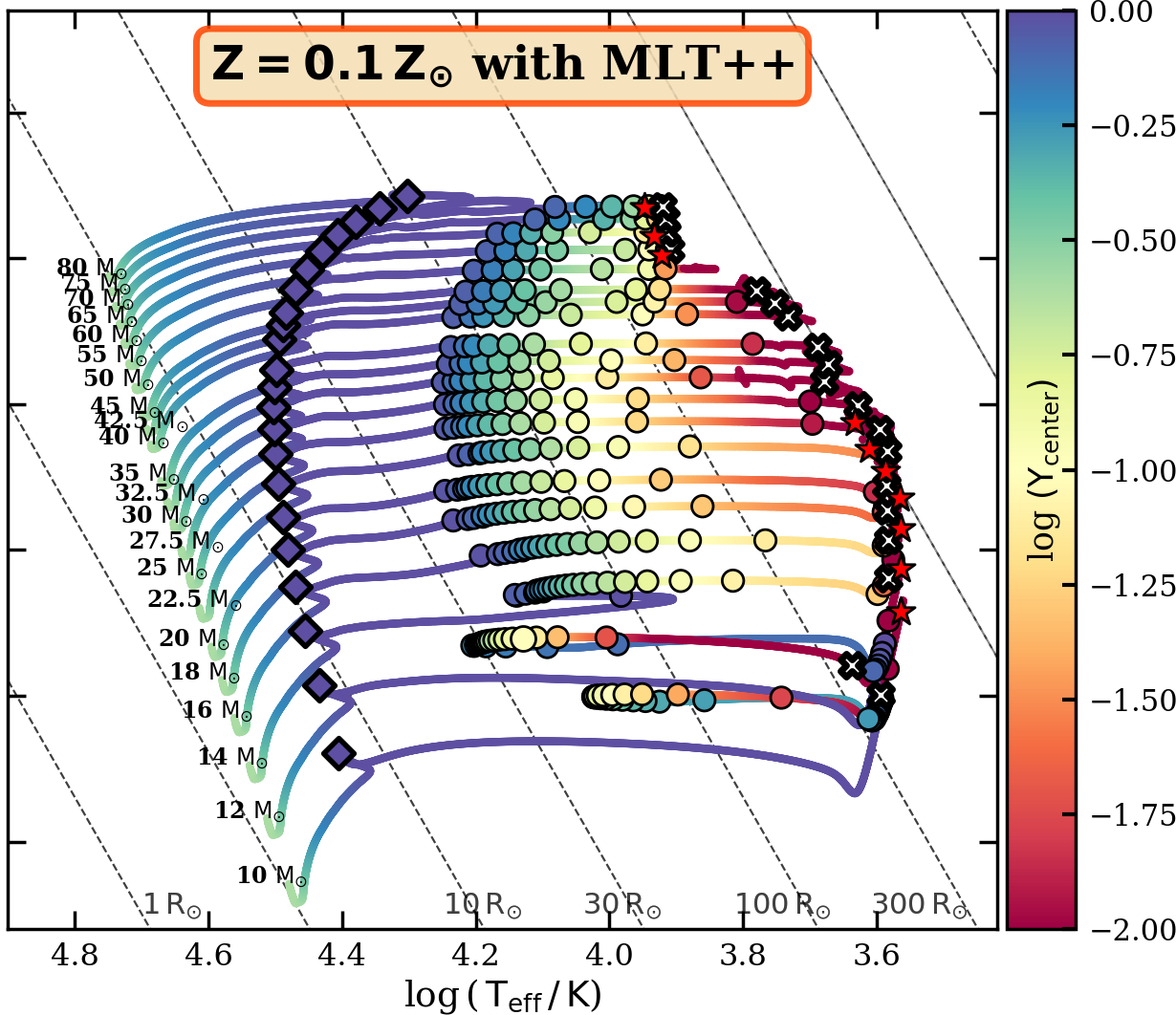} \\
\end{tabular}
\caption{Effect of MLT++ on the evolution of massive stars in the HR diagram. The various scatter points have the same 
meaning and color scale as in Fig.~\ref{fig.HRD} and \ref{fig.HRD_app}. In the 
most massive stars, in which the density in regions with $L_{\rm rad} \, / \, L_{\rm Edd} > 1$ is so small that it 
limits the efficiency of convection, a temperature gradient becomes superadiabatic and a 
density inversion develops in order to stabilize the structure (see App.~\ref{sec:App_mltpp}). The MLT++ approach 
gradually smooths the temperature gradient and reduces superadiabaticy, also removing density inversions and the 
$L_{\rm rad} \, / \, L_{\rm Edd}$ excess in the most massive stars. A direct consequence of the MLT++ treatment of 
convection is that the effective temperature becomes higher, as illustrated by the above figure. }
\label{fig.HRD_app_mltpp}
\end{figure*}

The model in Fig.~\ref{fig.bin_param_variations_e} includes the MLT++ approach in MESA, which was developed to model convection in 
radiation-dominated superadiabatic envelope layers \citep{Paxton2013}. In short, the MLT++ option enforces a 
reduction in the temperature gradient in radiation-dominated convective zones, which effectively reduces 
superadiabacity and prevents the formation of density inversions. A more detailed explanation is provided in 
Appendix~\ref{sec:App_mltpp}, where we also briefly discuss a link between density inversions in envelopes of 
massive stars and enhanced mass-loss rates beyond the Humphreys-Davidson limit. Crucially, 
MLT++ also increases the effective temperature of the model and thus reduces the radius expansion of massive models, 
which in turn reduces the parameter space for mass transfer in massive wide systems. This is shown the 
bottom left panel of Fig.~\ref{fig.bin_param_variations} and in the comparison of two HR diagrams of stellar 
tracks computed with and without MLT++ in Fig.~\ref{fig.HRD_app_mltpp}. Other methods of preventing density inversions
(e.g., using the density scale height or increasing the mass-loss rate) have a similar 
effect of increasing the effective temperatures of massive supergiants \citep[e.g.,][]{Maeder1987}.

Finally, in Fig.~\ref{fig.bin_param_variations_f} we show a model with evolution tracks of rotating stars with an 
initial angular velocity $\Omega / \Omega_{\rm crit} = 0.4$. This value was chosen to allow for a comparison with 
the GENEVA tracks \citep[e.g.,][]{Georgy2013,Groh2019}. Similarly to \citet{Schootemeijer2019}, we find that such 
moderate rotation does not have a significant effect on the post-MS radius evolution. However, we note that for a more 
rapid initial rotation rate $\Omega / \Omega_{\rm crit} \approx 0.6,$ the most massive models in our grid at $Z = 
0.1\zsun$ ($\gtrsim 65 \msun$) would enter the chemically homogeneous evolution regime.

Similar figures to Fig.~\ref{fig.bin_param_variations} for another four metallicities ($Z = 0.4\zsun$, $0.2\zsun$, 
$0.04\zsun$, and $0.01\zsun$) can be found in the Appendix in Fig.~\ref{fig.app_bin_param_variations}. In 
Sec.~\ref{sec:obs_b_vs_r} we show that for $Z = 0.2\zsun$ none of the 
model variations discussed above, except for the model with $\alpha_{\rm sc} = 1$, can be confidently excluded based on 
the observed samples of supergiants in the SMC 
($Z_{\rm SMC} \approx 0.2\zsun$). At even lower metallicities, the model uncertainties are all the more unconstrained.

\subsection{Convective envelope donors}

\label{sec.res_conv_env}

Whether or not a donor star has an outer convective envelope is particularly important for the question of mass 
transfer stability \citep[e.g.,][]{Hjellming1987,Soberman1997}. In short, convective stars respond to mass 
loss by adiabatically expanding\footnote{This may no longer be true 
in stars with a sufficiently well-developed superadiabatic layer in the outer envelope \citep{Woods2011,Pavlovskii2015}.}
, whereas stars with radiative envelopes respond by adiabatically contracting. For this 
reason, mass transfer from stars with outer convective envelopes is more likely to become dynamically unstable, 
possibly leading to a CE evolution \citep[][]{Webbink1984,Hjellming1987,Ivanova2013,Ge2015, Pavlovskii2015}.
To take this into account, population synthesis codes need to assume different 
stability criteria depending on whether the donor has a radiative or convective envelope.
Analytical fits to evolutionary tracks 
that many of such codes are based on \cite[e.g., the SSE/BSE method][]{Hurley2000,Hurley2002} do not usually 
include an explicit information about the type of the stellar envelope. For this reason, stability criteria are 
sometimes based on the evolutionary type of the star alone \citep[e.g.,][]{Belczynski2008, VignaGomez2018}. This can lead to large 
errors: as Fig.~\ref{fig.bin_param_ranges} shows, the occurrence of outer convective envelopes cannot be unambiguously 
linked to any particular evolutionary phase. In particular, the mass transfer instability (and CE evolution) may be 
substantially overpredicted in this calculation if CHeB donors are assumed to respond to mass loss as convective stars.

\begin{figure}
      \includegraphics[width=\columnwidth]{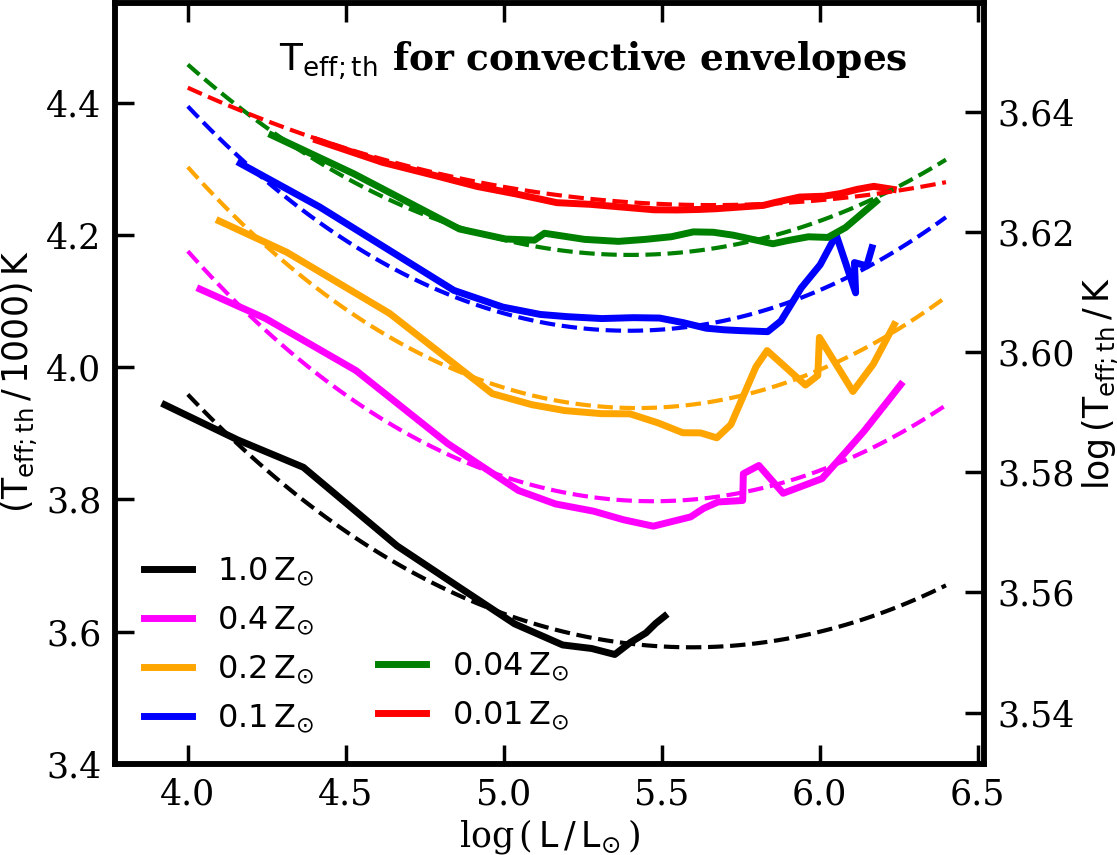}
    \caption{Threshold effective temperature $T_{\rm eff;th}$ below which at least 10\% of the outer mass in a star is 
convective. Solid lines correspond to our stellar models, and the dashes lines show the numerical fit described with 
Eqs.~\ref{eq:fit1} and \ref{eq:fit2}. } 
    \label{fig.conv_env_thresh}
\end{figure}

Instead, mass transfer stability criteria should include a relation to the effective temperature of the donor star, 
which is the primary factor that determines the occurrence of a convective envelope. In Fig.~\ref{fig.conv_env_thresh} we plot
as a function of luminosity and for various metallicities the threshold effective temperature $T_{\rm eff;th}$ 
below which at least 10\% of the outer mass of the star is convective. Solid lines correspond to the stellar tracks 
computed in this work (see Fig.~\ref{fig.HRD} and \ref{fig.HRD_app} for HR diagrams). Dashed lines are the result of a 
numerical fit described by the following relation:
\begin{equation}
 {{\rm log} (T_{\rm eff;th} / {\rm K})} = a_1 \, {{\rm log} (L/L_{\odot})}^2 + a_2  \,  {{\rm log} (L/L_{\odot})} + a_3 
\label{eq:fit1}
,\end{equation}
where the coefficients $a_i$ for $i$ in $\{1,2,3\}$ are given as a function of metallicity,
\begin{equation}
 a_i = b_{1i} \, {{\rm log} (Z/Z_{\odot})}^2 + b_{2i} \, {{\rm log} (Z/Z_{\odot})} + b_{3i}  \label{eq:fit2}
.\end{equation}
The values of the nine $b_{ij}$ coefficients are given in Table~\ref{tab.conv_fit}. Notably, $T_{\rm eff;th}$ in Fig.~\ref{fig.conv_env_thresh} depends 
on $Z$: the lower the metallicity, the higher the threshold temperature for convective envelopes. 
The reason is that the Hayashi line depends on metallicity for a given value of the mixing-length parameter, where fully convective stars of high $Z$ have larger 
radii (and lower effective temperatures) than their lower metallicity counterparts.
It should be noted, however, that the debate on whether the effective temperatures of RSGs (an indication for the Hayashi line)
do indeed depend on their metallicity is still ongoing.
Based on effective temperatures derived from spectral energy distribution fits, \citet{Davies2015} and \citet{Patrick2015} did not find any significant 
trend between $Z$ and $T_{\rm eff}$ of RSGs of various metallicities from several different host galaxies (although this conclusion may depend on 
the method for determining $T_{\rm eff}$, see Sec.~4.5 of \citealt{Chun2018}). \citet{Britavskiy2019} performed a similar analysis on a sample of RSGs 
from dwarf irregular galaxies from the Local Group, arguing in favor of a trend in which the minimum effective temperature increases toward lower metallicities. 
Another piece of the puzzle has recently been added by \citet{Chun2018}, who found that the inferred effective temperatures of RSGs in various local galaxies 
could be better matched by stellar models in which the mixing length increases with metallicity \citep[see also ][]{Tayar2017}.
With all these results in mind, we stress that our fits in Fig.~\ref{fig.conv_env_thresh} are primarily meant to serve as a simple and approximate method 
for determining the transition from radiative to convective-envelope giants in population synthesis.
We especially urge caution when a connection is made between the metallicity trend in Fig.~\ref{fig.conv_env_thresh} 
and expected effective temperatures of RSGs or maximum stellar radii of stars at different metallicities.

\begin{table}
\caption{Values of $b_{ij}$ coefficients in Eq. \ref{eq:fit2}.}              
\label{tab.conv_fit}      
\centering                                      
\begin{tabular}{c | c c c}          
\hline                     
$b_{ij}$ & $i = 1$ & $i=2$ & $i=3$ \\    
\hline
    $j = 1$ & $-0.0060$ & $-0.0066$ & $0.0173$ \\      
    $j = 2$ & $0.0596$ & $0.0587$  & $-0.1940$ \\
    $j = 3$ & $-0.1637$ & $-0.1967$  & $4.0962$ \\
\hline                                             
\end{tabular}
\end{table}

\section{Comparison with observations: clues for post-MS expansion}
\label{sec:obs_b_vs_r}

The question of the post-MS expansion and blue-to-red evolution of massive stars is a long-standing problem in 
stellar astrophysics, with models being very sensitive to the adopted assumptions \citep[see e.g.,][]{Langer1991, 
Eldridge2008,Meynet2013}. Most notably, it was shown that the degree of the 
HG expansion crucially depends on the efficiency of overshooting and semiconvective mixing 
\citep{Langer1989,Stothers1992,Langer1995}, with rotational mixing also playing a role, see \citet{Georgy2013}. The 
observable BSG-to-RSG ratio can be used as a diagnostic tool in an attempt to calibrate the models 
\citep[e.g.,][]{Eldridge2008,Georgy2013,Choi2016}. Traditionally, the BSG group comprises 
O, B, and A stellar types (${\rm log} T_{\rm eff} \gtrsim 3.9$) and the RSG group is composed of K and M supergiants 
\citep[${\rm log} T_{\rm eff} \lesssim 3.7$, e.g.,][]{Eggenberger2002,Massey2003}. In this way, the BSG group 
includes MS stars (O-type stars) and is dominated by them in terms of numbers.
This limits the accuracy of the method in constraining 
specifically the post-MS phase of evolution.

A slightly modified approach has recently been carried out by 
\citet{Schootemeijer2019}, who constructed a sample of BSGs  from the SMC by combining only the B- and A-type stars 
($3.85 \gtrsim {\rm log} T_{\rm eff} \lesssim 4.4$). In this way, they were able to probe predominantly the post-MS 
phase of evolution at the SMC metallicity (see Fig.~\ref{fig.HRD}). Notably, \citet{Schootemeijer2019} showed that models 
that assume the Schwarzschild criterion for convection (an assumption that is often made in previous calculations of 
massive-star models), which is equivalent to extremely efficient semiconvection, tend to underestimate the degree of the 
post-MS expansion and underpredict the number of RSGs. The disagreement of such models with observations has been pointed out by 
previous authors \citep[e.g.,][]{Georgy2013}. 

\subsection{SMC-like metallicity ($Z = 0.2 \zsun$)}

\label{sec:obs_02Z}

We took an approach similar to that of  \citet{Schootemeijer2019} and compared our $Z = 
0.2 \zsun$ models with the observed population of luminous ($\logL \geq 4.7$) supergiants in the SMC 
with effective temperatures $ \logteff \lesssim 4.35$. Following \citet{Drout2009}, we defined RSGs as supergiants cooler than a threshold 
effective temperature ${\rm log} (T_{\rm eff;thresh} / {\rm K}) = 3.68$.
We aim to compare the number of RSGs to the number of hotter supergiants from the 
effective temperature range between $\logteff = 3.68$ and $4.35$. 
In this way, we constrain the post-MS expansion of our models, which either expand all the way to the red giant 
branch during the HG phase and spend most of the CHeB lifetime as RSGs, or they complete the HG expansion 
as hotter stars and spend a significant fraction of the CHeB lifetime with $\logteff$ between $3.68$ and $4.35$.
We note that in our $Z = 0.2\zsun$ tracks the convective RSG branch is located at slightly lower temperatures
(at $\logteff \lesssim 3.6$, depending on luminosity, see Fig.~\ref{fig.HRD})
than the assumed threshold temperature ${\rm log} (T_{\rm eff;thresh} / {\rm K}) = 3.68$ and than the 
measured effective temperatures of most of the RSGs in the SMC \citep[e.g., obtained through spectral energy distrigution fitting][]{Davies2018}.
This is a likely sign that our assumed mixing length parameter $\alpha = 1.5$ is somewhat too small.
For example, \citet{Chun2018} reported that models with $\alpha \approx 2$ agree better with temperatures of cool supergiants from the SMC.
However, this discrepancy in temperatures of RSGs does not significantly affect the prediction for the number of RSGs from our models 
because for the majority of the models, the time that stars spend with $\logteff$ between $3.6$ and $3.68$ is far shorter than the CHeB lifetime.
We point out that other factors such as the effect of binary interactions
or the shortcomings of the mixing-length theory in massive convective-envelope giants (see App.~\ref{sec:App_mltpp})
are likely to be a higher-order source of uncertainty.

For the sample of RSGs, we relied on the most recent investigation of the population of cool supergiants in the SMC by 
\citet{Davies2018}. The authors combined a number of input catalogs in order to construct a highly complete 
sample of cool luminous SMC stars with effective temperatures $\logteff \lesssim 3.74$ \citep[based on the effective 
temperature scale of][]{Tabernero2018}. Out of 151 stars with $\logL \geq 4.7$ in their sample, 147 meet our RSG criteria ($\logteff < 
3.68$).

\begin{figure}
      \includegraphics[width=\columnwidth]{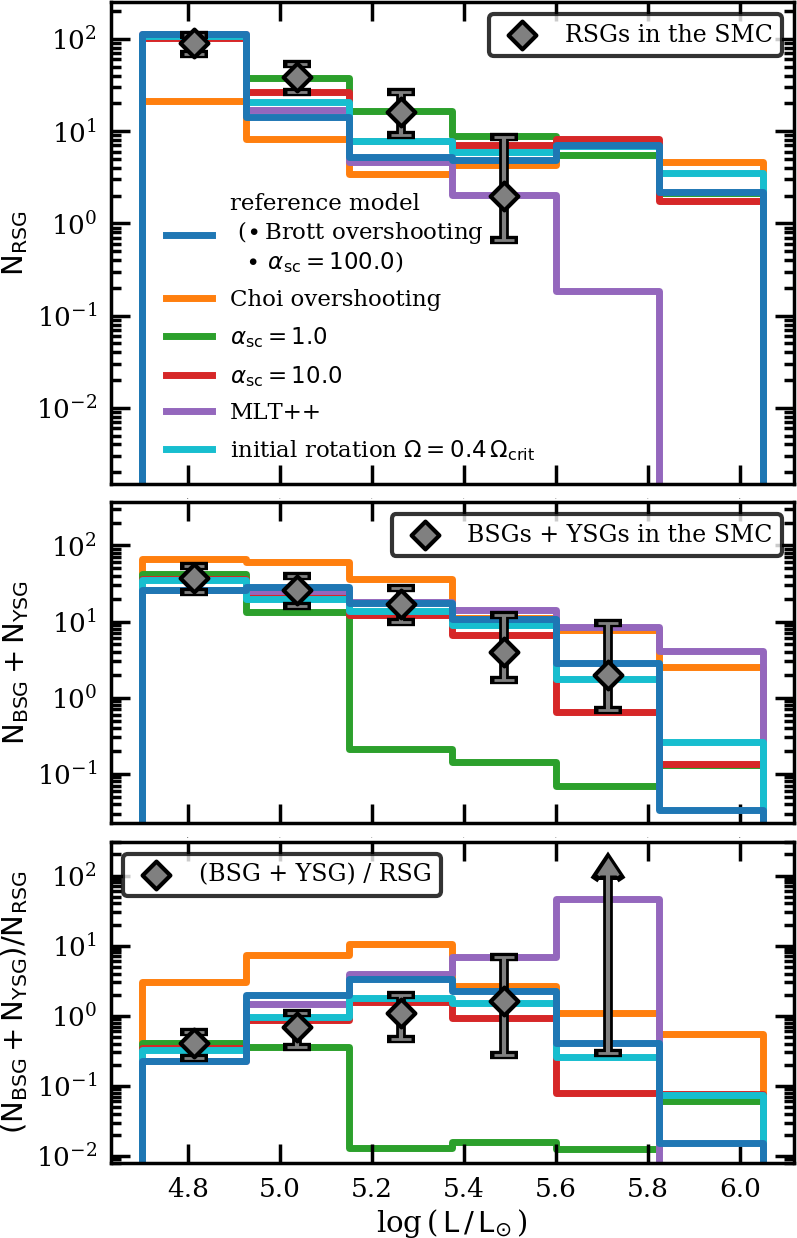}
    \caption{ Distributions of RSG and BSG plus YSG number counts (top and middle panels, respectively),
    as well as the number ratio of BSGs plus YSGs to RSGs (bottom panel) in logarithmic 
luminosity bins between $4.7$ and $6.05$ in $\logL$. Diamond scatter points correspond to observational samples of 147 
RSGs (with $\logteff < 3.68$) and 86 BSGs plus YSGs (with $\logteff$ between $3.68$ and $4.35$). 
Uncertainties correspond to 90\% confidence limits assuming a Poissonian distribution for the number of supergiants in each bin. 
Only a lower limit is available for the ratio of BSGs plus YSGs to RSGs in the luminosity bin $5.6 \lesssim \logL \lesssim 5.8$.
Solid-line histograms, normalized to 233 stars in total, correspond to several different sets of stellar tracks at $Z = 0.0034 = 0.2\Zsun$ 
metallicity.} 
    \label{fig.BvR_SMC}
\end{figure}

As explained above, we mean to compare the number of RSGs with stars in the temperature range $3.68 < \logteff < 4.35$.
This includes both yellow supergiants (YSGs) 
\citep[$3.68 \geq \logteff \leq 3.88$ ][]{Drout2009} and BSGs with $\logteff$ between 
$3.88$ and $4.35$. To construct a combined sample of YSGs and BSGs, we relied on several different 
studies. In the temperature range $3.68 < \logteff < 3.74$, four YSGs were reported by 
\citet{Davies2018}. A survey by \citet{Neugent2010}, targeting YSGs and cool BSGs and identifying their SMC membership 
based on their radial velocities, found 40 stars in the temperature range $3.74 < \logteff < 4.1$ (eight of which were YSGs at $\logteff < 3.88$). The authors estimated 
the completeness of their sample to be high and that only a few rather than tens of objects were missing, especially at the higher 
luminosity end that we are interested here (i.e., $\logL > 4.7$). Finally, 42 BSGs in the SMC with 
temperatures within $4.1 < \logteff < 4.35$ have recently been reported by \citet{Kalari2018} with luminosities that meet 
our criteria. In total, the authors provided physical parameters of 69 BSGs, which were selected from 
the OGLE-II I-band imagining survey based on availability of multi-epoch photometry. Notably, a total of 110 BSGs were 
identified in the OGLE-II I-band data, which in turn is a fraction of 179 SMC BSGs reported in the catalog of massive 
SMC stars by \citet{Bonanos2010}. 

The combined number of BSGs and YSGs in our sample is thus 86, and the  ratio of BSGs plus YSGs to RSGs is equal to $86/147 \approx 0.59$. As described above, the number of luminous BSGs in our sample is likely underestimated in the temperature range 
$4.1 < \logteff < 4.35$, possibly by a factor of a few. Nonetheless, we consider the samples of RSGs and BSGs plus YSGs
constructed in this way to be representative of the underlying population and sufficiently complete to compare with our 
models.

In Fig.~\ref{fig.BvR_SMC} we plot luminosity distributions of RSGs and BSGs plus YSGs (top and middle panels, respectively),
and the number ratio of BSGs plus YSGs to RSGs (bottom panel). 
Diamond points correspond to the observational samples described above, with Poissonian uncertainties marked as well (90\% confidence level).
Solid-line histograms correspond to several 
different sets of stellar tracks at $Z = 0.0034 = 0.2\Zsun$ (see Sec.~\ref{sec.res_model_var} for a 
description of various models). To construct the histograms, we assumed a constant star formation rate and the Salpeter 
initial mass function slope for massive stars \citep[$dN/dM \propto M^{-2.35}$][]{Salpeter1955}. We normalized the 
histograms to the total number of 233 supergiants in the observed sample. We stress that the comparison in 
Fig.~\ref{fig.BvR_SMC} is rather tentative because it attempts to compare single stellar tracks with a population of 
stars that are likely affected by binary evolution \citep[most massive stars are born in binaries;][]{Sana2012,Moe2017}. 
Additionally, the BSG, YSG, and RSG samples are not constructed in an unbiased and homogeneous way.

In the low-luminosity range ($\logL \lesssim 5.2$), most of the models agree reasonably well with 
observations, up to a factor of a few in BSG plus YSG and RSG counts. Because the character of the comparison is tentative, we deem 
these discrepancies insufficient to confidently reject the models. The possible exception is the 
model with Choi overshooting, for which the ratio of BSGs plus YSGs to RSGs appears to be significantly too high in all three low 
luminosity bins ($\logL < 5.4$). It seems challenging to explain this discrepancy by the unaccounted-for binary effects 
because binary interactions would likely further reduce the theoretically expected number of RSGs (e.g., stars 
that lose mass in mass transfer become bluer). Unless the observed sample is missing most of the SMC BSGs or YSGs and the real 
ratio of BSGs plus YSGs to RSGs is a few times higher, the model with Choi overshooting is most likely disfavored.

In the high-luminosity range the discrepancies between models and observations become larger. In particular, no RSGs are 
observed in the SMC with luminosities above $\logL \approx 5.6$ (the highest luminosity reported by \citealt{Davies2018} 
is $\logL = 5.55 \pm 0.01$), while all models but one predict at least a few of them in every 233 SMC supergiants. At 
the same time, several BSGs are observed with luminosities between $5.6$ and $5.8$ in $\logL$.
This may indicate that stars with masses $\gtrsim 40 \msun$ at $Z = 0.2\zsun$ either do not expand to become RSGs at 
all or that the RSG phase is extremely short in this case. This might simply be due to rotation. For instance, 
\citet{Schootemeijer2019} found fewer RSGs in the upper part of the HR diagram in their models with rotation than in 
the nonrotating tracks (initial velocity of $300 \rm \, km \, s^{-1}$, see their Fig.~11). A short-lived (and therefore 
difficult to observe) RSG phase of $M \gtrsim 40 \msun$ stars might be a result of increased mass-loss rates due to 
dust-driven winds \citep{Chieffi2013} or proximity to the Eddington limit \citep{Chen2015}.

Alternatively, the lack of 
RSGs above $\logL \approx 5.6$ may suggest that density inversions that would form in envelopes of such 
stars (in 1D models at least) either do not form at all because of an additional energy transport mechanism that
prevents superadiabacity, or that such inversions are prone to instabilities and extensive mass loss (possibly the luminous blue variables 
phenomenon). For this reason, the model in which the formation of density inversions is prevented (the MLT++ approach) 
does not have solutions in the upper right corner of the HR diagram (see Fig.~\ref{fig.HRD_app_mltpp}) and is 
statistically consistent with the lack of RSGs above $\logL = 5.6$ in Fig.~\ref{fig.BvR_SMC}. On the other hand, the 
MLT++ model predicts about 15 BSGs above $\logL = 5.6,$ while only 2 such stars are present in the observed samples. 
Finally, the lack of RSGs from $M \gtrsim 40\msun$ stars might arise because binary interactions dominate the evolution of 
massive stars and prevent them from becoming as large as $\gtrsim 2000 \rsun$. 
Regardless of the reason, it seems unlikely that it has much to do with the tendency of some of the low-metallicity stars 
to begin the CHeB phase already as BSGs. 
We note that there appears to be a general lack of very massive (luminous; $\logL > 5.8$) stars in the observed samples 
compared to our theoretical predictions. This might be an indication that very massive stars ($M \gtrsim 60 \msun$) are rare in 
the SMC for reasons that are yet to be understood.

The middle panel of Fig.~\ref{fig.BvR_SMC} shows that most of the models agree reasonably well with 
observations. The only exception are tracks that are computed with the lowest efficiency of semiconvective mixing, that is, 
$\alpha_{\rm sc} = 1$. In this case, the model predicts no BSGs or YSGs with 
luminosities $\logL > 5.2$ in the observed sample, while in reality, there are 23 such stars. 
Consequently, the ratio of BSGs plus YSGs to RSGs at $\logL > 5.2$ in the model with $\alpha_{\rm sc} = 1$ is strongly inconsistent with 
the observations (see the bottom panel).

One possible explanation for this discrepancy that should be considered is the effect of mass transfer evolution in binary systems. 
As we show in our upcoming paper, Klencki et al. (in prep), 
donor stars in stable mass transfer events are unlikely to help alleviate the issue because 
in the case of the $\alpha_{\rm sc} = 1$ model, they would quickly become too hot to qualify for our BSG or YSG categories. 
On the other hand, sufficiently rejuvenated accretor stars or products of stellar mergers from early case B (or very late case A) mass transfer
might produce CHeB supergiants that evolve as BSGs or YSGs with luminosities $\logL > 5.2$ \citep{Justham2014}. 
Because the parameter space for this type of evolution is limited, however, it is not clear whether such mass gainers or stellar mergers can fully account 
for the underabundance of BSGs plus YSGs from single stellar evolution in the model with $\alpha_{\rm sc} = 1$ \citep[although mass gainers in general may be 
quite common among apparently single stars;][]{deMink2014}.
With this caveat in mind, we consider the model with $\alpha_{\rm sc} = 1$ (and similar models in which most of the CHeB evolution at $Z = 
0.2\zsun$ takes place during the RSG stage) to be disfavored by the observations.

\subsection{Solar metallicity}
\label{sec:obs_10Z}

\begin{figure}
      \includegraphics[width=\columnwidth]{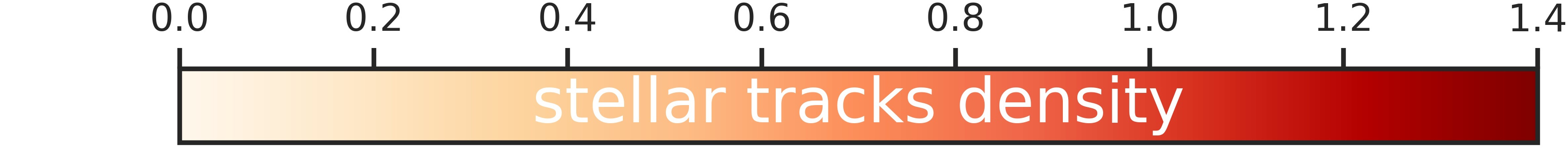}
      \includegraphics[width=\columnwidth]{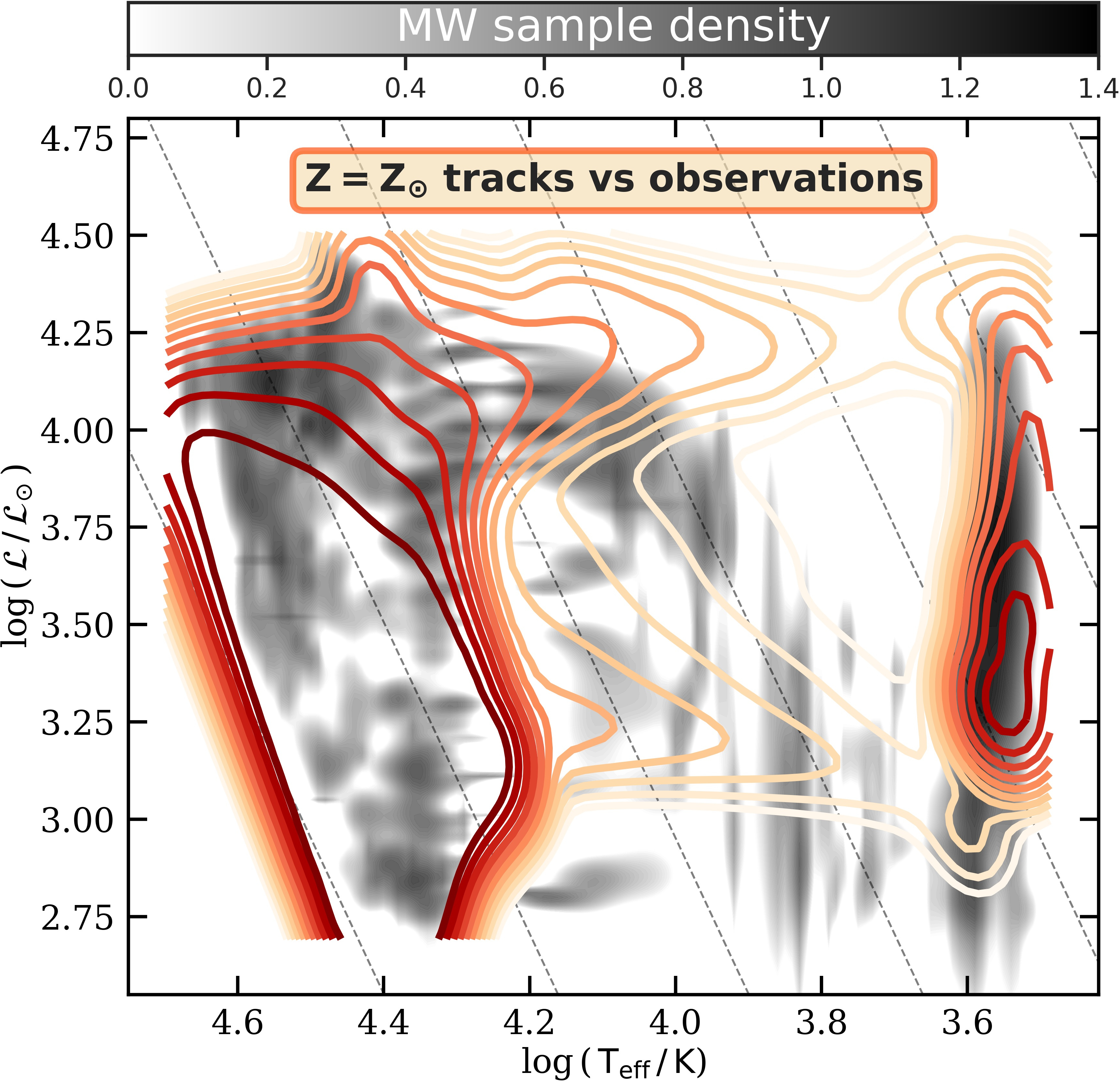}
    \caption{Comparison of our stellar tracks computed at solar metallicity with the observed sample of Galactic 
massive stars collected by \citet{Castro2014}. The observations are plotted as a density map in a 
logarithmic grayscale. We overplot contours of the theoretical distribution of stars 
obtained from our reference model at $Z = \zsun$ in color (also in a logarithmic scaling). The lack of observed stars in the parameter range of $3.7 \lesssim 
{\rm log} (T_{\rm eff}/K) \lesssim 3.9$ and $3.5 \lesssim {\rm log} (\mathcal{L}/\mathcal{L}_{\odot}) \lesssim 4.0$ is a 
tentative piece of evidence that solar metallicity massive stars (mass range of about $16$ to $40 \msun$) expand  until the red giant branch during the HG phase.}
    \label{fig.sHR_solar}
\end{figure}

In the solar metallicity case, it is not feasible to use the ratio of BSGs plus YSGs to RSGs to constrain the post-MS radius 
evolution because of the larger (and more mass-dependent) sizes of MS stars. This makes it difficult to probe the 
post-MS phase of evolution alone. Additionally, it is more challenging to obtain a reasonably complete sample of supergiants 
in the Milky Way. Instead, we take a different approach to compare our $Z = \Zsun$ tracks with observations.
Recently, \citet{Castro2014} compiled a Galactic sample of massive stars in the spectroscopic HR diagram 
\citep[sHR,][]{Langer2014} and compared them to single stellar tracks from \citet{Brott2011} and \cite{Ekstrom2012}. 
Their sample is not complete;  it was collected from various surveys that 
often had different scientific goals, therefore it is biased in several ways. This leads to an overpopulation of stars in certain 
regions such as the instability strip or the O- and early B-type temperature ranges. Nonetheless, certain sharp 
features of stellar tracks, such as the position of the TAMS or the red-giant branch, can be constrained.

In Fig.~\ref{fig.sHR_solar} we compare the sample collected by \citet{Castro2014} with our reference model in the sHR 
diagram (where $\mathcal{L} = T_{\rm eff}^4 / g$). The observations are plotted in the background as a grayscale 
density map (reproduced after \citealt{Castro2014} with permission from the authors). We overplot contours of 
the theoretical distribution of stars obtained from our single stellar tracks at $Z = \zsun$ in color. The theoretical 
prediction was obtained under the assumption of the Salpeter initial mass function for $M \geq 10 \msun$ \citep{Salpeter1955}. We also 
assumed that all such massive stars are formed in binary systems \citep[e.g.,][]{Duchene2013,Moe2017} and that 
their evolution can be approximated with our single tracks only until the eventual RLOF. For the initial orbital period 
distribution, we assumed $dN/d(\rm log \,P) \propto {(\rm log \, P)}^{-0.55}$ for ${\rm log \, P}$ in range $[0.15, 4.0]$ 
\citep{Sana2012}, and for simplicity, we assumed a fixed initial mass ratio of $0.5$ and circular orbits. We assumed that 
when a star initiates RLOF and starts losing mass from its envelope, then it immediately becomes a hot helium star that 
does not fit into the $T_{\rm eff}$ range in Fig.~\ref{fig.sHR_solar}, and we did not include it in the theoretical 
distribution of stars in the figure. In our upcoming paper with binary models (Klencki et al. in prep.), we find 
that this is a good assumption in the case of stellar models in which the rapid HG expansion continues 
until the red giant branch (e.g., in the $Z = \zsun$ case), but not in the case of models with much smaller HG 
stars (e.g., the $Z = 0.2\zsun$ case). The 2D probability density function of stars in the sHR diagram was estimated from discrete stellar tracks 
computed in this study through a kernel-density estimation using Gaussian kernels \citep[method \texttt{stats.gaussian\_kde} 
from the SciPy package][]{Scipy2019}.

The agreement of our theoretical distribution with observations is not perfect, although it is certainly not worse than 
in the case of the tracks by \citet{Brott2011} or \citet{Ekstrom2012}, see \citet{Castro2014}. Perhaps the most significant 
discrepancy is
that we predict many more RSGs ($\logteff \lesssim 3.68$) and B-type stars ($\logteff$ between $4$ and $4.2$) in the 
high-$\mathcal{L}$ region above $4.25$ than what is observed. This might be an indication that the line-driven wind 
mass-loss prescriptions used in our models overestimate the mass-loss rates at solar metallicity (when a star loses 
mass from its envelope, it tends to move upward in the sHR diagram because $\mathcal{L} \propto L/M$). In the case of 
B-type massive stars, this is supported by some empirical evidence \citep[e.g.,][]{Oskinova2011,Hainich2018}.

From the point of view of the post-MS expansion and blue to red evolution, the interesting region of the sHR diagram is 
the parameter range of about $3.7 \lesssim {\rm log} (T_{\rm eff}/K) \lesssim 3.9$ and $3.5 \lesssim {\rm log} 
(\mathcal{L}/\mathcal{L}_{\odot}) \lesssim 4.0$. This corresponds to the YSG position of massive stars in the range 
of about $16$ to $40 \msun$. If, similarly to the subsolar metallicity case, these stars were to slow down their 
HG expansion and continue with the CHeB evolution as BSGs at first and later as YSGs, 
then we would expect them to populate this region in the sHR diagram. 
The lack of observed stars in this parameter range in the sample compiled by \citet{Castro2014} (which
notably includes stars in the same $T_{\rm eff}$ range, but at lower $\mathcal{L}$) is a tentative piece of evidence 
that solar metallicity massive stars expand to the red giant branch during the HG phase.

\section{Discussion}
\label{sec:discussion}

The stellar tracks computed in this work show a gradual metallicity trend in the post-MS radius expansion of stars: the lower the metallicity, the stronger the tendency of massive stars to limit the rapid HG expansion and 
regain thermal equilibrium as helium-burning supergiants already in the blue part of the HR diagram, as much more compact 
stars than RSGs. In other words, the maximum size of massive HG stars decreases with metallicity. This trend has previously been described by \citet{Brunish1982} and \citet{Baraffe1991} and also appears in some of the more recent single stellar tracks 
computed by other authors \citep[e.g.,][]{Marigo2001,deMink2008,Ekstrom2012,Georgy2013,Tang2014,Choi2016,Groh2019}. 
In the following sections we discuss the origin and the robustness of this result, as well as some of its implications 
in the context of massive binary evolution.

\subsection{Why does the size of massive HG stars decrease with metallicity?}

\label{sec.disc_Z_trend}

Some low-metallicity models begin the CHeB phase as more compact stars than their 
higher metallicity counterparts because of the complicated interplay between at least two metallicity-dependent factors: 
higher central temperatures and central densities at TAMS of low-$Z$ stars (resulting in a less pronounced mirror 
effect during the HG phase) and lower opacities in their envelopes (enabling smaller radius solutions) (see 
\citet{Brunish1982,Baraffe1991,Langer1991,Groh2019}).
The wind mass-loss rate also plays a role. In general, a giant star with stronger winds has a more extended envelope 
than an otherwise very similar model with weaker winds, resulting in an increased rate of redward evolution 
\citep{Brunish1982}. A notable exception are the models of MS and WR stars with inflated envelopes, for which the 
inflated envelope solution was found to disappear for mass-loss rates above a certain critical value 
\citep{Petrovic2006}.

However, the way in which a star of a given mass and metallicity expands after the end of MS is not a robust result of 
evolutionary computations. In some cases, it depends sensitively on rather uncertain input parameters. 
For example, even at a metallicity as low as $Z = 0.0017,$ it is possible to have stellar tracks in which the 
most massive stars reach the RSG stage during the short-lived HG phase, similarly to solar metallicity models
(see the model with $\alpha_{\rm sc} = 1$ in Fig.~\ref{fig.bin_param_variations}). 

The key factor that essentially determines the post-MS radius expansion in cases sensitive to model 
assumptions is the H/He abundance profile in the layer just above the helium core. Depending on the 
efficiency of mixing, it can vary from a shallow linear slope (i.e., left at the end of the MS by 
the retreating convective core) to an extremely steep H/He gradient, almost a step function 
\citep[e.g.,][]{Schootemeijer2018}. The abundance profile is important because it determines the location and strength of 
hydrogen-shell burning, which in turn determines the size of HG stars (see  \citealt{Kippenhahn1990} and 
also \citealt{Georgy2013}).
In Appendix~\ref{sec:app_mixing} we show in detail how different efficiencies of semiconvective mixing ($\alpha_{\rm sc} 
= 1$ or $100$) result in different H/He abundance profiles above the helium cores of our $25\msun$ models, leading to 
vastly different degrees of HG radius expansion \citep[see also][]{Schootemeijer2019}.

In addition to semiconvection, other uncertain mixing processes affect the H/He abundance in the layer above the helium 
core. For example, convective core overshooting \citep{Langer1985,Stothers1992,Langer1995}, rotational mixing and 
instabilities \citep[][]{Georgy2013}, and most likely, 
shell overshooting. Purely numerical aspects such as the determination of convective boundaries in stellar evolution 
codes also play a role \citep[see, e.g., the convective pre-mixing approach in][]{Paxton2018,Paxton2019}. The effect of all these factors on the abundance profile above the helium core is uncertain and most likely degenerate 
\citep{Meynet2013}

\subsection{How robust is the trend with the metallicity?}

\label{sec.disc_robustness}

Because of the uncertainties in modeling the radial expansion of post-MS massive stars, it is vital to turn toward 
observations. The relative numbers of blue and yellow supergiants compared to red supergiants in the SMC indicate that already at $Z = Z_{\rm 
SMC} \approx 0.2 \zsun$ , some massive stars (roughly between $15$ and $40 \msun$) spend most of their CHeB lifetimes 
as BSGs or YSGs, see Sec.~\ref{sec:obs_02Z}. The same conclusion has been reached previously by \citet{Schootemeijer2019}. This 
allows us to reject models with relatively inefficient mixing in the zone above the helium core, for instance, the model with 
$\alpha_{\rm sc} = 1.0$. On the other hand, the lack of YSGs with spectroscopic luminosities 
$\mathcal{L/L_{\odot}} \gtrsim 3.6$ in the spectroscopic HR diagram of massive Galactic stars \citep{Castro2014} 
indicates that stars of the same mass but higher metallicity (e.g., $Z \sim \zsun$) burn helium predominantly as RSGs, 
see Sec.~\ref{sec:obs_10Z}. 

These two pieces of evidence increase the credibility of our reference set of stellar tracks in terms of predicting the 
radius evolution of massive stars. Notably, the metallicity trend in the degree of post-MS expansion is present in all 
our models with variations in the input physical parameters as well as in the stellar tracks with rotation (see 
Sec.~\ref{sec.res_model_var} and Fig.~\ref{fig.app_bin_param_variations} in the appendix). While the exact boundaries 
between evolutionary stages of the 
donor stars in interacting binaries in Fig.~\ref{fig.bin_param_ranges} should be considered rather uncertain, it 
appears that the trend with metallicity and its implications are a robust prediction of stellar models.

\subsection{WR stars formed through binary interactions}

\label{sec.disc_WR_formation}

As we illustrated in Sec.~\ref{sec:res_binpar}, the lower the metallicity, the more evolved are massive stars in 
binary systems at the point when they initiate mass transfer through RLOF. At $Z =\zsun$ and 
$0.4\zsun$, a giant donor is almost always a star that has just completed its MS evolution and is rapidly 
expanding during the HG phase. During mass transfer, such a star very quickly ($\lesssim 10^4\yr$, i.e., on a thermal 
timescale) loses almost its entire envelope and reveals its helium core \citep{Kippenhahn1967}. In this way, a WR star can 
be formed by means of binary interactions \citep[e.g.,][]{Maeder1994,Bartzakos2001}. The lack of a hydrogen envelope 
means that the helium core cannot grow in mass because of hydrogen-shell burning. Instead, the naked core loses mass in 
strong WR winds during the subsequent evolution. Recently, \citet{Woosley2019} quantified these two 
effects on the final core masses and other core properties with respect to single stellar evolution. They did so by 
computing models of single naked helium stars, starting from the onset of helium burning. This choice of 
the starting point assumes that binary interactions strip the hydrogen envelope during the HG evolution (i.e., immediately 
before the onset of CHeB), which is a well-justified assumption only in the high-metallicity case. 

However, low-metallicity massive stars tend to remain more compact during the HG phase. As a 
result, at $Z \leq 0.2 \zsun,$ the scenario in which the hydrogen envelope is lost during the HG evolution is no 
longer the dominating case. Instead, in somewhat wider systems (orbital periods $> 10-50$ days, depending on 
metallicity), the RLOF mass transfer is initiated only after a significant fraction 
of the helium-burning lifetime, see Fig.~\ref{fig.bin_param_ranges}. In such cases, the hydrogen envelope is 
preserved for longer and the helium core of the donor star has already grown in mass since the end of the MS as a result of 
hydrogen-shell burning. The remaining lifetime of such a donor after the start of RLOF is 
shorter than in the case of a HG donor star (in some cases, by more than an order of magnitude). This means that WR stars
originating from intermediate- to long-period interacting binaries at low metallicity have much shorter 
lifetimes than stars in higher-metallicity environments, which likely plays a role in the WR demographics in 
different host galaxies. Additionally, this shorter duration of the WR stage (after the RLOF) and the longer-lasting 
hydrogen-shell burning (before the RLOF and the resulting loss of the H-rich envelope) both work in favor of higher final core masses 
compared to stars that lose their hydrogen envelopes at the onset of CHeB \citep[as assumed in][]{Woosley2019}.

Interestingly, the time interval between the end of the MS and the onset of mass transfer through RLOF may affect the H/He abundance gradient at the surface of the resulting WR star. The longer the envelope is preserved, the more 
time there is for internal mixing (mainly semiconvection) to steepen the H/He gradient at the boundary of the helium 
core \citep[see Appendix~\ref{sec:app_mixing} and also][]{Schootemeijer2018}. These layers then become the surface layers 
of the WR star after the envelope is stripped. The surface H/He abundance gradients of WR stars might be indirectly 
inferred from observations and used to constrain the evolution of their progenitors \citep{Schootemeijer2018}.

\subsection{Mass transfer evolution and slow envelope stripping}

\label{sec.disc_mass_transfer}

As the metallicity decreases, increasingly more cases of mass transfer are initiated by slowly expanding CHeB 
giants (Fig.~\ref{fig.bin_param_ranges}). Such donors do not necessarily become stripped of their entire envelopes 
on a short thermal timescale as is traditionally predicted for evolved giant donors \citep{Kippenhahn1967} 
and implemented in population-synthesis codes \citep[e.g.,][]{Kruckow2018}. \citet{Klencki2019} showed an example of a 
detailed model of an interacting binary system comprising a $25 \msun$ donor star (a CHeB giant) and a $10 \msun$ BH 
accretor at subsolar metallicity $Z = 0.2 \zsun$, in which mass transfer continues on a nuclear timescale of helium 
burning until the end of the CHeB phase. Throughout the entire mass transfer phase, the donor star maintains several 
solar masses of the hydrogen-rich envelope ($\sim 2 \msun$ at the point of core-helium depletion). We discuss this type of 
mass transfer evolution in much more detail in Klencki et al. (in prep.).

When we assume that mass transfer from relatively compact CHeB stars could last for significantly longer than mass 
transfer from rapidly expanding HG donors (i.e., a nuclear timescale instead of a thermal timescale), then according to 
Fig.~\ref{fig.bin_param_ranges} we expect to find significantly more mass-transferring systems per unit of 
star formation rate in a low-metallicity galaxy compared to a high-metallicity one. This might be particularly 
interesting in the case of binaries with compact accretors, such as high-mass X-ray binaries. Observationally, X-ray 
binaries are indeed significantly more common in low-metallicity galaxies \citep[e.g., by a factor of $\sim 10$ 
in blue compact dwarf galaxies;][]{Brorby2014}.

\subsection{Late cases of RLOF: progenitors of type II supernovae, long GRBs, and binary BH mergers?}

\label{sec.disc_SN_GRB_BBH}

An interesting case is formed by systems in which the first RLOF takes place already very late into the evolution of 
the donor star, after most (or all) of the helium has been burned, for example, $Y_{\rm C} < 0.05$. Such a case 
of mass transfer is sometimes referred to as case C mass transfer and corresponds to donors at the final stages of CHeB 
and to HeHG and SHeB donors in Figs.~\ref{fig.bin_param_ranges} and \ref{fig.bin_param_variations}. 
We find that the parameter space for late case C mass transfer is a strong function of metallicity. 
At $Z \gtrsim 0.2 \zsun$ , such systems are very rare compared to all the other cases. At metallicities $Z \leq 0.1 
\zsun$ , the probability of late RLOF becomes significantly higher, up to $\sim 40\%$ of all systems at $Z = 0.01 \zsun$, see 
Fig.~\ref{fig.bin_param_ranges}.

In many systems that evolve through such a case of late mass transfer, only several thousand years are left from 
the point of RLOF to the collapse of the donor's core. This is not enough time to fully strip the 
hydrogen-rich envelope of a giant star, even if mass transfer is proceeding on a thermal timescale. 
Thus, a supernova from such a star would most likely appear as a hydrogen-rich type II supernova with a blue or 
yellow supergiant progenitor. In higher metallicity environments, a similar star would have been stripped of its 
hydrogen-rich envelope because mass transfer would have occurred far earlier during its evolution, and would explode as a type Ib 
or Ic supernova. Notably, any mass lost from a binary as a result of such a late case C mass transfer (stable or 
unstable) is expected to still be present in the proximity of the system at the time of supernova \citep[$\sim 0.1$ pc for material 
ejected with velocity of $\sim 100 \kms$ traveling for $\sim 1000$ yr, see, e.g., ][]{Sun2020}.

Systems in which late case C mass transfer becomes unstable and leads to a merger are among the promising 
candidates for the origin of long GRBs \citep[][]{Fryer2007}. Because of the short remaining lifetime of the 
merger product, its CO core maintains rapid rotation until the collapse without losing much of the angular momentum in 
winds, as is required in the collapsar model \citep{Woosley1993}. Notably, long GRBs are typically found in 
low-metallicity hosts \citep{Fruchter2006,Wolf2007}.

Evolution through unstable mass transfer and successful CE ejection is one of the most promising formation channels for 
compact binary mergers. Importantly, mass transfer is more prone to become unstable and to lead to CE evolution if the 
donor has a convective (rather than radiative) outer envelope (see Sec.~\ref{sec.res_conv_env} and references therein). 
Additionally, in the case of BH accretors, convective-envelope donors are more likely to satisfy the energy budget for 
the CE ejection \citep[i.e., avoiding a merger within the CE;][]{Kruckow2016}. 
Here, we find that the parameter space for mass transfer from convective-envelope donors is very small at all 
metallicities, and that it disappears completely for $M \gtrsim 40\msun$ donors at $Z \leq 0.04 \zsun$ metallicity. In 
reality, this parameter space may be even smaller than predicted in our models: the observational lack of RSGs above 
$\logL \approx 5.6$ in the nearby galaxies \citep[][, see also Sec.~\ref{sec:obs_02Z}]{Davies2018,Chun2018} may indicate 
that also in the higher metallicity environments ($Z \geq 0.2 \zsun$) stars with masses above $\sim 40 \msun$ never 
develop outer convective envelopes. 
For the CE evolution scenario to explain the observed population of binary BH mergers, we either need this channel to 
work also when the mass transfer is initiated by a radiative envelope donor, or the parameters of progenitor binaries 
need to be fine-tuned to the small parameter space for mass transfer from convective envelope donors.

Notably, we find that almost all cases of mass transfer from convective envelope donors with masses $\lesssim 60 \msun$ and 
metallicities $ \leq 0.2 \zsun$ are associated with the late case of RLOF from a core-helium depleted star, that is, a HeHG 
donor (see Fig.~\ref{fig.bin_param_ranges}). The remaining lifetime of the donor star (up to $10^4$ years) is an upper 
limit for the duration of the subsequent BH-WR stage. This is at least an order of magnitude shorter than the full duration 
of the CHeB phase, that is, the lifetime of a WR star, if the envelope is stripped shortly after the end of the MS. The 
duration of the BH-WR stage is relevant for the degree of tidal spin-up of the WR star, which may be of importance for the spin 
of the second formed BH \citep[e.g.,][]{Kushnir2016,Hotokezaka2017a,Zaldarriaga2018}.

\section{Conclusions}
\label{sec:conclusions}

We showed that metallicity has a strong influence on the type of mass transfer evolution that is expected in massive 
binaries. To do so, we computed a set of evolutionary tracks of stars between $10$ and $80\msun$ for six different 
metallicities ranging from $Z = 0.017 = \zsun$ to $Z = 0.00017 = 0.01\zsun$ (both nonrotating and rotating models 
with $\Omega_{\rm init}/\Omega_{\rm crit} = 0.4$). We explored several variations of factors known to affect the radial
expansion of massive stars (e.g., semiconvection and overshooting) and compared our models with observations.
Our conclusions are summarized below.
\begin{itemize}

 \item The lower the metallicity, the stronger the tendency of massive stars to remain relatively compact during 
the phase of rapid HG expansion ($\sim 100 \rsun$) and only reach the red giant branch towards the end of 
helium burning (CHeB phase) or later, during contraction of the CO core (HeHG phase). At solar 
metallicity, no models behave in this way, and a post-MS donor in an interacting binary is almost exclusively a 
rapidly expanding HG star. At $Z = 0.1\zsun$ , stars 
in the mass range between $\sim 16$ and $50\msun$ remain relatively compact before the onset of CHeB, and a 
slowly expanding CHeB star is the typical donor in binaries with periods above $\sim 40$ days.
At $Z = 0.01\zsun$ , stars of all masses in our grid remain compact during the HG expansion. As a result, CHeB or HeHG 
stars are the most likely post-MS donors at very low metallicity.

\item This metallicity trend is a relatively robust prediction of stellar models that is supported by the observations 
of blue and red supergiants in the Milky Way and in the SMC. The exact mass range for the compact HG evolution at any 
given metallicity is considered highly uncertain. Models that predict the HG expansion to reach the red giant 
branch for all masses at $Z = 0.2 \zsun$ are ruled out by the observations (e.g., the model with a relatively inefficient 
semiconvection $\alpha_{\rm SC} = 1.0$).

\item At low metallicities ($\leq 0.2\zsun$), most massive stars can become stripped of their envelopes to form helium 
stars only after a substantial fraction of their helium-burning lifetime, as opposed to losing their envelopes shortly 
after the end of MS \citep[as is often 
assumed, e.g.,][]{Woosley2019}. This implies a shorter duration of the WR phase and a more massive final core.

\item As the metallicity decreases, increasingly more massive stars engage in their first mass transfer episode only very 
late in their evolution, after the core is almost completely depleted of helium ($Y_{\rm C} < 0.05$) and less then 
$10^4$ yr remains until the collapse of the core. The short remaining lifetime is unlikely to be long enough to strip
the giant star of its entire hydrogen-rich envelope before the possible supernova. 

\item The binary parameter space for RLOF from convective envelope donors is very small, and it decreases with 
metallicity. For example, in the $0.1\zsun$ metallicity case, it is only about 0.2 dex wide in log($P$/day), see 
Fig.~\ref{fig.bin_param_ranges}. At metallicities $Z \leq 0.04 \zsun$ , we find no mass transfer from convective envelope
donors with masses above $\sim 40 \msun$ in our models. The lack of red supergiants with luminosities above $\logL 
\approx 5.6$ in local galaxies may indicate that the same is true for higher metallicities $Z \geq 0.2 \zsun$.

\item No particular evolutionary stage can be used as a proxy for determining whether a star has a convective or a radiative 
envelope in population synthesis calculations. Instead, we provide fits for the threshold effective 
temperature below which stars develop outer convective envelopes (of at least 10\% in mass), see Eqn.~\ref{eq:fit1} 
and Fig.~\ref{fig.conv_env_thresh}. 

\end{itemize}

\begin{acknowledgements}
The authors would like to thank the referee for suggestions and comments that helped to improve the paper.
We would also like to thank Selma de Mink, Stephen Justham, Abel Schootemeijer, Rob Farmer, Ylva G\"{o}tberg, Martyna Chruslinska, Eva Laplace, 
and Manos Zapartas for valuable discussions and suggestions. We are grateful to Norberto Castro for sharing the density map of the sample of 
Galactic massive stars in the spectroscopic HR diagram. The authors acknowledge support from the Netherlands 
Organization for Scientific Research (NWO).
\end{acknowledgements}

\bibliographystyle{aa}
\bibliography{ULX_bib.bib}

\begin{appendix}

\section{Density inversions, MLT++, and the Humphreys-Davidson limit}

\label{sec:App_mltpp}

In massive stars, subsurface 
convective zones can develop as a consequence of iron and helium opacity peaks \citep{Cantiello2009}. At the same time, 
the low density and temperature in the outermost envelope layers of extended supergiants makes the convective energy 
transport ineffective. This leads to superadiabatic temperature gradients and requires most of the energy to be 
transported through radiation \citep[e.g.,][]{Pavlovskii2015}. In the case of stars that evolve near their Eddington limit, 
the subsurface opacity peaks and inefficient convection lead to a situation in which the radiative luminosity locally 
exceeds the Eddington limit in the envelope. The structure can become stabilized by a density inversion 
\citep{Grafener2012}. In this way, the formation of density inversions in 1D stellar models is a natural consequence 
of sufficiently high luminosities in massive stars, with the metallicity and opacities playing an important role 
\citep[see also ][]{Paxton2013}. 

On the one hand, density inversions and inefficient superadiabatic convection pose a numerical challenge that can lead to 
prohibitively short time steps. At the same time, many authors have considered such inversions unphysical in the first 
place and applied ad hoc solutions in order to remove them from stellar models, for example, by substituting pressure 
scale height with density scale height \citep[e.g.,][]{Stothers1973,Ekstrom2012,Yusof2013} or by capping the temperature 
gradient at an imposed upper limit \citep[e.g.,][]{Bressan1993,Fagotto1994}. Others would argue in favor of stability of 
density inversions \citep{Glatzel1993,Sanyal2015,Sanyal2017} with the caveat that this inversion might be suppressed 
by extremely strong winds  \citep{Asplund1998}. In any case, the fact that in superadiabatic layers the convective 
velocity approaches the speed of sound indicates that the standard mixing-length theory is beyond its domain of 
applicability, and that most likely 3D hydrodynamical simulations are required in order to advance the ongoing debate. 
Similarly to other ad hoc solutions mentioned above, MLT++ is a stellar engineering solution to prevent the formation 
of density inversions 
in MESA. It gradually reduces the temperature gradient and thus the superadiabaticy in some 
radiation-dominated convective zones \citep[see ][]{Paxton2013}, which not only eliminates density 
inversions, but also alleviates many numerical difficulties. Crucially, MLT++ also increases the effective temperature of 
the model, similarly to other methods of preventing density inversions \citep[e.g., using the density scale height or 
increasing the mass-loss rate][]{Maeder1987}.
The increase in effective temperature, even though somewhat artificial in the ad hoc methods such as MLT++, mimics 
an effect that an increased mass-loss rate would have on stars in the upper right corner of the HR diagram.
It has been proposed that exceeding the Eddington limit in subsurface layers, the associated density 
inversions, and hydrodynamical turbulence and shocks in superadiabatic convective zones might be responsible for extreme 
mass loss of luminous supergiants and the luminous blue variable phenomenon 
\citep[e.g.,][]{Owocki2004,vanMarle2008,Quataert2016}. 
This finds support in the fact that the location of the Eddington limit in the HR diagram coincides with the empirical 
HD limit \citep{Humphreys1979, Humphreys1994,Ulmer1998}, a line in the upper right side of the HR 
diagram beyond which almost no stars are observed in the Milky Way and in the Large Magellanic Cloud. Moreover, the 
Eddington factor $\Gamma$ has been identified as the key parameter determining the mass-loss rates of massive WR stars 
\citep{Vink2011,Grafener2011}. \citet{Chen2015} assumed that the same $\Gamma$ dependence of the mass-loss rates 
applies to all stars (together with the metallicity scaling proposed by \citealt{Grafener2008}), and 
obtained stellar tracks that agree reasonably well with the HD limit at solar and LMC metallicities. A similar 
effect could be achieved through the use of MLT++ or the density scale height in place of the pressure scale height 
in superadiabatic zones. In this sense, stellar tracks computed with MLT++ might be more accurate in estimating the 
binary parameter space for mass transfer from the most massive and largest stars in our grid ($\gtrsim 50 \msun$). 
On the other hand, it is currently unknown whether the empirical HD limit also appears in low-metallicity environments 
($Z \lesssim 0.2 \zsun$). In the case of solar 
metallicity, for which the HD limit is best evidenced, our non-MLT++ tracks also reproduce the lack of stars in the 
upper right corner of the HR diagram. Additionally, \citet{Chun2018} found a better agreement with the Galactic sample 
of luminous RSGs when MLT++ was not used. It remains certain that any numerical results for the maximum radii of 
massive stars ($\gtrsim 50 \msun$) based on 1D stellar evolution computations should be considered highly uncertain. 

\section{Internal mixing during the Hertzprung gap phase}
\label{sec:app_mixing}

During the MS evolution of a massive star, the shrinking convective core leaves a region with a composition 
gradient behind. The corresponding gradient of mean-molecular weight stabilizes this region against convection, and at 
least initially, the composition is primarily being mixed by a slower process: semiconvection \citep{Langer1985}. The 
semiconvective zone becomes most extended immediately after the end of the MS when the star rapidly expands during the HG phase 
and hydrogen-shell burning becomes important. The mixing that occurs during this short-lived phase ($\sim 10^4$ 
yr) has a key effect on the stellar radius at the onset of helium burning and the blue to red evolution in the HR 
diagram \citep[e.g.,][]{Langer1991}.

\begin{figure}
      \includegraphics[width=0.95\columnwidth,center]{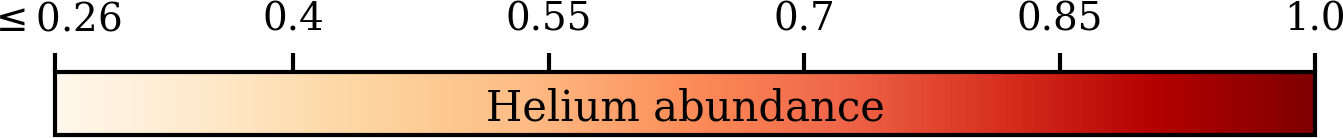}
      \includegraphics[width=0.95\columnwidth,center]{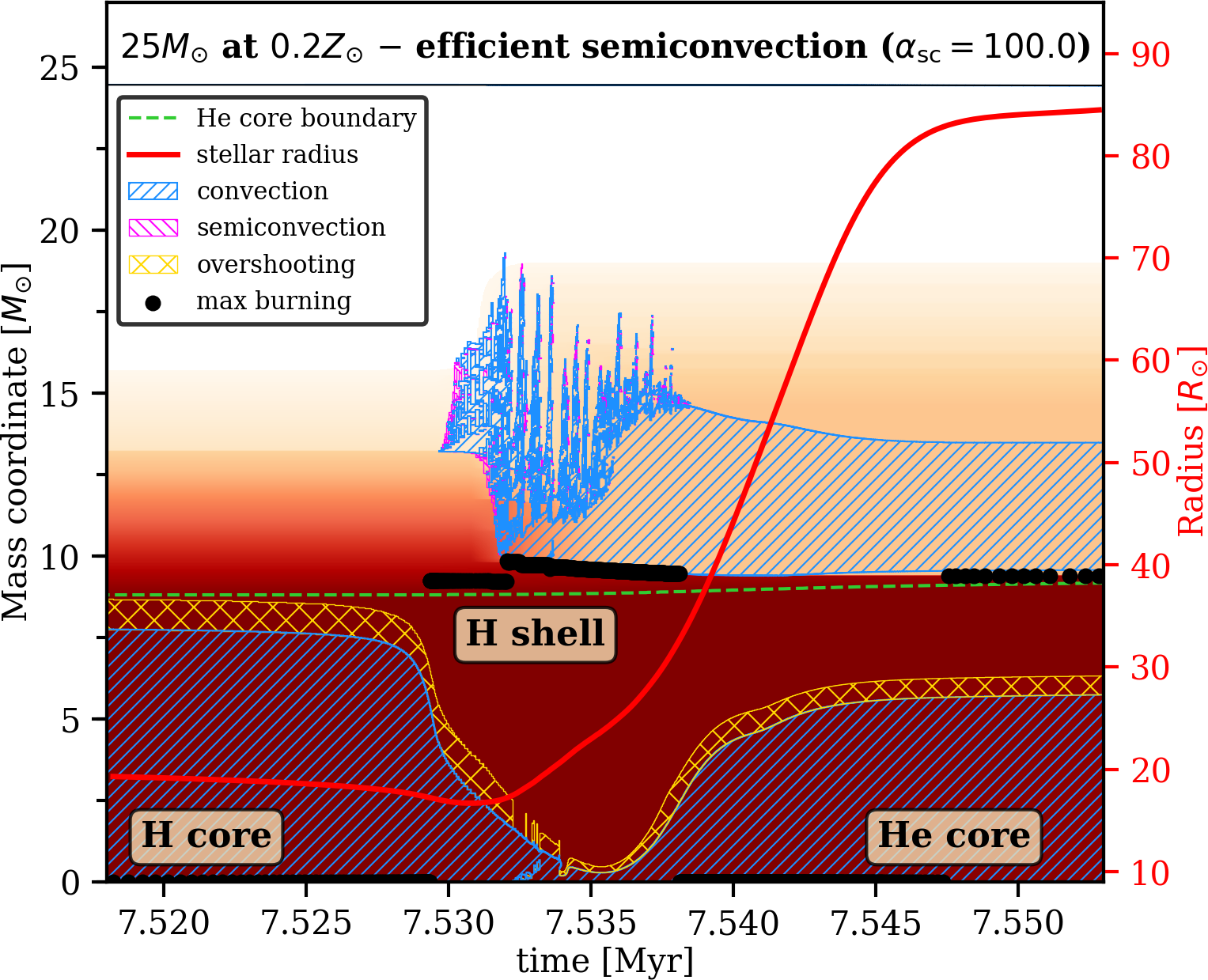}
      \includegraphics[width=0.95\columnwidth,center]{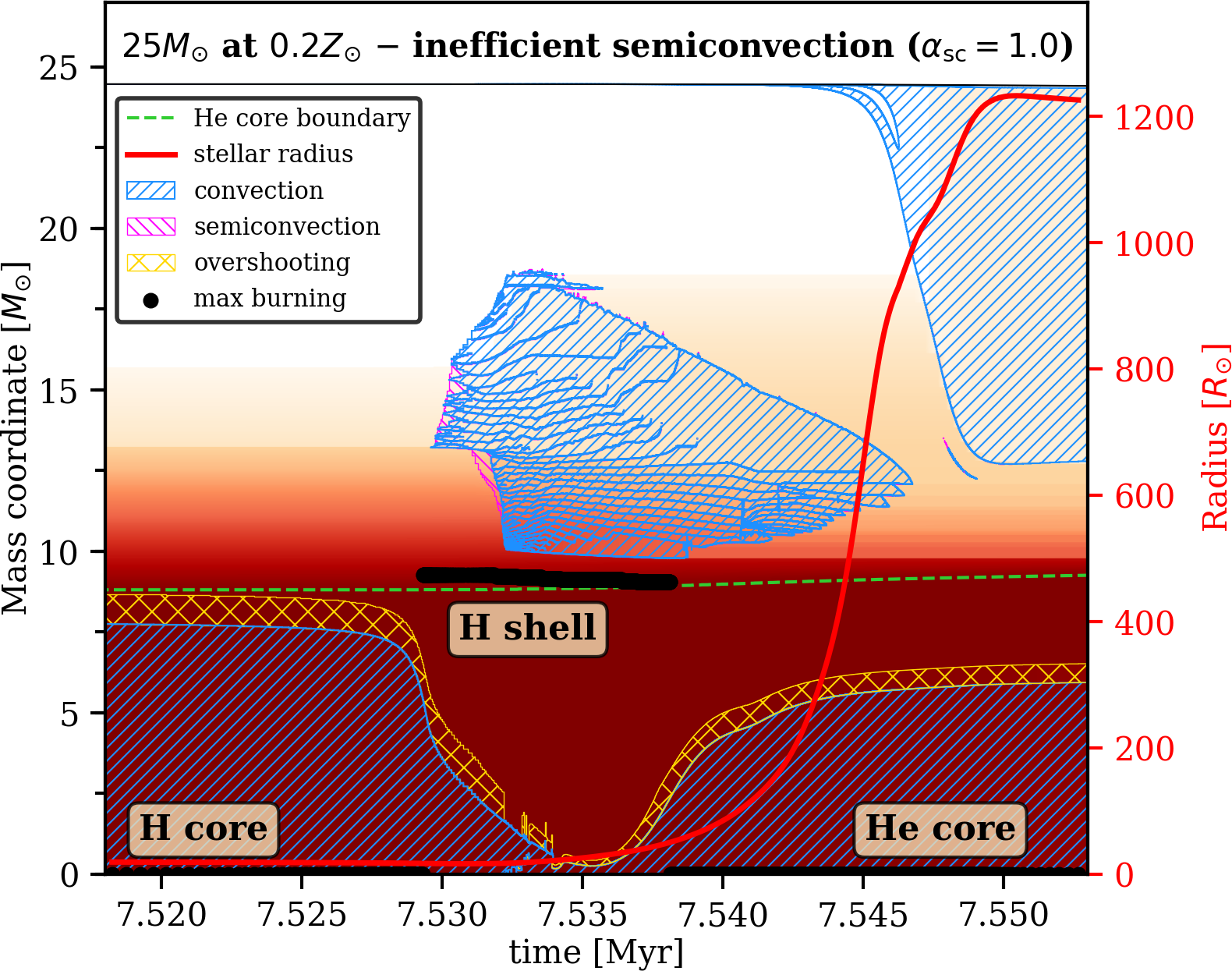}
      \includegraphics[width=0.95\columnwidth,center]{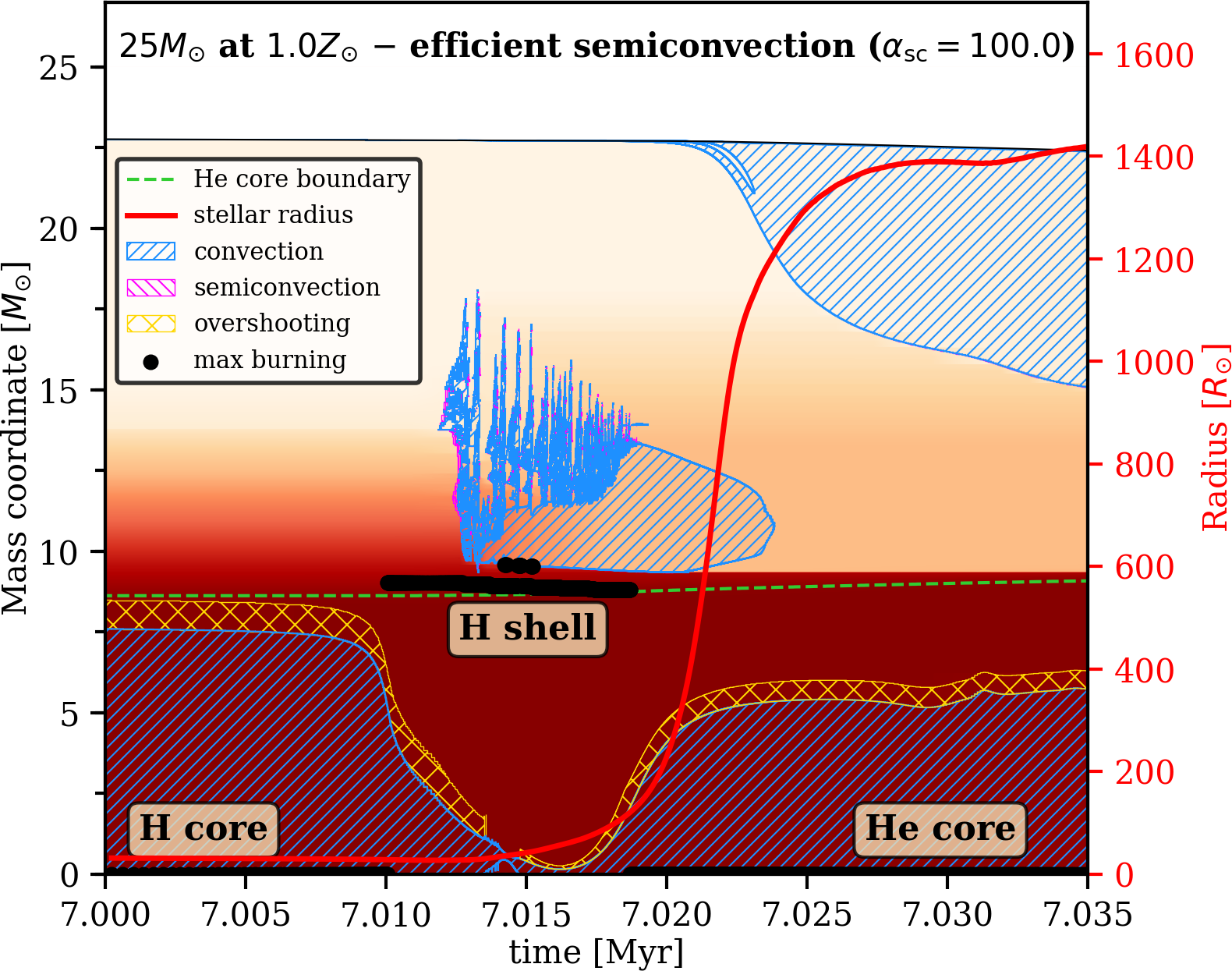}
    \caption{Kippenhahn diagrams of three models of a single $25 \msun$ star: at $0.2 \zsun$ 
    with efficient semiconvection ($\alpha_{\rm sc} = 100$, top panel), at $0.2 \zsun$ but with inefficient 
    semiconvection ($\alpha_{\rm sc} = 1$, middle panel), and at $1.0 \zsun$ with efficient semiconvection 
($\alpha_{\rm sc} = 100$, bottom panel). We also mark the stellar radius (red solid line, labeled on the right-hand 
side) to show   the influence of mixing on the evolution in the HR diagram. In the solar metallicity case, even if the mixing is efficient, the model expands to $R>1000\rsun$ during the HG phase. The resulting 
helium abundance profiles are shown in Fig.~\ref{fig.app_Yprof}.}
    \label{fig.app_mixing}
\end{figure}

In Fig.~\ref{fig.app_mixing} we show Kippenhahn diagrams of three models of a single $25 \msun$ star: at $0.2 \zsun$ 
with efficient semiconvection, at $0.2 \zsun$ but with less efficient semiconvection, and at $1.0 \zsun$ with efficient 
semiconvection. The diagrams are centered around the short-lived phase between 
the end of MS and the onset of helium burning, see also Fig.~5 of 
\citet{Schootemeijer2019}. If semiconvection is 
efficient and quick to sufficiently flatten the composition gradient for convective mixing to fully kick in, then a 
single extended convective zone is eventually formed at the top of the hydrogen shell (the top and bottom panels). In 
the case of less efficient semiconvection, multiple separate convective zones are 
formed instead, with semiconvective regions in between them \citep[see, e.g., Fig.~5 of][]{Langer1985}. This can be understood 
in the following way: with less efficient semiconvection, the composition gradient is not flatenned quickly enough 
for an extended convective zone to form during the short-lived HG phase. However, when any small zone becomes locally 
unstable to convection, its composition is mixed very quickly. This eliminates any composition gradient within that 
region, which helps to maintain convective mixing. At the same time, the gradient at the edges of the small convective 
region becomes steeper, which increases the stability of neighboring zones against convection, and they can only be 
mixed further by semiconvection. If semiconvection in such neighboring zones is not efficient, then the small 
convective zone becomes isolated and cannot merge with other similar small convective regions. An 
onion-like structure with multiple convective and semiconvective zones emerges, as in the middle panel of 
Fig.~\ref{fig.app_mixing}.

\begin{figure}
      \includegraphics[width=\columnwidth]{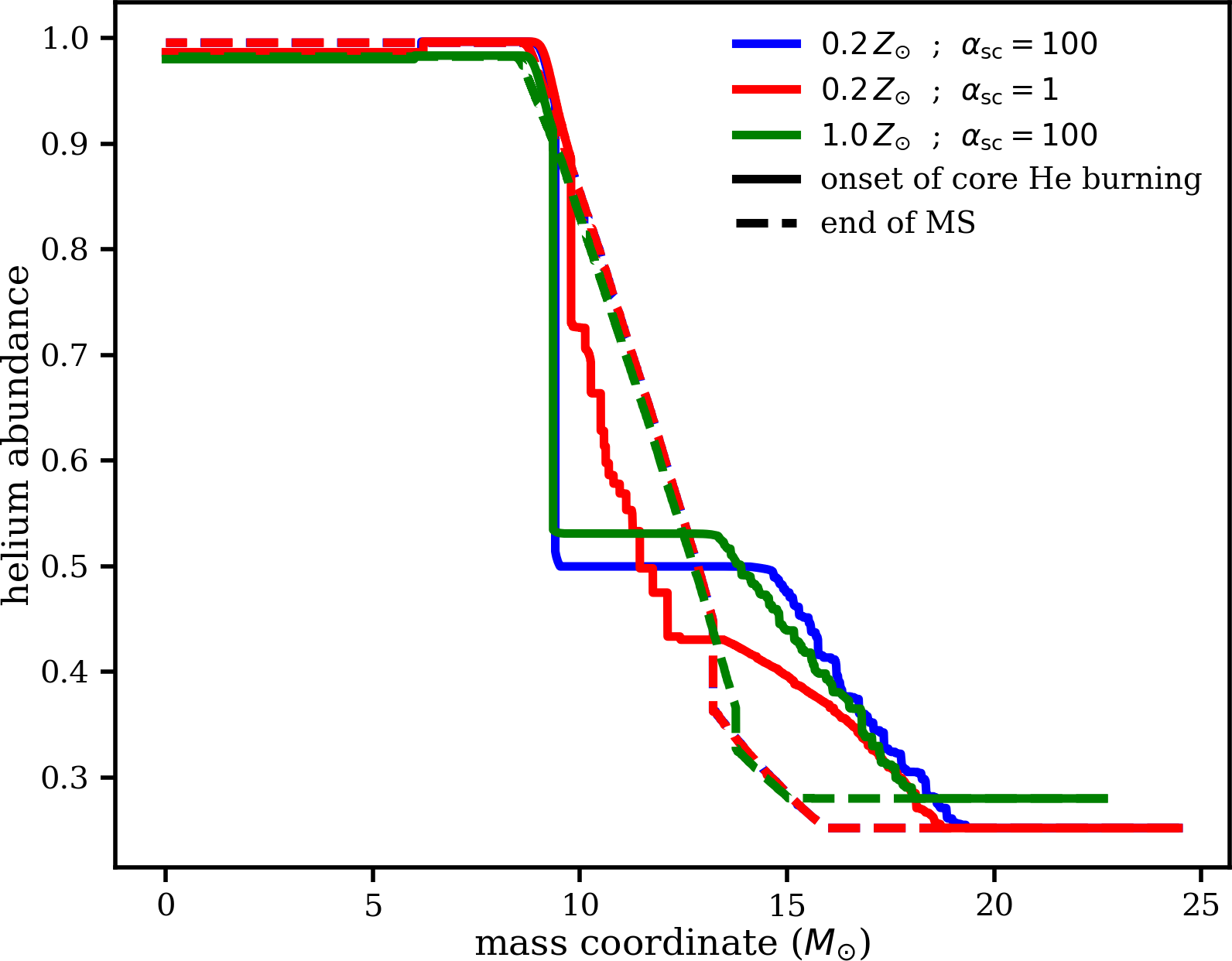}
    \caption{Helium abundance profiles of models shown in Fig.~\ref{fig.app_mixing}, taken from the point at the 
end of the MS and before the mixing (dashed lines) and after a convective helium burning core fully develops (solid lines).}
    \label{fig.app_Yprof}
\end{figure}

The interplay between semiconvective and convective mixing ultimately determines the abundance profile in the 
region above the hydrogen shell. In Fig.~\ref{fig.app_Yprof} we show helium abundance profiles taken at the end of MS 
(before the mixing) and after the convective helium core develops (after the mixing) for the three models shown in 
Fig.~\ref{fig.app_mixing}. Efficient mixing ($\alpha_{\rm sc} = 100$) results in a plateau of helium abundance at the 
bottom of the envelope, much different from the step-like profile that forms in the case of less efficient mixing 
($\alpha_{\rm sc} = 1$). This plateau is a common feature of all our post-MS donor stars in binary models as well. 
See 
also \citet{Schootemeijer2018} for a discussion of how the resulting H/He gradient is connected to surface 
abundances of WR stars. In the efficient mixing case, the envelope has become more enriched in helium, and at the same 
time, a larger amount of hydrogen has been brought down as fuel for the shell burning. In the solar 
metallicity case, even if the mixing is efficient, the model expands to $R>1000 \rsun$ during the HG phase.

We comment that the forest-like structure of short-lived extended convective zones that form in models with 
efficient semiconvection has been found in MESA models by other author as 
well \citep[e.g.,][]{Farmer2016,Schootemeijer2018}. Its exact behavior appears very chaotic and is subject to numerical 
settings and resolution. This mixing affects the composition in the region above the single convective zone: 
it creates a step-like decrease in $Y$ between mass coordinates $\sim 15$ and $20\msun$ in Fig.~\ref{fig.app_Yprof}, 
the details of which are considered highly uncertain. Notably, we have found that models running with convective 
premixing \citep{Paxton2019} without any limits on the velocity of a convective boundary advance 
(\textsc{conv\_premix\_time\_factor} = 0.0 in the MESA jargon) are effective in forming a single extended convective 
zone even with $\alpha_{\rm sc} = 1$.

\section{Additional HR diagrams}
\label{sec:App_HR}

\begin{figure*}
\centering
    \includegraphics[width=0.75\textwidth]{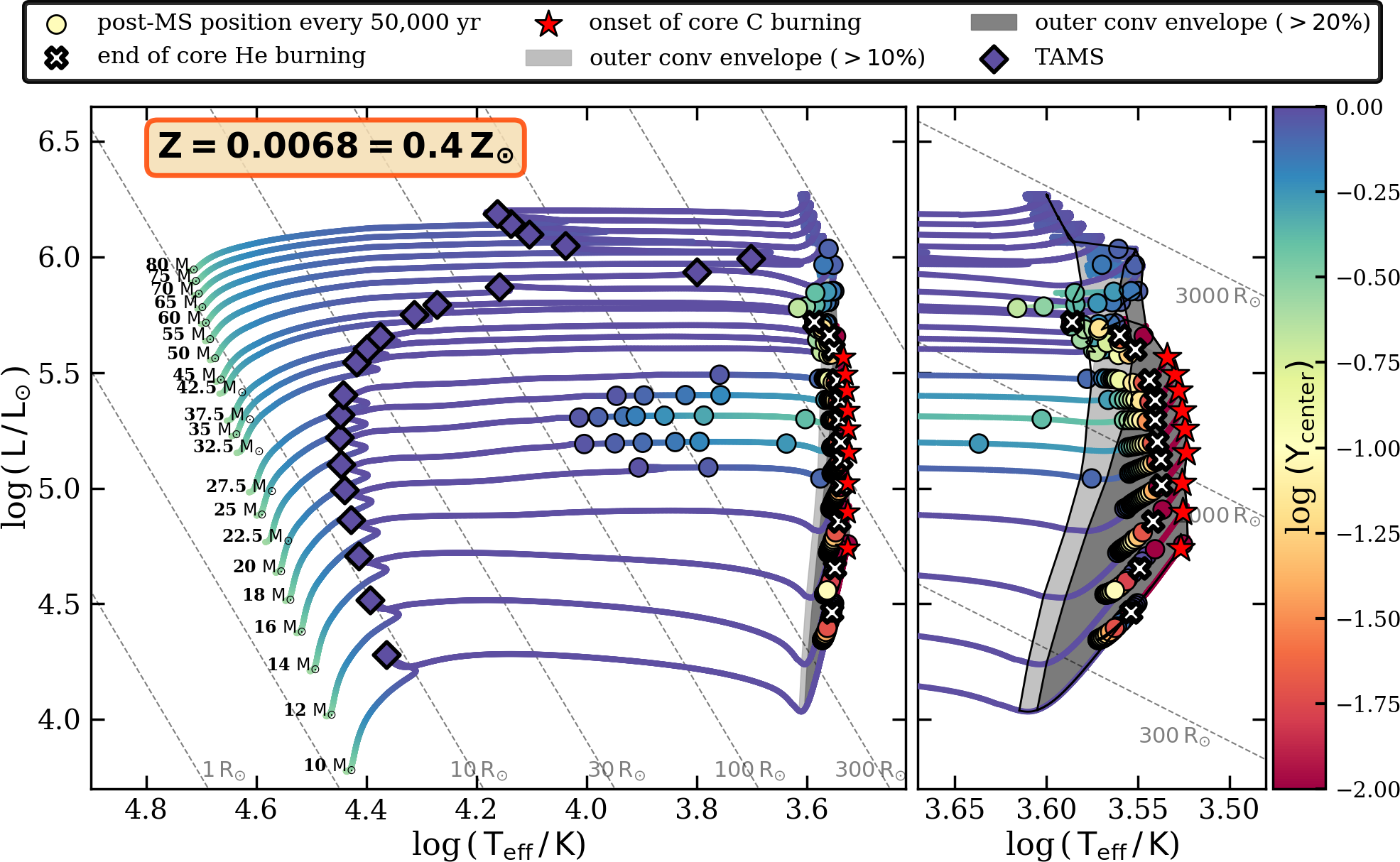}
    \includegraphics[width=0.75\textwidth]{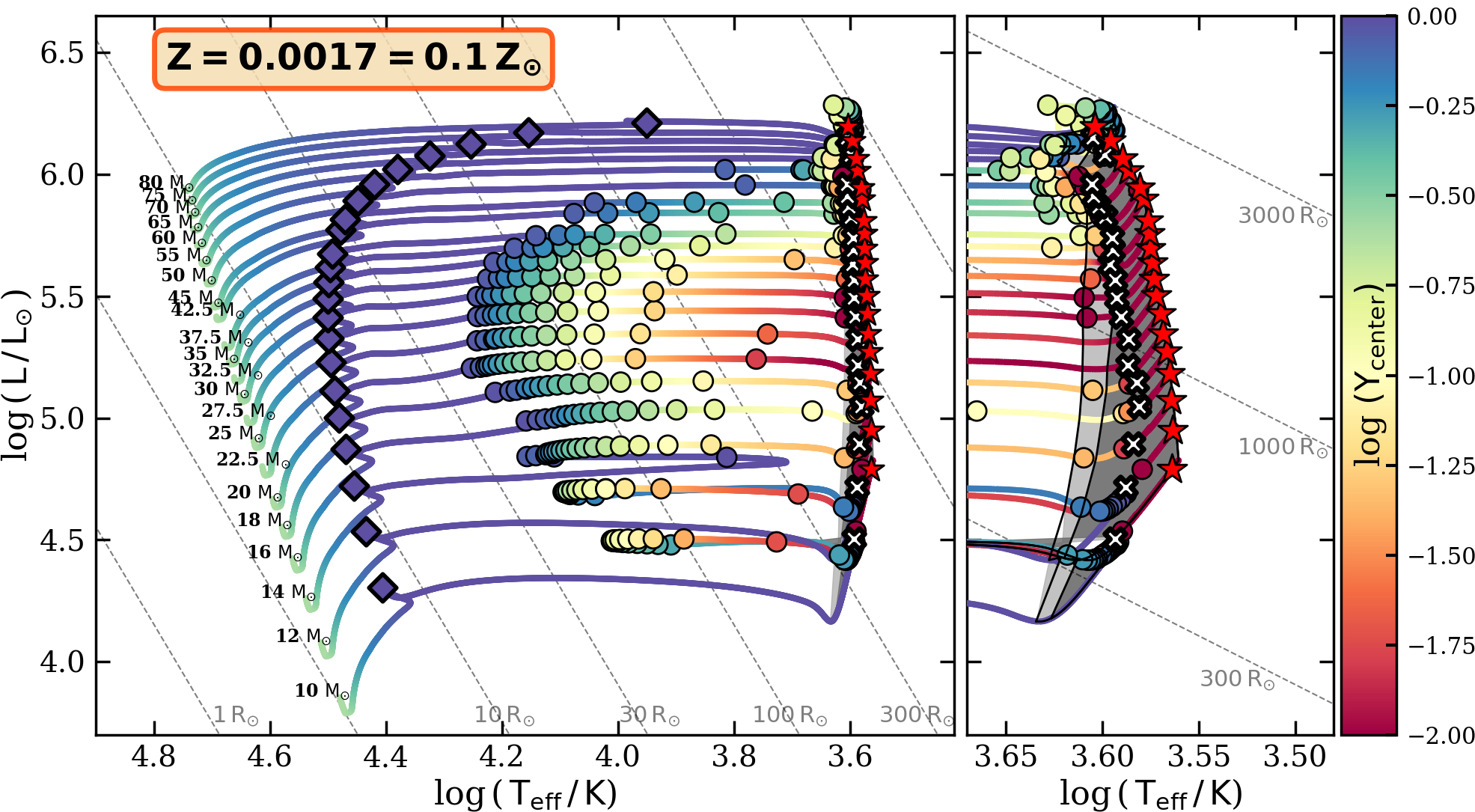}
    \includegraphics[width=0.75\textwidth]{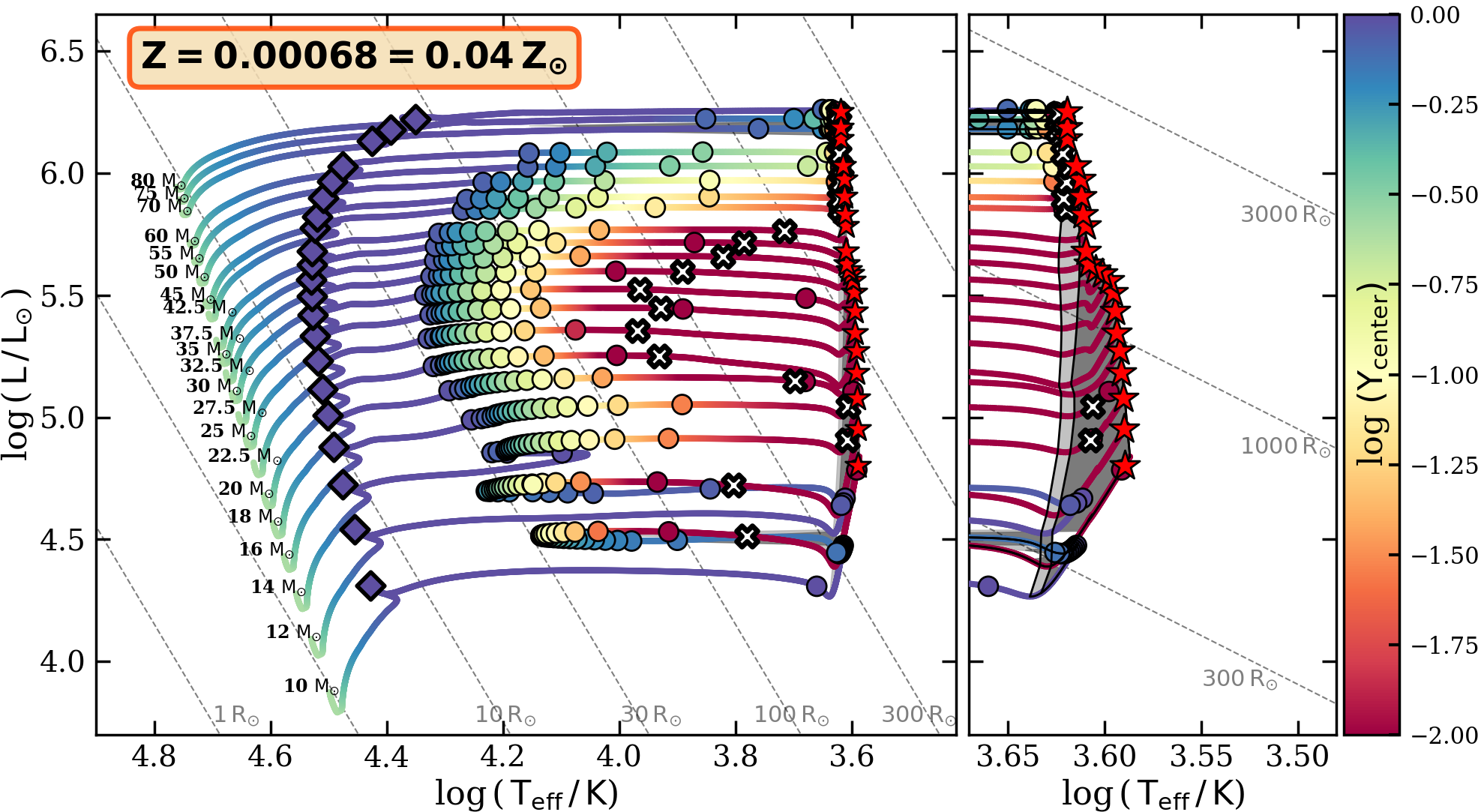}
    \caption{Same as Fig.~\ref{fig.HRD} but for 3 additional metallicities.}
    \label{fig.HRD_app}
\end{figure*}

\begin{figure*}
\begin{tabular}{C{4.7cm}  C{4.1cm}  C{4.1cm} C{4.1cm}}
\includegraphics[width=0.27\textwidth,height=115px]{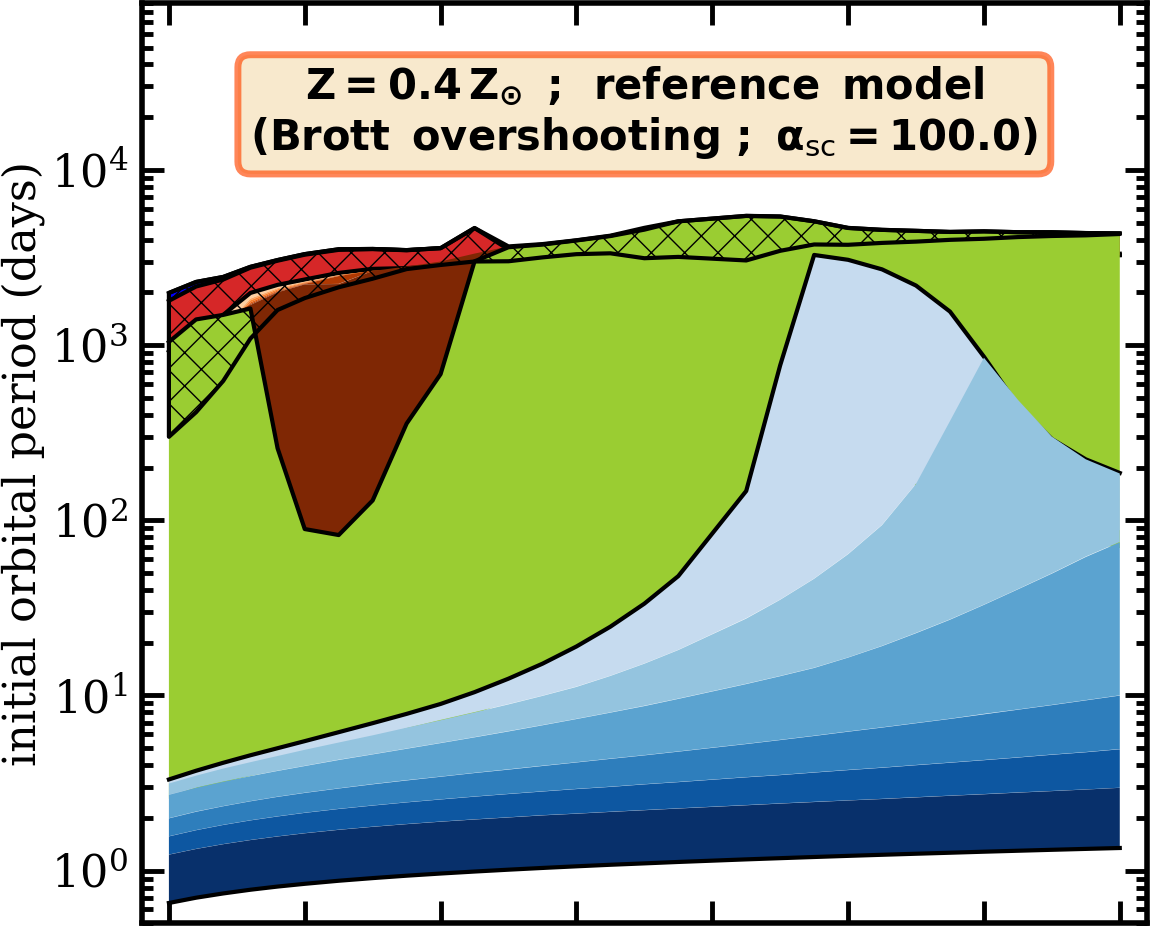} &
\includegraphics[width=0.24\textwidth,height=115px]{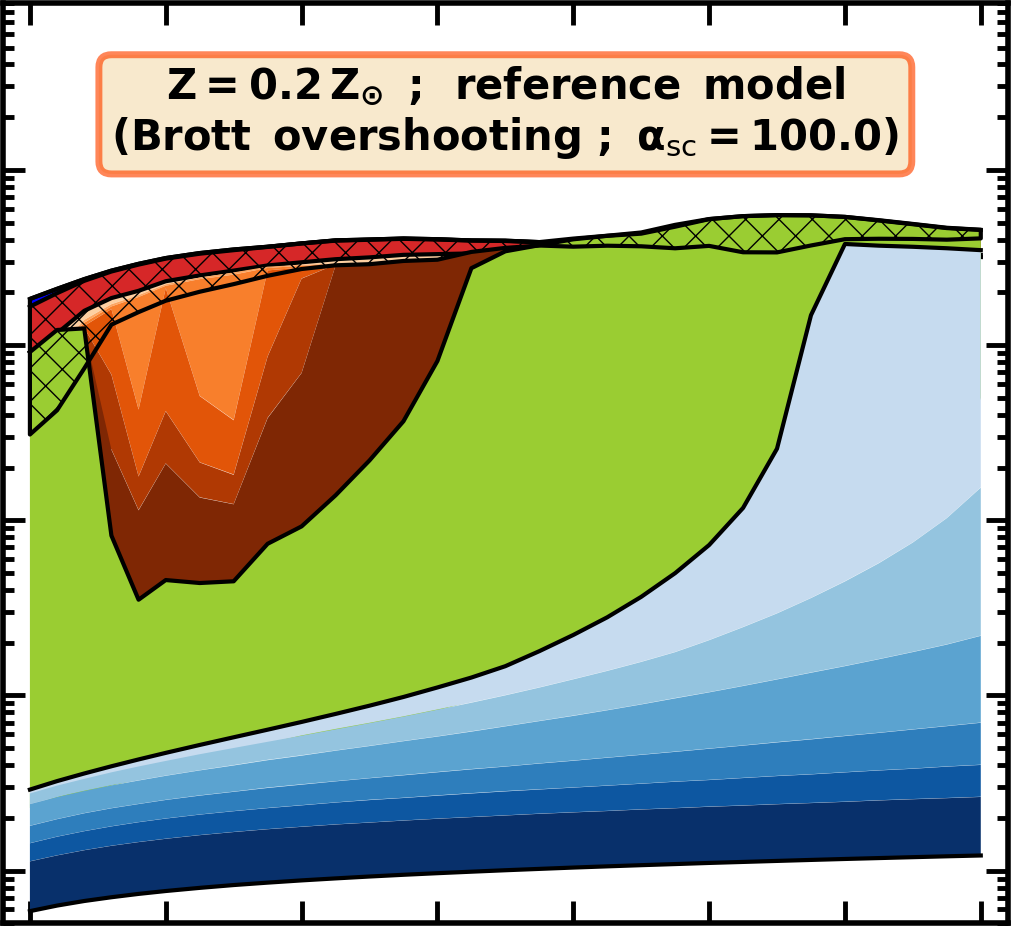} & 
\includegraphics[width=0.24\textwidth,height=115px]{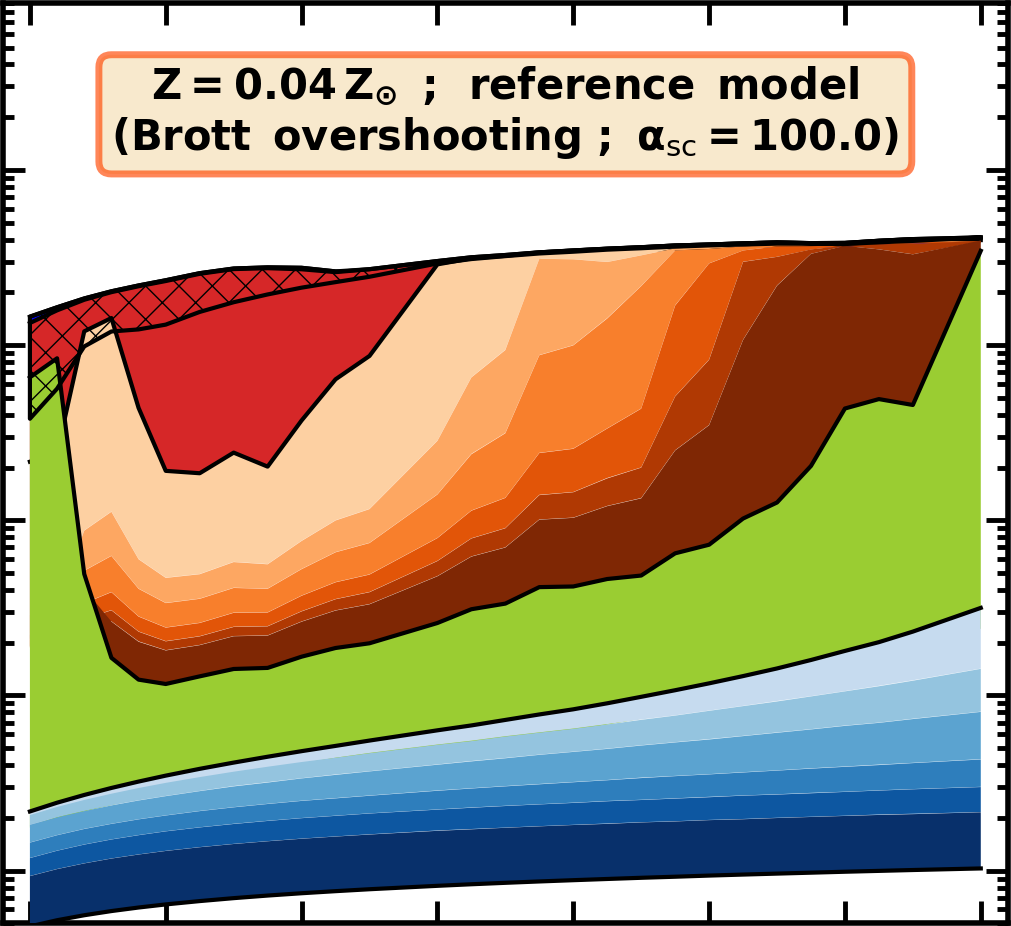} &
\includegraphics[width=0.24\textwidth,height=115px]{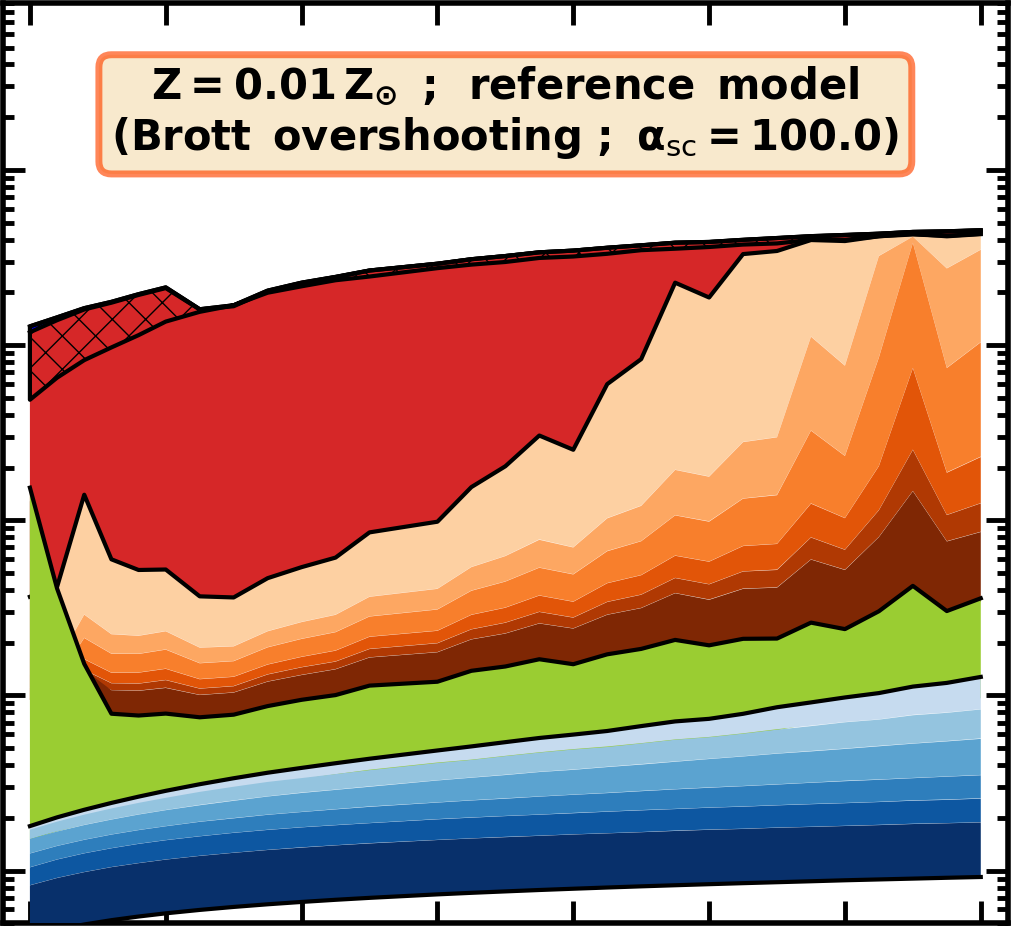} \\

\includegraphics[width=0.27\textwidth,height=115px]{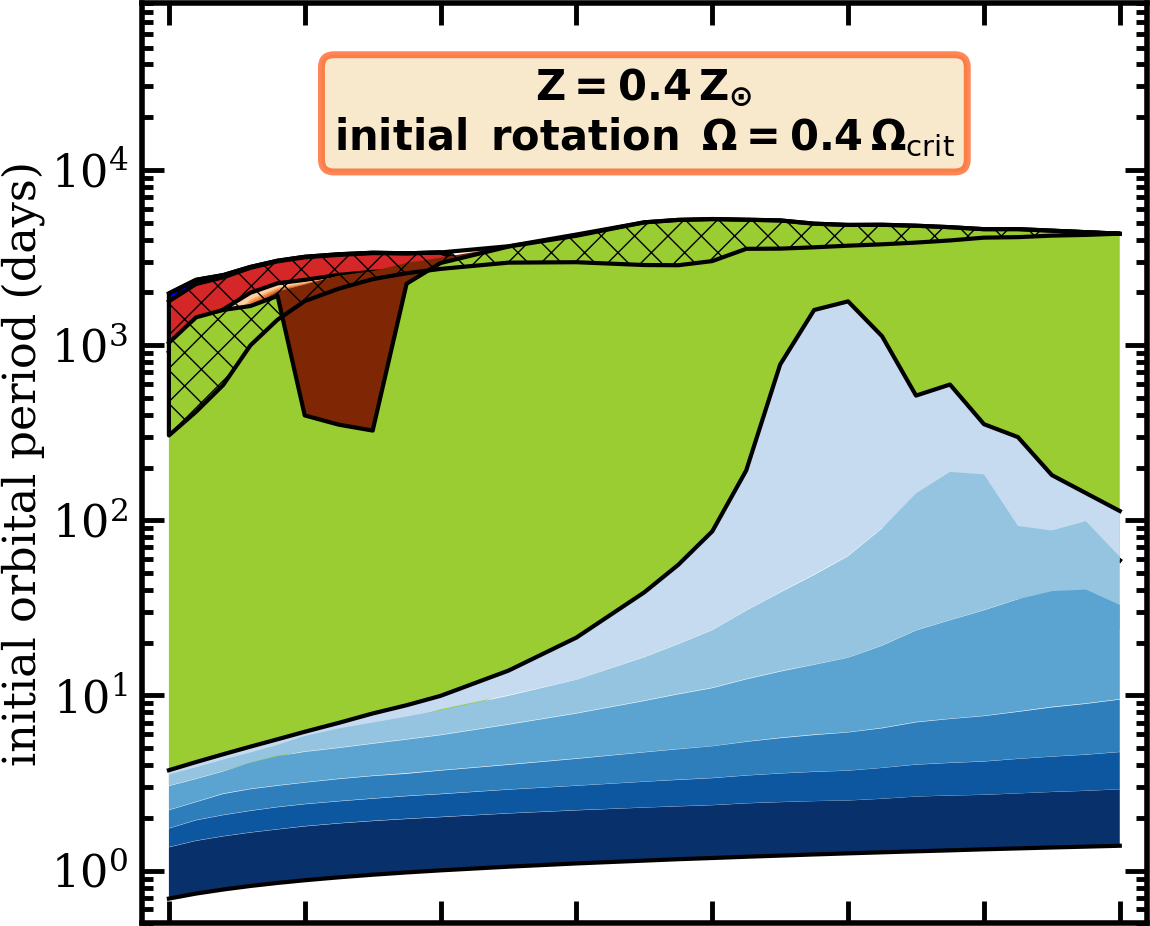} &
\includegraphics[width=0.24\textwidth,height=115px]{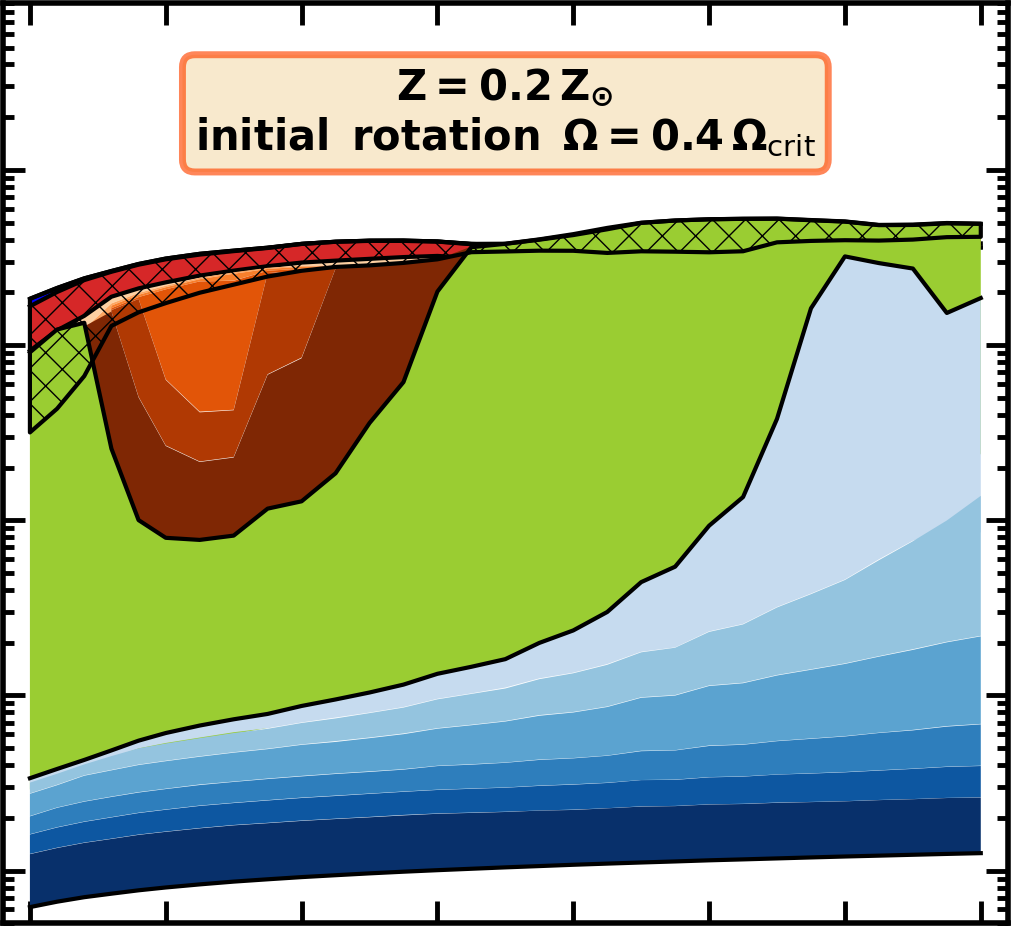} & 
\includegraphics[width=0.24\textwidth,height=115px]{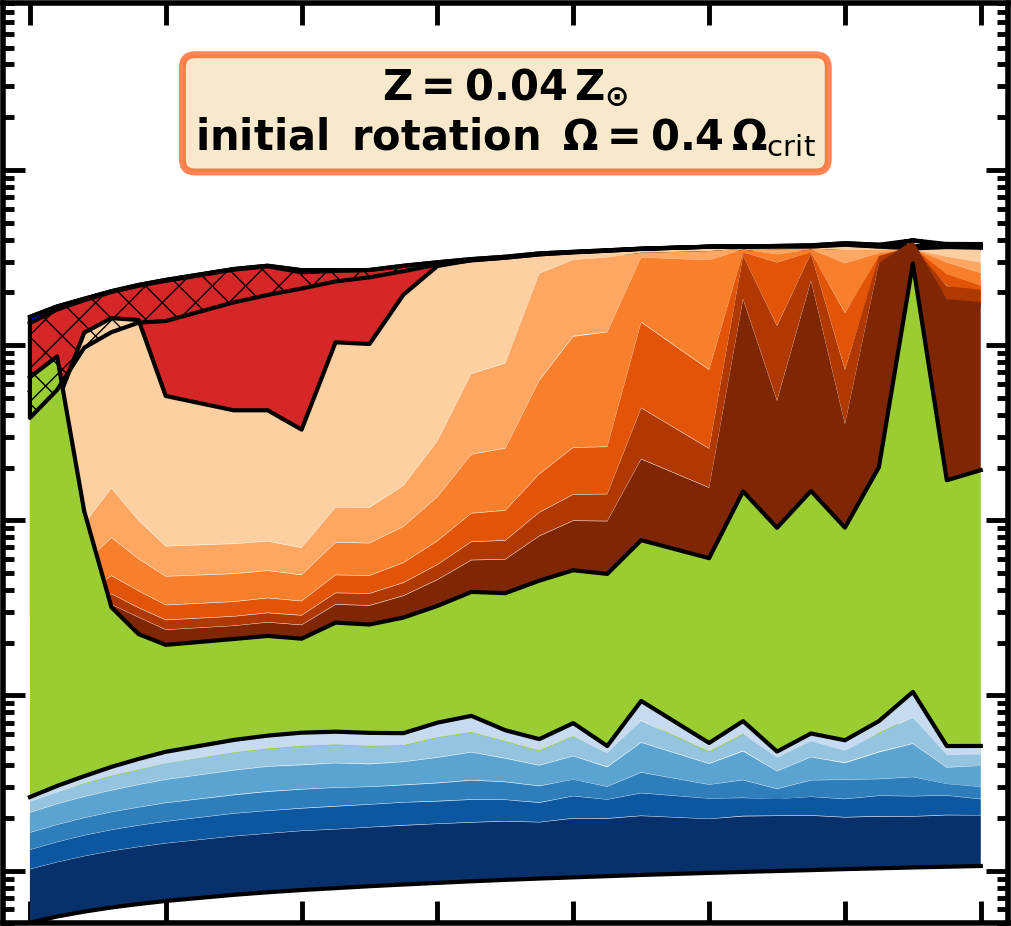} &
\includegraphics[width=0.24\textwidth,height=115px]{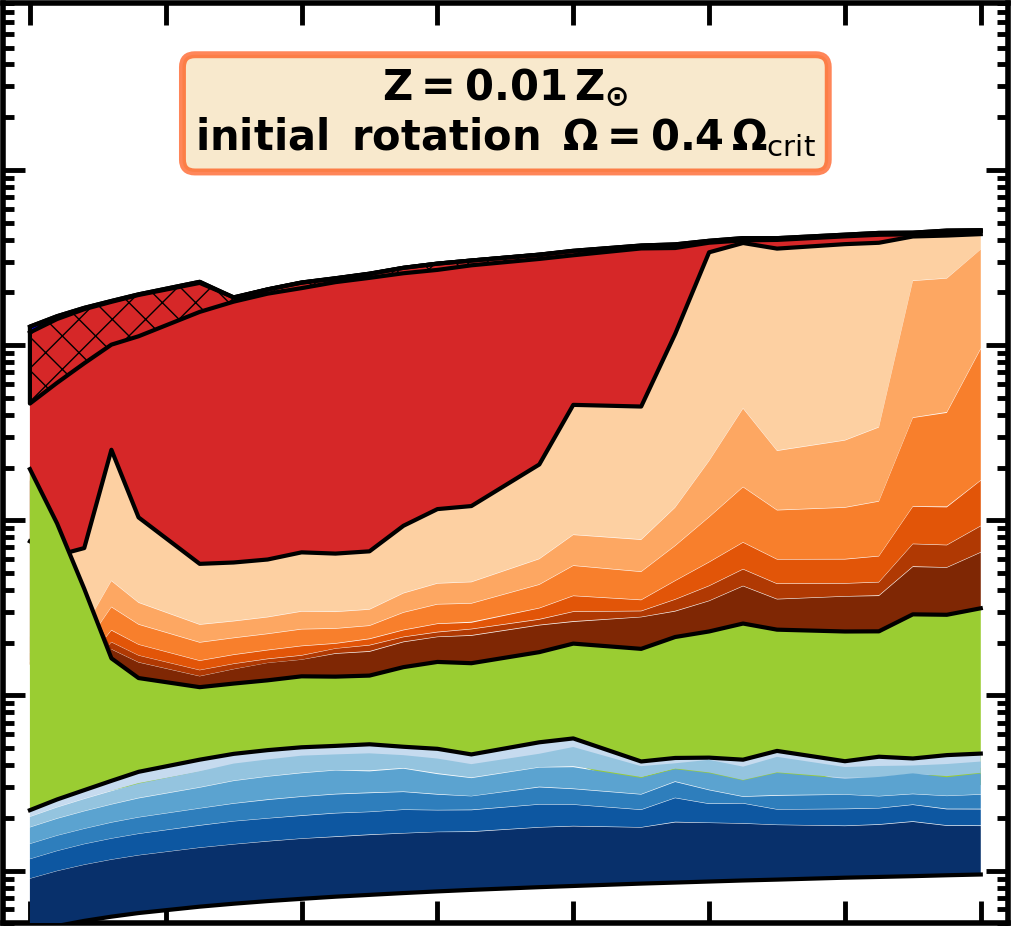} \\

\includegraphics[width=0.27\textwidth,height=115px]{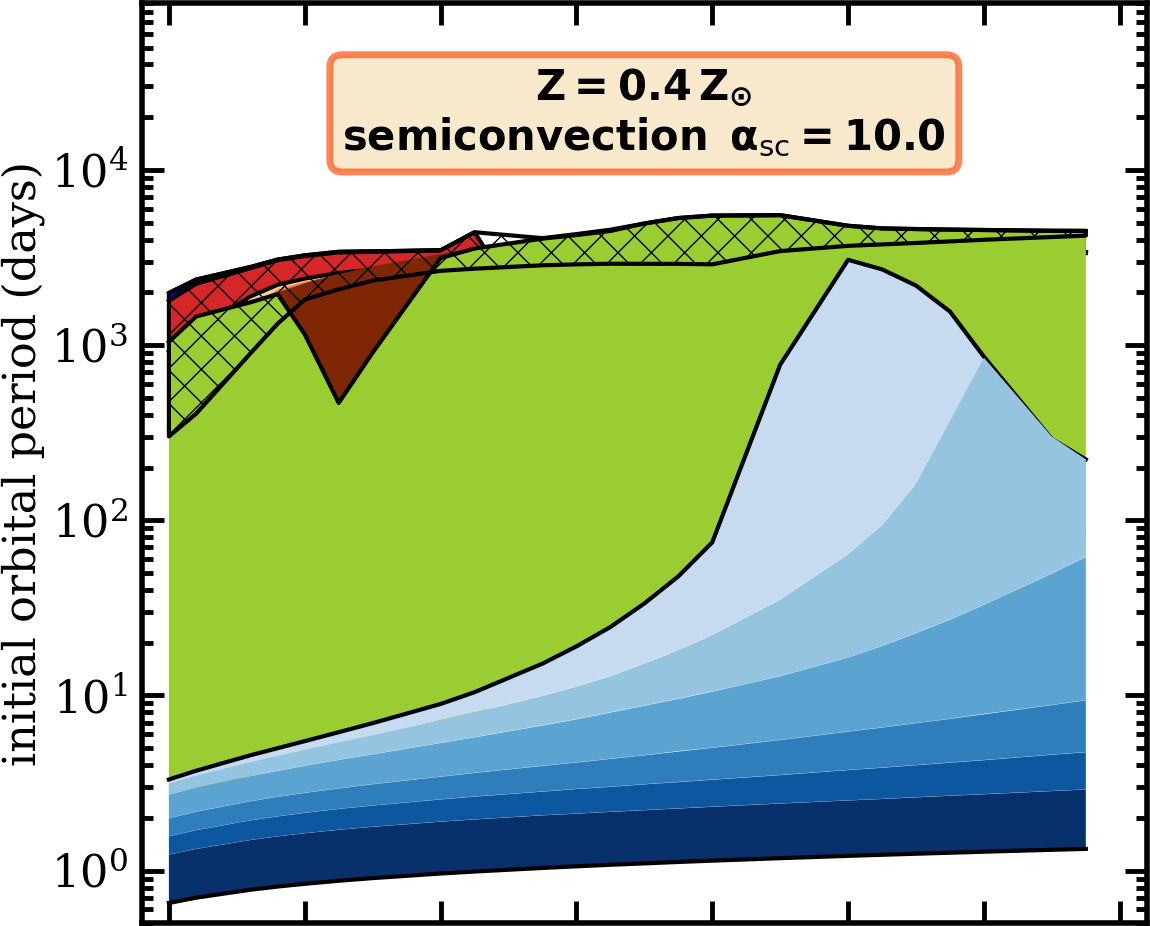} &
\includegraphics[width=0.24\textwidth,height=115px]{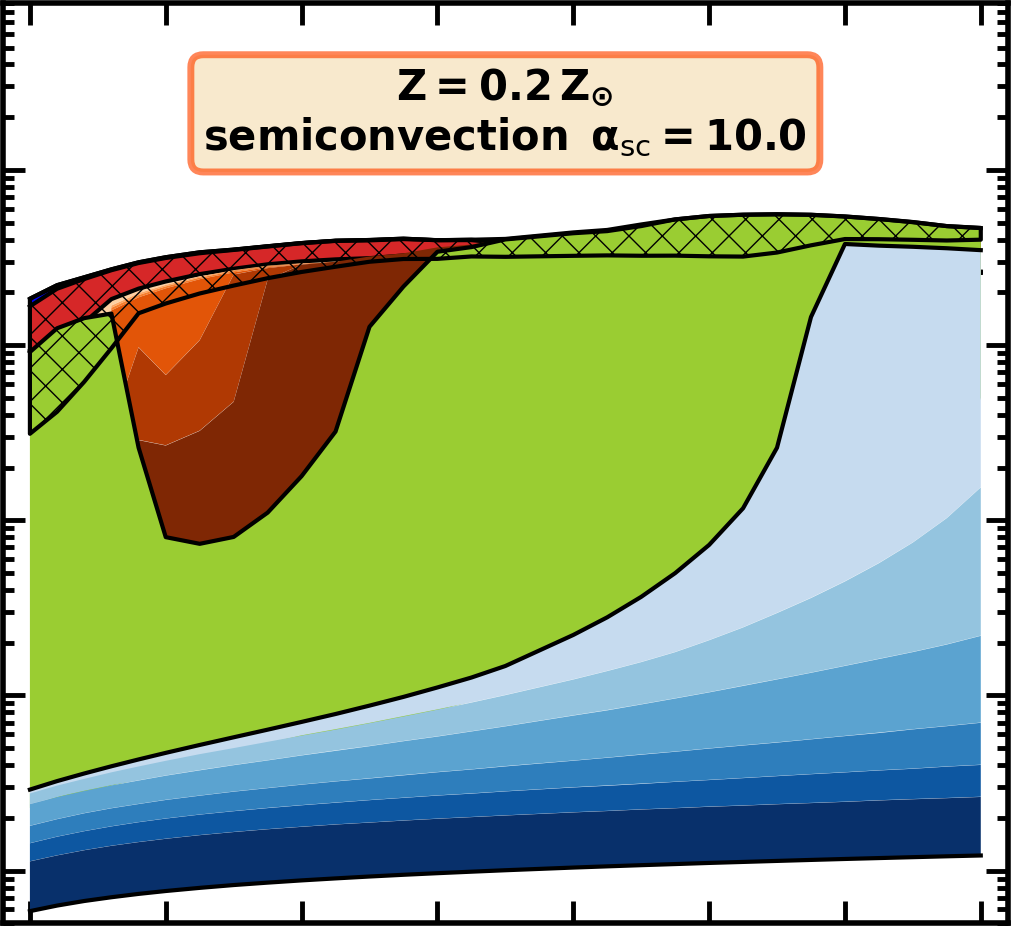} & 
\includegraphics[width=0.24\textwidth,height=115px]{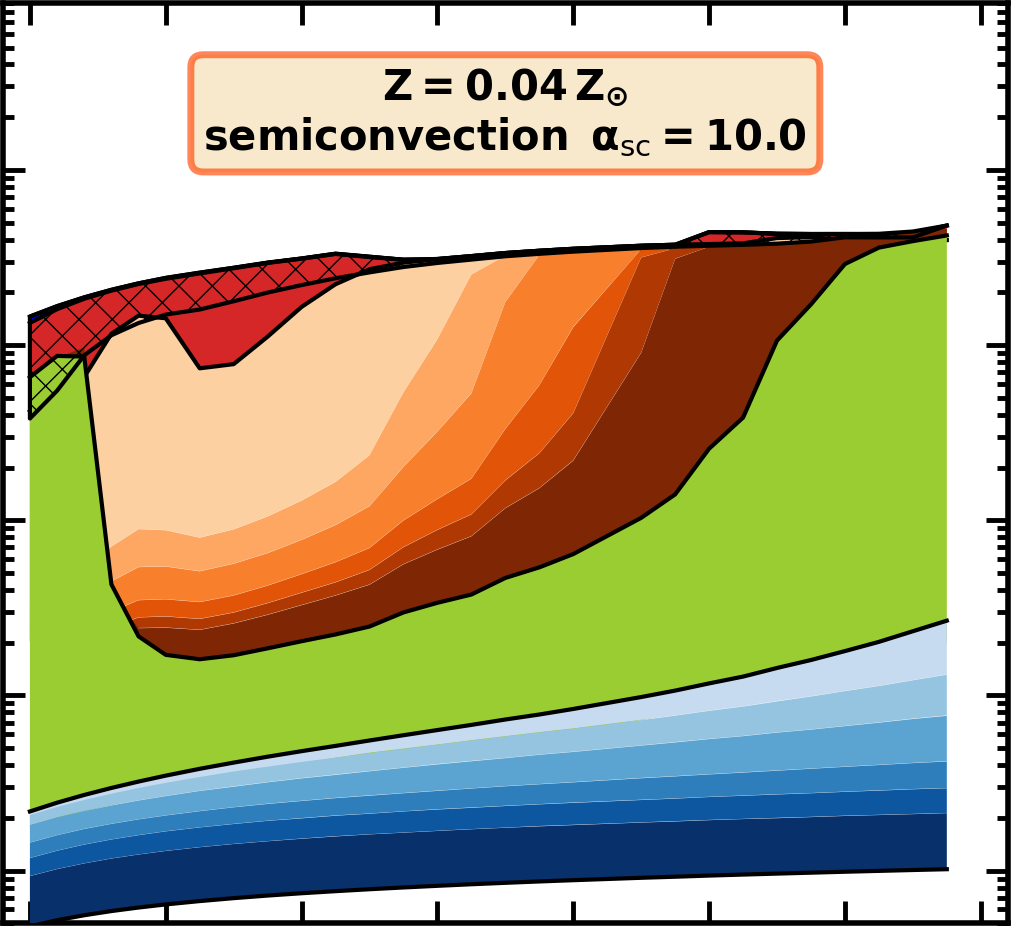} &
\includegraphics[width=0.24\textwidth,height=115px]{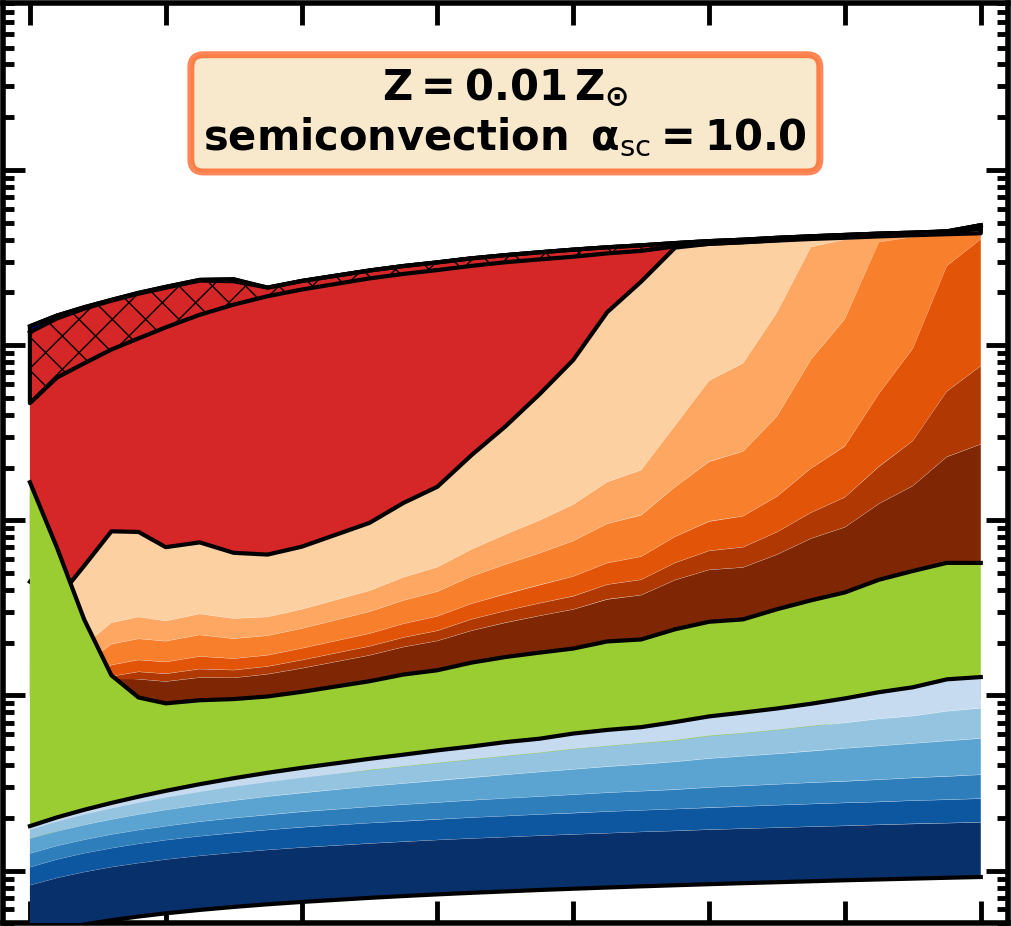} \\

\includegraphics[width=0.27\textwidth,height=115px]{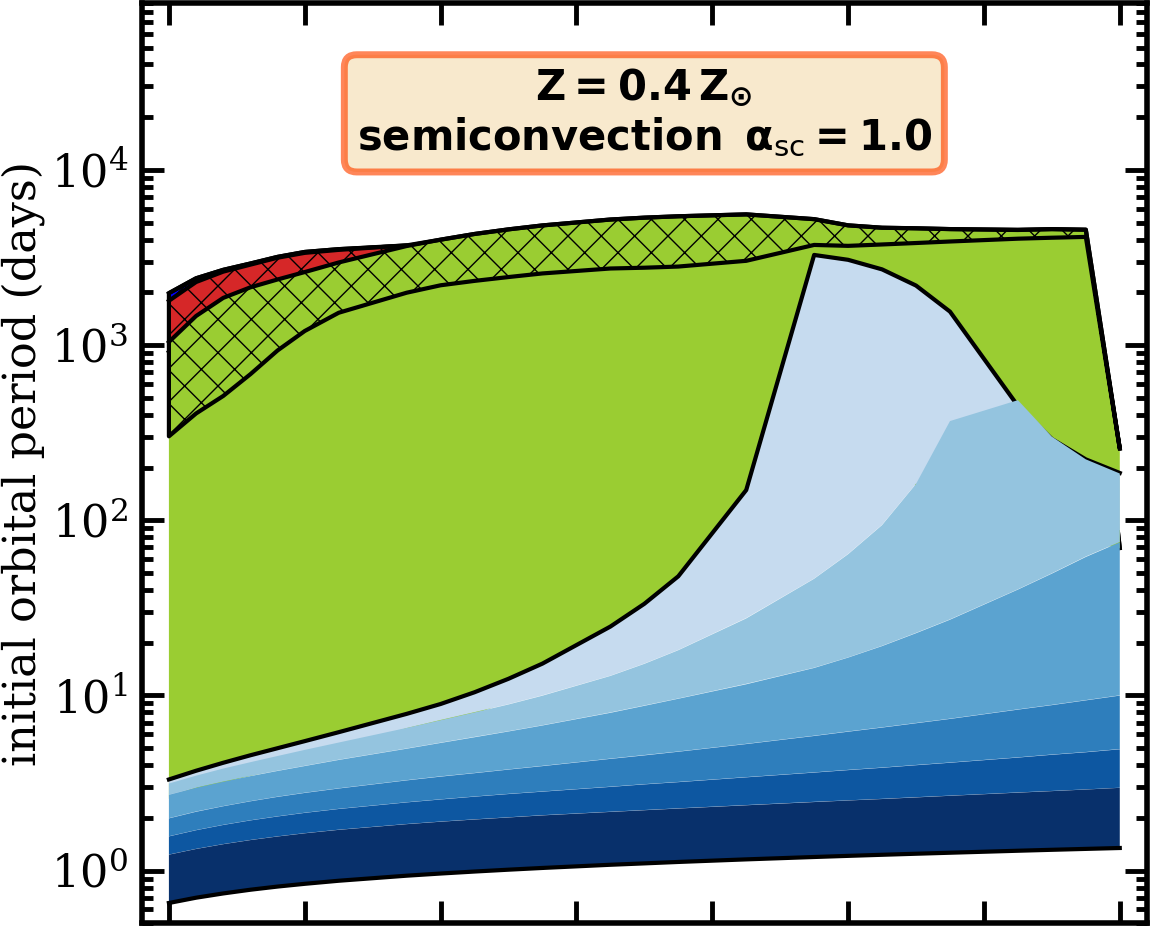} &
\includegraphics[width=0.24\textwidth,height=115px]{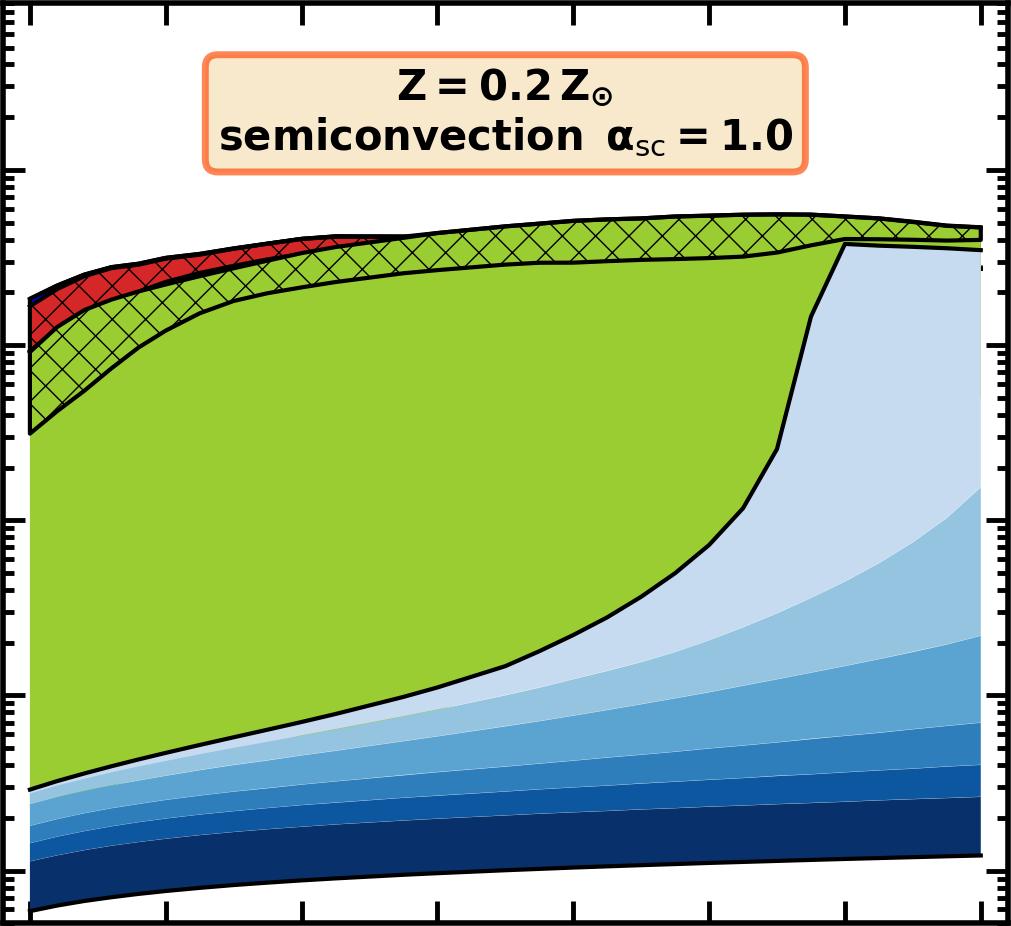} &
\includegraphics[width=0.24\textwidth,height=115px]{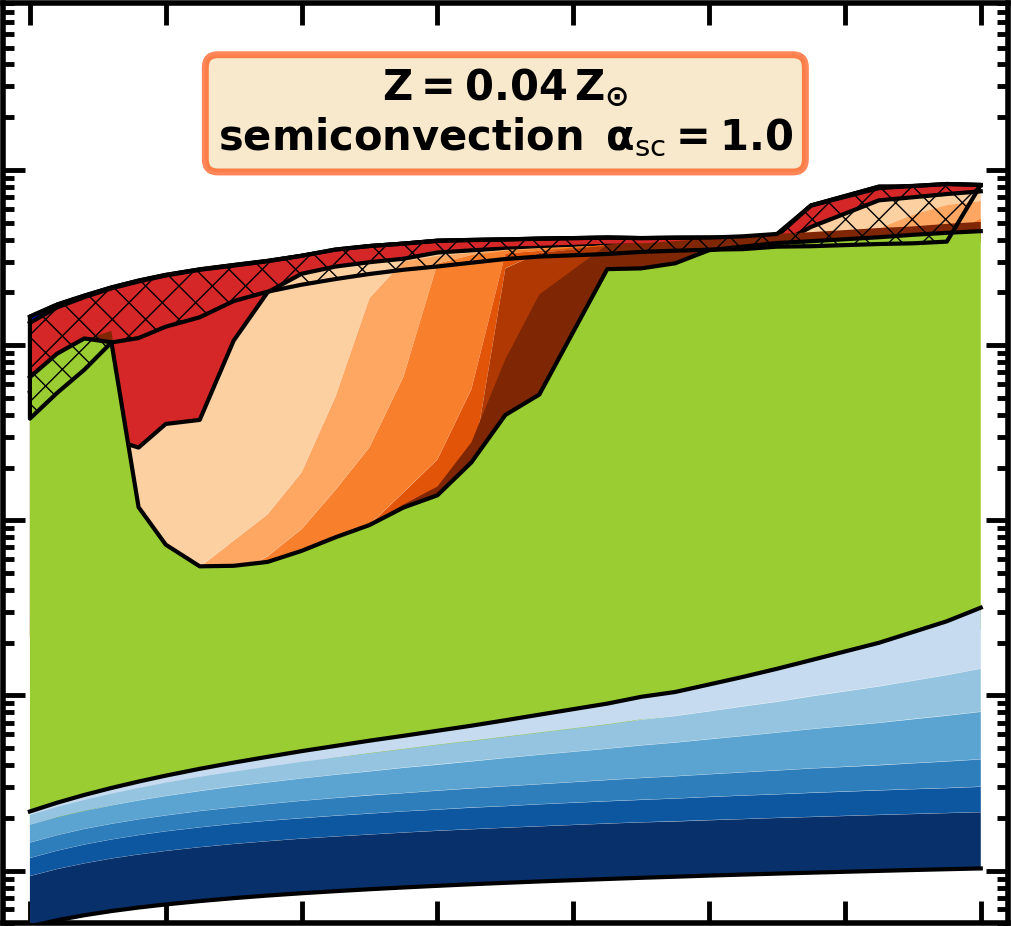} & 
\includegraphics[width=0.24\textwidth,height=115px]{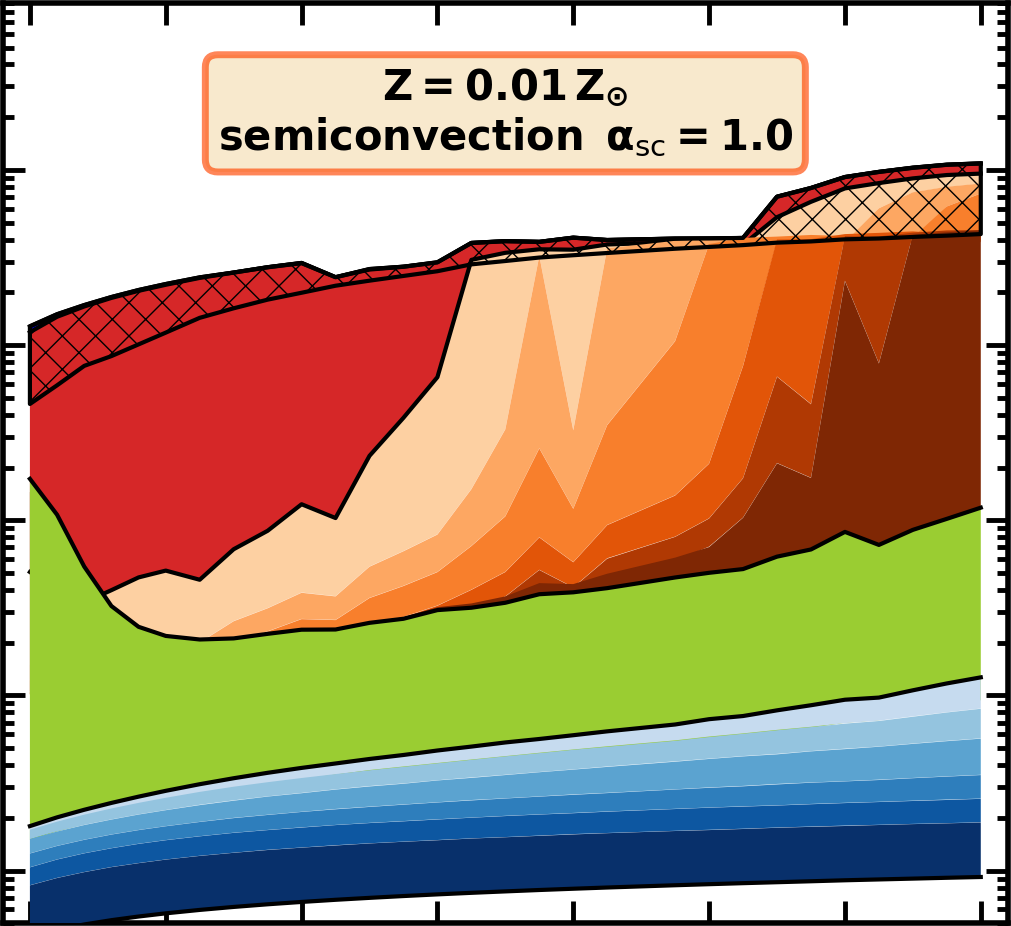} \\

\includegraphics[width=0.27\textwidth,height=125px]{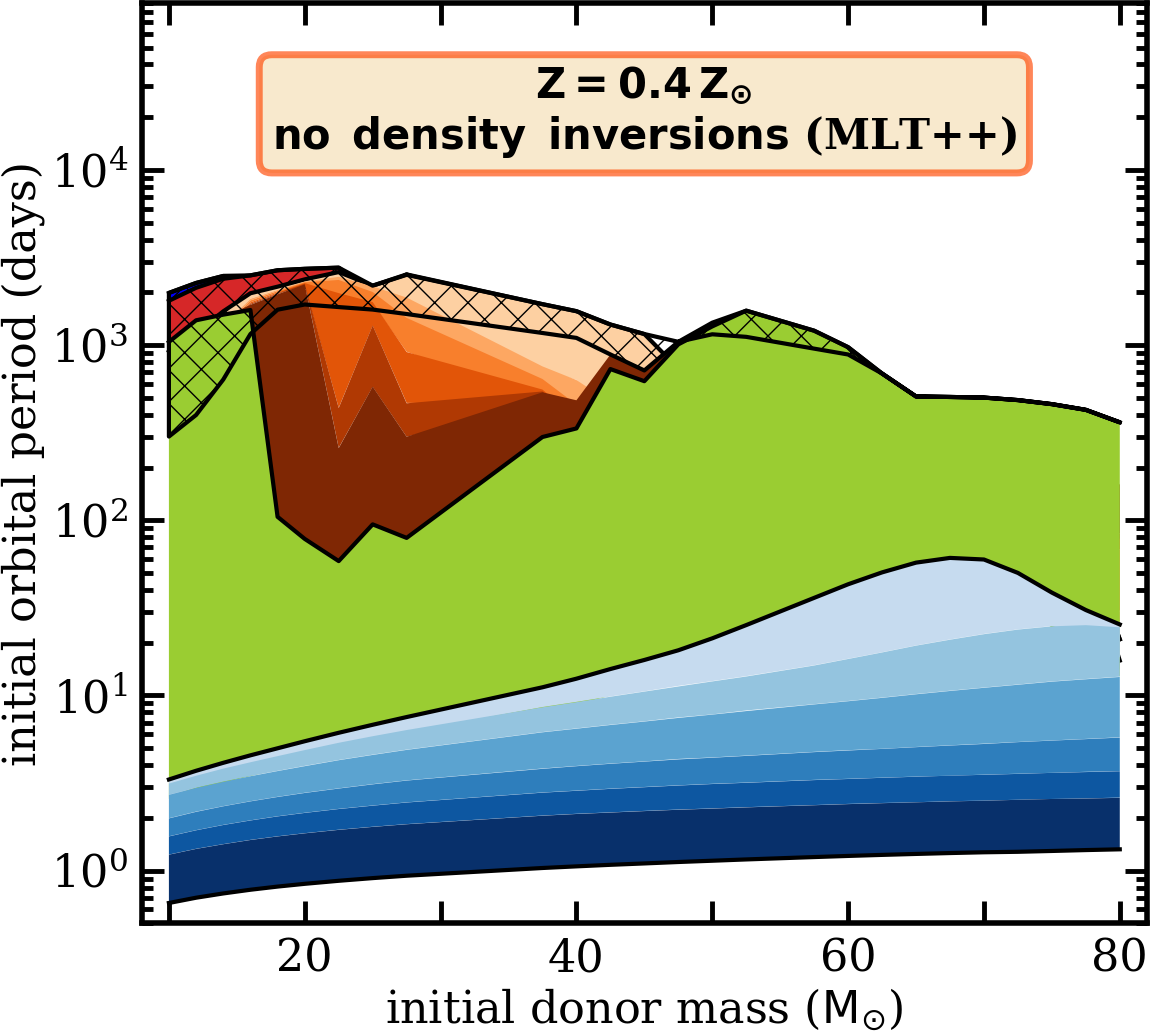} &
\includegraphics[width=0.24\textwidth,height=125px]{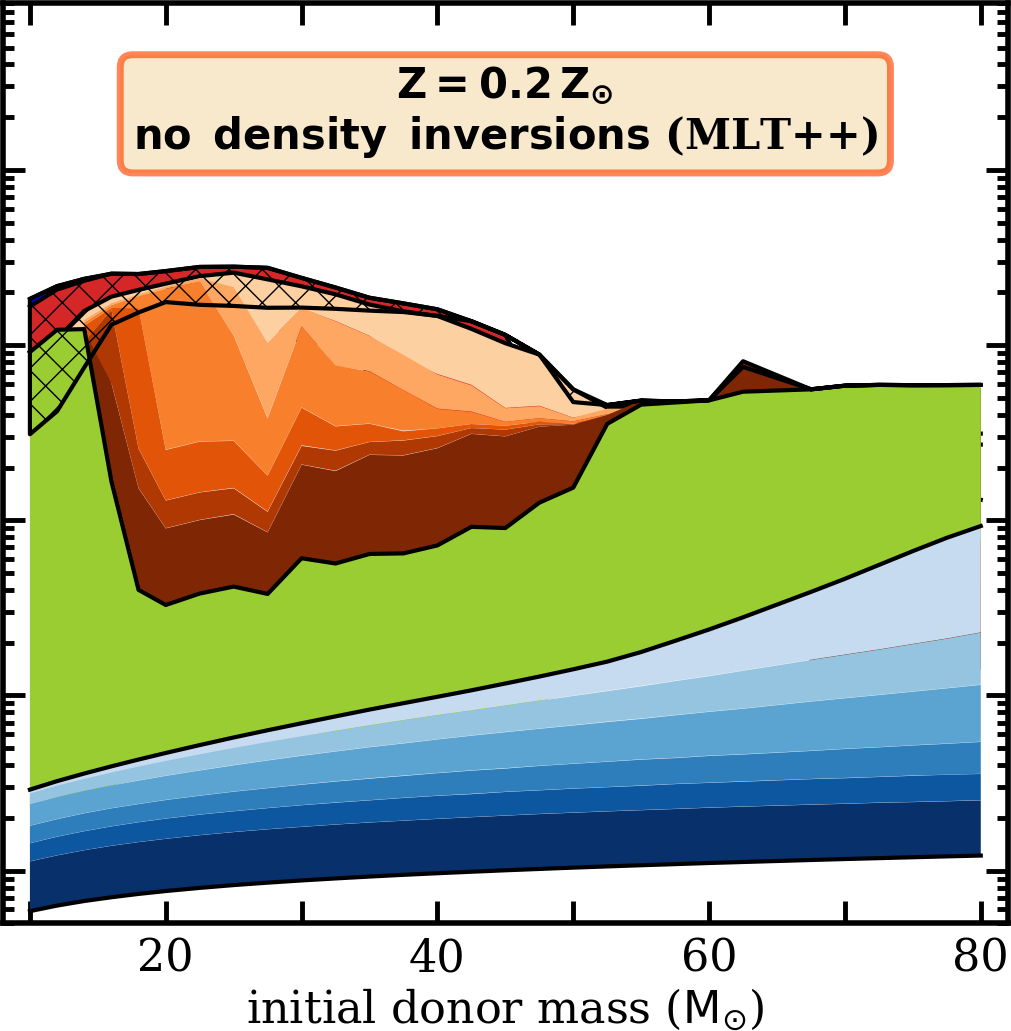} & 
\includegraphics[width=0.24\textwidth,height=125px]{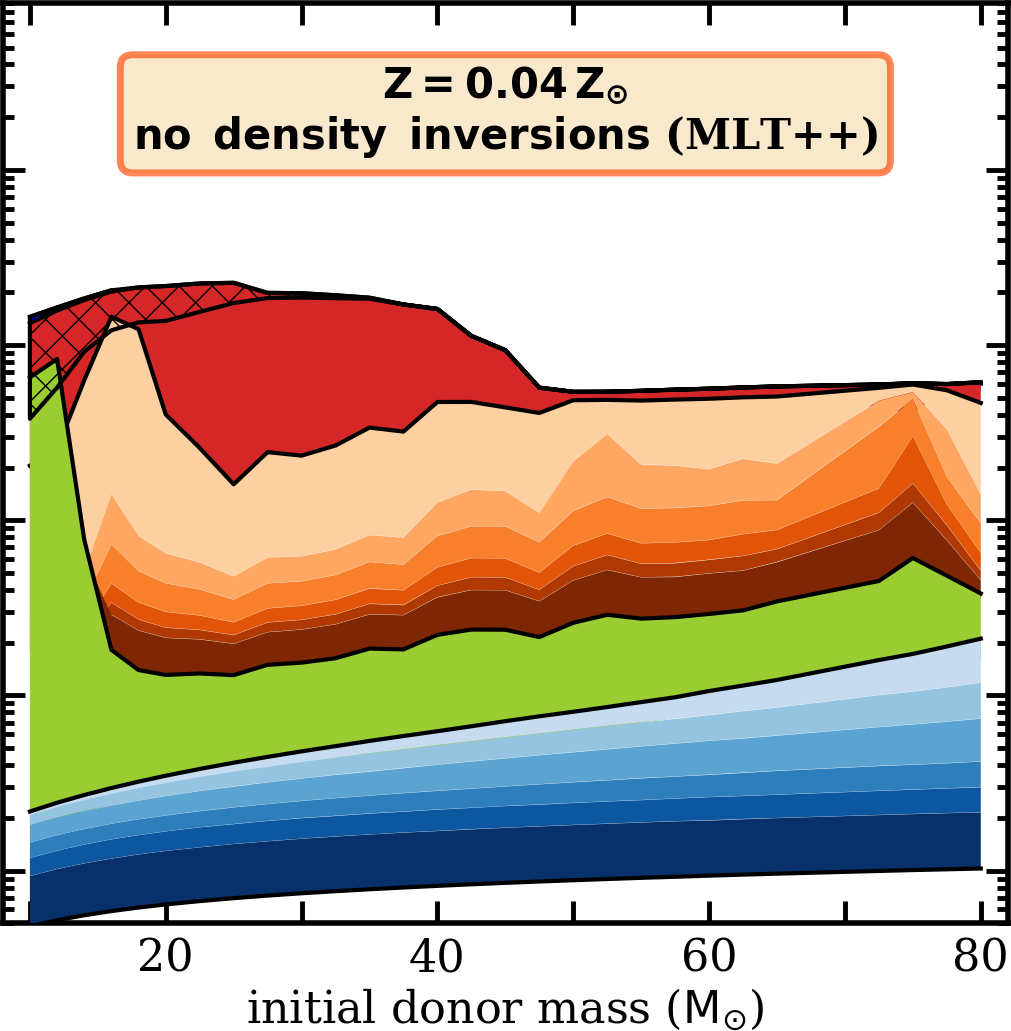} &
\includegraphics[width=0.24\textwidth,height=125px]{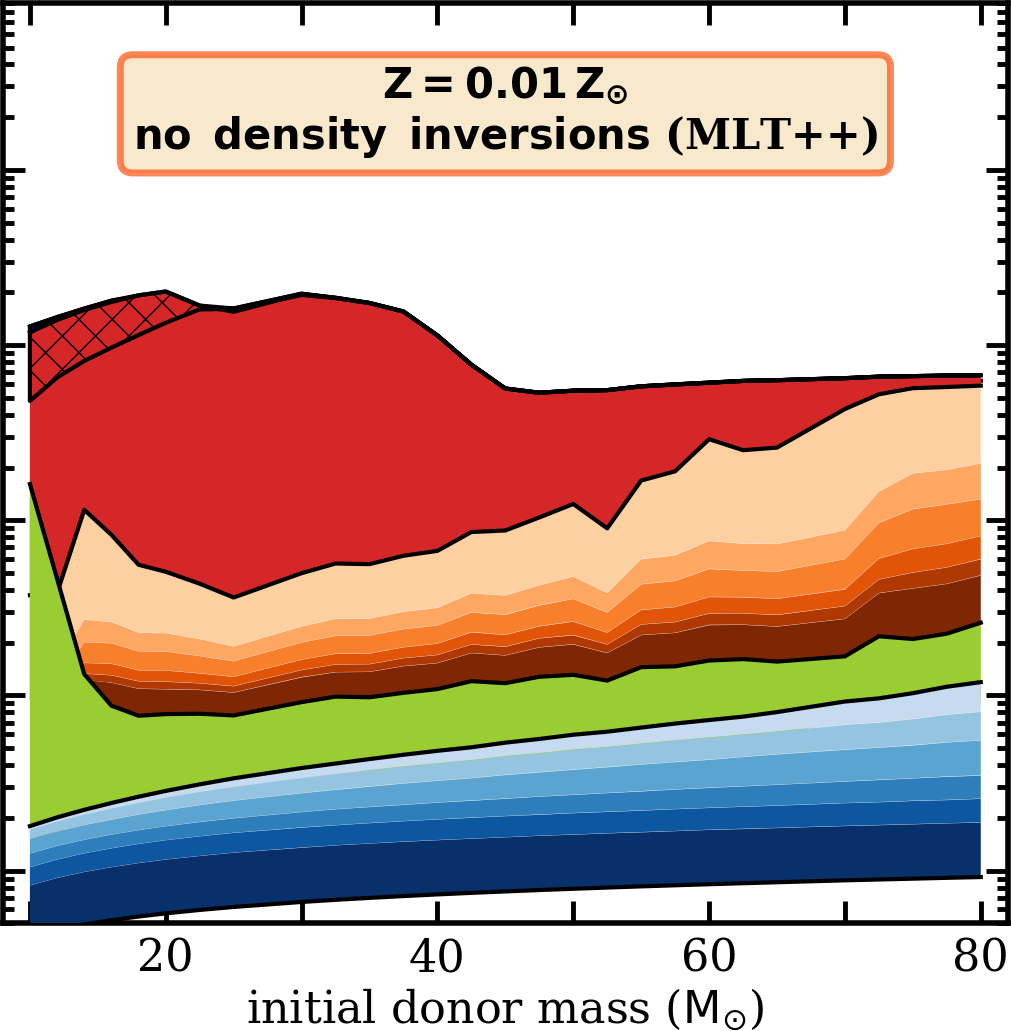} \\

\end{tabular}
\centering
\vspace{0.1cm}
\includegraphics[width=0.92\textwidth]{Figures_used/binary_par_ranges_legend.png}
\caption{Same as Fig.~\ref{fig.bin_param_variations}, but for other metallicities.} 
\label{fig.app_bin_param_variations}
\end{figure*}

\end{appendix}

\end{document}